\documentclass[structabstract]{aa}  
\usepackage{graphicx}
\usepackage{txfonts}
\usepackage{multirow}
\usepackage{amssymb}
\usepackage{natbib}
\def\kms{km\,s$^{-1}$}
\def\Ha{H{$\alpha$}}
\def\Hb{H{$\beta$}}

\def\ni{$^{56}$Ni}
\def\co{$^{56}$Co}
\def\fe{$^{56}$Fe}
\def\a{SN~1987A}
\def\em{SN~1999em}

\def\h{SN~1992H}
\def\od{SN~2007od}
\def\bw{SN~2009bw}
\def\dd{SN~2009dd}
\def\cs{SN~2005cs}

\def\s{SN~1998S}
\def\pk{SN~2007pk}
\def\aj{SN~2010aj}

\def\ad{SN~1995ad}

\def\ww{SN~1996W}

\def\mcento{mag\,(100d)$^{-1}$}

\def\M{M$_{\odot}$}

\def\ebv{E(B--V)}

\begin{document}
   \title{
Moderately luminous type II Supernovae
\thanks{This paper is based on observations made with the following facilities: the Italian Telescopio Nazionale Galileo, the Liverpool Telescope, the North Optical Telescope, the William Herschel (La Palma, Spain),  the Copernico telescope (Asiago, Italy), the Calar Alto Observatory (Sierra de los Filabres, Spain),  the orbital Telescope SWIFT (NASA), the Hale Telescope at the Palomar Observatory, and the ESO Telescopes at the La Silla and Paranal Observatories.}}

 %  \subtitle{I. Overviewing the $\kappa$-mechanism}

   \author{C. Inserra
             \inst{1,2,3}\thanks{E-mail:
c.inserra@qub.ac.uk (CI)}
          \and
          A. Pastorello\inst{4}
          \and
          M. Turatto\inst{4}
          \and
          M. L. Pumo\inst{4,5,6}
          \and
          S. Benetti\inst{4}
          \and
          E. Cappellaro\inst{4}
          \and
          M. T. Botticella\inst{7}
          \and
          F. Bufano\inst{8}
          \and
          N. Elias-Rosa\inst{9}
          \and
          A. Harutyunyan\inst{10}
          \and
          S. Taubenberger\inst{11}
          \and
          S. Valenti\inst{4}
          \and
          L. Zampieri\inst{4} }

   \institute{Astrophysics Research Centre, School of Mathematics and Physics, Queen's University Belfast, Belfast BT7 1NN, United Kingdom
\and
Dipartimento di Fisica ed Astronomia, Universit\`a di Catania, Sezione Astrofisica, Via S.Sofia 78, 95123, Catania, Italy
   \and
   INAF - Osservatorio Astrofisico di Catania, Via S.Sofia 78, 95123, Catania, Italy
\and
INAF - Osservatorio Astronomico di Padova, Vicolo dell'Osservatorio 5, 35122, Padova, Italy
\and
Bonino-Pulejo Foundation, Via Uberto Bonino 15/C, I-98124 Messina, Italy
\and
Universit\`a di Padova, Dipartimento di Fisica ed Astronomia ÒG. GalileiÓ, Vicolo dellÕOsservatorio, 3, 35122 Padova
\and
INAF - Osservatorio astronomico di Capodimonte, Salita Moiariello 16, I- 80131 Napoli, Italy
\and
Departamento de Ciencias Fisicas, Universidad Andres Bello, Avda. Republica 252, Santiago, Chile
\and
Institut de Ci\`encies de lÕEspai (IEEC-CSIC), Facultat de Ci\`encies, Campus UAB, 08193 Bellaterra, Spain
\and
Fundaci\'on Galileo Galilei-INAF, Telescopio Nazionale Galileo, Rambla Jos\'e Ana Fern\'andez P\'erez 7, 38712 Bre\~na Baja, TF - Spain
\and
Max-Planck-Institut f\"ur Astrophysik, Karl-Schwarzschild-Str. 1, 85741 Garching, Germany
\\}

   \date{Received ...; accepted ...}

% \abstract{}{}{}{}{} 
% 5 {} token are mandatory
 
  \abstract
  % context heading (optional)
  % {} leave it empty if necessary  
   {Core-collapse Supernovae (CC-SNe) descend from progenitors more massive than about 8 \M. Because of the young age of the progenitors,  the ejecta may eventually interact with the circumstellar medium (CSM)
via highly energetic processes detectable in the radio, X--ray, ultraviolet (UV) and, sometimes, in the optical domains.}
  % aims heading (mandatory)
   {In this paper we present ultraviolet,  optical and near infrared observations  of five type II SNe, namely SNe 2009dd, 2007pk, 2010aj, 1995ad, and 1996W. Together with few other SNe they form a group of moderately luminous type II events. We investigate the photometric similarities and differences among these bright objects. We also attempt to characterise them by analysing the spectral evolutions, in order to 
find some traces of CSM-ejecta interaction.} 
  % methods heading (mandatory)
   {We collected photometry and spectroscopy with several telescopes in order to construct well-sampled light curves and spectral evolutions from the photospheric to the nebular phases. Both photometry and spectroscopy indicate a degree of heterogeneity in this sample.
   Modelling the data of SNe 2009dd, 2010aj and 1995ad allows us to constrain the explosion parameters and the properties of the progenitor stars.}
   %and compare the inferred estimates with those available for SNe 2007od and 2009bw.}
  % results heading (mandatory)
   {The light curves have luminous peak magnitudes ($-16.95<M_{B}<-18.70$).  The ejected masses of \ni\/ for three SNe span a wide range of values ($2.8\times10^{-2}$\M$<$M(\ni)$<1.4\times10^{-1}$\M), while for a fourth (\aj) we could determine a stringent upper limit ($7\times10^{-3}$\M).
 Clues of interaction, such as the presence of high velocity (HV) features of the Balmer lines, are visible in the photospheric spectra of SNe 2009dd and 1996W. 
 %not for 1995ad !!! MT 
 For \pk\/ we observe a spectral transition from a type IIn to a  standard type II SN. Modelling the observations of SNe 2009dd, 2010aj and 1995ad with radiation hydrodynamics codes, we infer kinetic plus thermal energies of about 0.2--0.5 foe, initial radii of 2--5$\times10^{13}$ cm and ejected masses of $\sim$5.0--9.5 \M\/. }
  % conclusions heading (optional), leave it empty if necessary 
   {These values suggest moderate-mass, super-asymptotic giant branch (SAGB) or red super-giants (RSG) stars as SN precursors, in analogy with other luminous type IIP SNe 2007od and 2009bw.} 
%AP: questa la rimuoverei -Thefore, the SNe analysed here share only the high luminosity, consistent with the bright branch of type II SNe. }
   %In addition to the bright peak luminosity there are not common features characterizing the objects of the sample here studied. }
   
   \keywords{ supernovae: general - supernovae: individual: SN 2009dd - supernovae: individual: SN 2007pk - supernovae: individual: SN 2010aj - supernovae: individual: SN 1995ad - supernovae: individual: SN 1996W }

   \maketitle
%
%________________________________________________________________

\section{Introduction}

Type II Supernovae (SNe) are a very heterogeneous class of stellar explosions that stem from the collapse of the core of massive stars \citep[ZAMS mass $\gtrsim$8 \M, e.g.][and reference therein]{sm09,pumo09}, in most cases a red supergiant (RSG). 
Stars with H-rich envelope at the explosion are thought to produce type II ``plateau" (IIP) SNe \citep{bar79}, which show a nearly constant luminosity (plateau phase) lasting up to 4 months, during which the envelope recombines, releasing the internal energy. The length of the plateau primarily depends on the envelope mass \citep[e.g. ][]{pumo11}. If the H envelope mass is very low, the light curve shows a linear, uninterrupted decline after maximum. These SNe are historically  called as type II ``linear''  \citep[SNe IIL,][]{bar79}.
Intermediate cases have been found with light curves  showing less pronounced plateaus, e.g. SNe 1992H \citep{cl96}. 

A common feature of SN IIP and IIL is the linear tail of the late light curve powered by the energy release of the radioactive decay of \co\/ to \fe\/ with the characteristic slope of 0.98 mag/100d, indicating complete $\gamma$-ray and e$^+$ trapping.
In some cases the observed decline rate is significantly modified by 
dust formation within the ejecta, which absorbs light at optical wavelengths and re-emits photons in the near infrared
(NIR), by the interaction of the ejecta with the circum-stellar medium (CSM) which converts kinetic energy into radiation, {or by incomplete $\gamma$-ray trapping}.
%. This increases the opacity and makes optical light curves to decline faster.
%A second effect is governed by the presence of circumstellar medium (CSM) may interact with the 
%fast SN ejecta transforming part of the kinetic energy in to radiation thus providing additional energy to the luminosity slowing down the luminosity decline.

During the first few days after the explosion the spectra of most SNe~IIP and IIL approximate a black-body from
UV through IR wavelengths. The spectra become progressively dominated by broad P-Cygni profiles of Balmer lines with a
strong \Ha\/ emission, while metal lines arise during the plateau. The late-time spectra are dominated by relatively narrow emission lines of H and prominent forbidden transitions of Ca~{\sc ii}, O~{\sc i}, Fe~{\sc ii} and MgI \citep[e.g.][]{tu03a}.

A different subclass of type II SNe is constituted by objects showing narrow emission lines already at early phases \citep[SNe IIn,][]{sc90}.
Their spectral appearance and slow luminosity evolution are attributed to the interaction  of the fast ejecta
with a slowly expanding, dense CSM which generates a forward shock in the CSM
and a reverse shock in the ejecta. The shocked material emits energetic radiation
whose spectrum depends primarily on the density and velocity of both the CSM and the
ejecta \citep{ch94}. 
%Sometimes the pressure and
%temperature behind the shock are sufficiently high that the post-shock ejecta
%and CSM become powerful X-rays emitters. At the same time synchrotron
%radiation is generated by electrons accelerated up to relativistic energies at the shock front. 
Thus the study of SNe IIn provides clues to the mass-loss history of their progenitors.
Typical mass loss rates are of the order of $10^{-6}-10^{-5}$ \M\/yr$^{-1}$, but this value can increase significantly
and exceed $10^{-4}$ \M\/yr$^{-1}$ \citep[e.g. in SNe 1988Z and 1995N, ][]{chugai94,za05}.

Sometimes normal SNe II, most often of the linear sub-type, e.g. SNe 1979C \citep[][ and references therein]{mili09}, 
1980K \citep[][ and references therein]{mili12}, 1986E \citep{cap95} show the onset of ejecta-CSM interaction at late stages, after an otherwise normal evolution from the photospheric to nebular phase. This is interpreted as evidence that the ejecta, after a phase of free expansion, reach a dense gas shell ejected by the progenitor a few $10^{2}-10^{3}$ yr before the explosion.
Recently also a few SNe with overall normal spectral features have shown weak but unequivocal evidence of ejecta-CSM interaction from early times, reviving the interest for the studies of interacting SNe.  %, e.g. %type Ia SNe 2002ic \citep{h03}, 2003du \citep{ge04}, 2002bo \citep{maz05} and 
Among them we recall the type IIP SNe 1999em and 2004dj \citep{chu07} or the atypical type II 
 %\gl\/ \citep{gy07,gy09} {\bf NON LA DEFINIREI "WEAK"; E' UNA FAST-EVOLVING, MA PUR SEMPRE UNA IIn}, 
 \od\/ \citep{07od} and \bw\/ \citep{09bw}, the last two belonging to the bright tail of the type IIP SN luminosity distribution.

In this paper we present the spectroscopic and photometric observations of five bright objects, the type II SNe 2009dd, 2007pk and 2010aj, plus unpublished archival data of SNe~1995ad and 1996W. %Together with other four SNe (SNe 1992H, 2004et, 2007od and 2009bw) these make up almost all what %is available for bright, but otherwise normal,  type II SNe.
%The common feature of these objects are the high peak luminosity and the short plateau duration. % which is shorter than that of other type II  SNe \citep{pa94,ri02}. 
%Some of the SNe here presented should be added to the number of overall normal core-collapse SNe with signature of early but not dominant ejecta-CSM interaction.
The plan of the paper is the following:
in Sect.~\ref{sec:sne} we introduce the SNe and their host galaxy properties, estimating distances and reddening. Photometric data, light and colour curves as well as the estimates of \ni\/ masses are presented in Sect.~\ref{sec:phot}; in Sect.~\ref{sec:spec} we describe and analyse the spectra; a discussion is presented in Sect.~\ref{sec:dis}, while a short summary follows in Sect.~\ref{sec:final}.

Throughout the paper we adopt H$_{0}$~=~73~\kms\/Mpc$^{-1}$, $\Omega_{\rm m}=0.27$ and $\Omega_{\rm \lambda}=0.73$.

%__________________________________________________________________

\section{SNe and their host galaxies}\label{sec:sne}
In this Section the selected SNe and their host galaxies are presented individually.

%The main data of each SN and the host galaxy are summarised in Tab.\ref{table:main}.

\begin{figure}
%\vspace{174pt}
\includegraphics[width=\columnwidth]{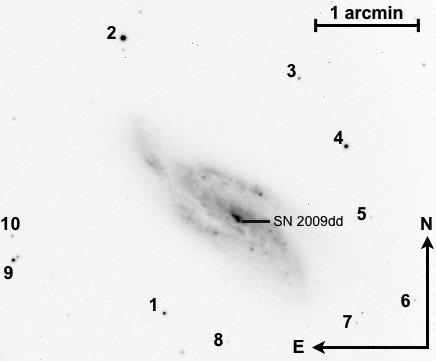}
\caption{R band image of SN 2009dd in NGC 4088 obtained with  CAHA+CAFOS on November 19th, 2009. The sequence of stars used to calibrate the optical and NIR magnitudes is indicated.} 
\label{fig:09dd}
\end{figure}

\begin{enumerate}
\item \dd\/ (Fig.~\ref{fig:09dd}) was discovered in the Sbc galaxy NGC~4088 \citep{hak12} by \citet{co09} on 2009 April 13.97. \citet{ner09a} classified the object as a young type II SN with strong Na~ID interstellar features suggesting significant absorption inside the parent galaxy. 

Prompt observations with Swift+XRT revealed an X-ray source at the optical position of the SN with $4.5 \sigma$ significance \citep{im09}. Over the course of the Swift XRT observations, which lasted a month, the X-ray source continuously brightened from $8 \times 10^{38}$ erg s$^{-1}$ to $1.7 \times 10^{39}$ erg s$^{-1}$ \citep[in the range 0.2-10 keV,][]{im09}. During the same period no radio emission was detected at the SN position with $3\sigma$ upper limits of 0.35 mJy at 1.3 cm, and of 0.15 mJy at 3.5 cm \citep{stoc09}.

NED provides the velocity of the host galaxy corrected for Virgo infall, $v_{\rm Virgo}=1025\pm15$ \kms\/  \citep[from][] {mo00},
%an heliocentric radial velocity of $V_{hel}$(NGC4088)$=757\pm1$ \kms\/ and .
%Adopting H$_{0}$=73 \kms\/Mpc$^{-1}$, we obtain 
corresponding to a distance modulus $\mu=30.74\pm$0.15 mag (d$\sim$14.0 Mpc). 

The coordinates of \dd, measured  on our astrometrically calibrated images on two different epochs, are $\alpha = 12^{h}05^{m}34^{s}.10 \pm0^{s}.05$, $\delta = +50\degr32\arcmin19\arcsec.40 \pm0\arcsec.05$ (J2000). The object is located in the galaxy inner region, $1\arcsec.5$ West and $4\arcsec$ South of the nucleus of  NGC~4088. This position, slightly revised with respect to the previous determination \citep{co09}, corresponds to a linear distance of $\sim$0.3 kpc from the nucleus, deprojected as in \citet{hak09}.

The Galactic reddening toward NGC 4088 is E$_{\rm g}$(B-V)~=~0.02 mag \citep[A$_{\rm g}$(B)~=~0.085 mag,][]{sc98}. In our best resolution optical spectra (cfr. Sect.~\ref{ss:09dd}), the interstellar Na~{\sc iD} ($\lambda\lambda$5890,5896) lines of the Galaxy are seen with average EW$_{\rm g}$(Na~{\sc iD})$\sim$0.13 \AA.
According to \citet{tu03b} this corresponds to a galactic reddening E$_{\rm g}$(B-V)$\sim$0.02 mag, exactly the same as the \citet{sc98} estimate.
With the same method we estimate the reddening inside the parent galaxy.
The corresponding interstellar Na ID components have an average equivalent width EW$_{\rm i}$(Na~{\sc iD})$\sim2.7$ \AA, providing an E$_{\rm i}$(B-V)$\sim$0.43 mag or A$_{\rm i}(B)\sim$1.81 mag.
We may notice, that while there have been conflicting reports on the reliability of the NaID line as tracer of dust extinction, a most recent analysis of \citet{poz:12}, on a large sample of SDSS galaxy spectra, basically confirmed the strong correlation between the two quantities.
Therefore we have adopted a total reddening to \dd\/ E$_{\rm tot}$(B-V)$=0.45$ mag, consistent with the position of the SN inside the parent galaxy and what reported in \citet{ner09a}.

A reasonable assumption to estimate the metallicity  is to consider that the SN has the same metallicity of the closest H~{\rm ii} region.
Extracting the spectrum of the region close to SN %along the slit of the spectrograph 
from the latest, deep observation of \dd, we have determined the N2 index \citep{pp04} ~to be N2= $-0.54$.  The relation (1) of \citet{pp04} then provides the O abundances which turns out to be $12+log(O/H)=8.59\pm0.06\pm0.41$ (where the first error is statistical and the second is the 95\% spread of the N2 index calibration relation), close to the solar abundance \citep[8.69,][]{asp09}. 
%%  MT:INUTILE E COMPLESSO>   Following \citet{loes10} this is equivalent to $Z\sim0.019$.

\item  \pk\/ (Fig.~\ref{fig:07pk}), discovered in the Scd galaxy NGC~579 \citep{hak12} on 2007 November 10.31 UT,  has been classified as a young ``peculiar" type IIn SN resembling SN 1998S at early phases \citep{par07}. \citet{im07a} reported a bright X-ray source within $23\arcsec.5$ from the SN position although, due to the large point-spread-function of the XRT instrument ($18\arcsec$ half-power diameter at 1.5 keV) the error box include the galaxy nucleus. An X-ray flux of ($2.9\pm0.5)\times10^{-13}$ erg cm$^{-2}$ s$^{-1}$ and a luminosity of ($1.7\pm0.3)\times10^{40}$ erg s$^{-1}$ have been calculated. No radio emission has been detected with VLA in the 8.46 GHz band \citep{cha07}.

\begin{figure}
%\vspace{174pt}
\includegraphics[width=\columnwidth]{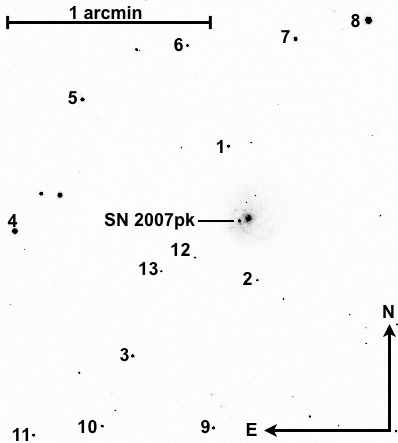}
\caption{R band image of SN 2007pk in NGC 579 obtained with NOT+ALFOSC on January 11th, 2008. The sequence of stars in the field used to calibrate the optical magnitude of the 2007pk is indicated.} 
\label{fig:07pk}
\end{figure}

NED provides a recession velocity of NGC~579 corrected for Virgo infall of $v_{\rm Virgo}=5116\pm16$ \kms\/  \citep[from][]{mo00} 
%$V_{hel}$(NGC579)$=4993\pm4$ \kms
%Adopting H$_{0}$=73 \kms\/Mpc$^{-1}$,
corresponding to a distance modulus $\mu=34.23\pm$0.15 mag.

The coordinates of \pk, $\alpha=01^{h}31^{m}47^{s}.07\pm0^{s}.04$ and $\delta= +33\degr36\arcmin54\arcsec.70\pm0\arcsec.04$ (J2000) measured on our astrometrically calibrated images are in fair agreement ($\Delta(\delta) = 0\arcsec.6$) with those provided by \citet{par07}. The object is located in an inner region of the spiral parent galaxy, $7\arcsec.4$ East and $1\arcsec.6$ South  \citep[slightly revised with respect to the determination of][]{par07} from the nucleus of NGC 579. The position of SN corresponds to a linear deprojected distance of $\sim$2.5 kpc from the nucleus  \citep[cfr.][]{hak09}.

The Galactic reddening toward NGC 579 was estimated as E$_{\rm g}(B-V)=0.05$ mag \citep[A$_{\rm g}(B)=0.22$ mag,][]{sc98}. We have measured the intensity of the interstellar Na~{\sc iD} lines of the Galaxy in our best resolution spectra, finding an average EW$_{\rm g}$(Na~ID)$\sim$0.57 \AA. This corresponds to a galactic reddening of E$_{\rm g}$(B-V)$\sim$0.09 mag (A$_{\rm g}$(B)$\sim$0.38 mag) according to \citet{tu03b}, 1.8 times larger than the above-mentioned estimate but still within the large uncertainty of the method. In analogy we estimated the reddening inside the parent galaxy with the Na~{\sc iD} components of the host galaxy. The derived EW$_{\rm i}$(Na~{\sc iD})$\sim$0.33 \AA\/ corresponds to reddening E$_{\rm i}$(B-V)$\sim$0.05 mag or A$_{\rm i}$(B)$\sim$0.22 mag, about three times less than the admittedly crude estimate of \citet{pri12}.
%Missing other methods of reddening determination, 
Throughout this work we have adopted a total reddening to \pk\/  E$_{\rm tot}(B-V)=0.10$ mag.

As for \dd\/ we have measured the emission lines of the region adjacent to the SN along the slit and
determined the index $N2=-0.70$ corresponding to $12+log(O/H) = 8.50\pm0.05\pm0.41$, again close to the solar value. %\citep[Z$\sim$0.018,][]{loes10}.

\item  \aj\/ was discovered in the Sc galaxy \citep{hak12} MGC-01-32-035 by \citet{ne10} on 2010 March 12.39 UT and was classified as a young type II SN resembling the type IIP SN 2006bp near maximum brightness \citep{ce10}.
The recession velocity of MGC-01-32-035 corrected for the Virgo infall is $v_{\rm Virgo}=6386\pm20$\kms \citep[][from NED]{mo00}, corresponding to a
distance modulus $\mu=34.71\pm0.15$ mag.

\begin{figure}
%\vspace{174pt}
\includegraphics[width=\columnwidth]{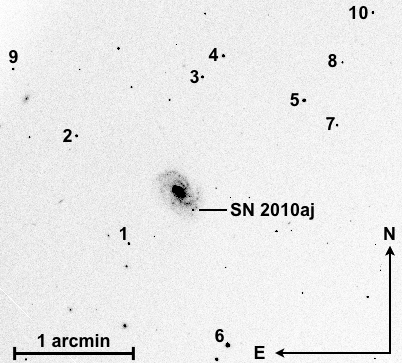}
\caption{R band image of SN 2010aj in MGC-01-32-035 obtained with TNG+DOLORES on May 22th, 2010. The sequence of stars in the field used to calibrate the optical and NIR magnitudes of SN 2010aj is indicated.} 
\label{fig:10aj}
\end{figure}

The coordinates of \aj\/ have been measured on our images at $\alpha= 12^{h}40^{m}15^{s}.16\pm0^{s}.05$, $\delta= -09\degr18\arcmin14\arcsec.30\pm0\arcsec.05$ (J2000). The object is located $12\arcsec.4$ West and $11\arcsec.7$ South of the centre of the SABbc: parent galaxy, (Fig.~\ref{fig:10aj}). The SN is centred on an H~II region that becomes dominant after 350d, as clearly seen from the SN spectral evolution in Sec.~\ref{sec:spec}.
The linear deprojected distance is $\sim7.2$ kpc from the nucleus.% \citep[cfr.][]{hak09}.

The Galactic reddening in the direction to MGC -01-32-035 was estimated as E$_{\rm g}$(B-V)$=0.036$ mag \citep[A$_{\rm g}$(B)$=0.148$ mag,][]{sc98}. 
The available spectra do show neither the Na~{\sc iD} lines of the parent galaxy nor those of the Galactic component.
%Maybe due to a combination of resolution and presence of metal lines. 
Throughout this paper, we will adopt a total reddening to \aj\/ of E$_{\rm tot}$(B-V)$=E_{\rm g}$(B-V)$=0.036$ mag, entirely due to the Galaxy.

Also for this SN we measured the N2 index, N2(\aj)$=-0.47$, providing  a metallicity of $12+log(O/H) = 8.63\pm0.06\pm0.41$, close to the solar value.
%corresponding to Z$\sim0.02$.

\item  \ad\/ (Fig.~\ref{fig:95ad}) was discovered by \citet{ev95} on 28.8 UT of September in the SBc galaxy NGC~2139 \citep{hak09a}. Based on a spectrum collected the day after with the ESO 1.5-m telescope in La Silla it was classified as a type II close to maximum because of broad P-Cygni profiles of Balmer and He I lines lying on a blue continuum \citep[T$_{\rm bb}\sim13000$ K,][]{ev95}. 
NED provides a heliocentric radial velocity of NGC~2139 corrected for the Virgo Infall of $v_{\rm Virgo}=1674\pm14$ \kms, from which we infer a distance modulus $\mu=31.80\pm0.15$ mag.

%The coordinates of the SN have been provided by \citet{mcn95} as  $\alpha=06^{h}01^{m}06^{s}.13$ and $\delta=-23^{\degr}40\arcmin29\arcsec.0$ (J2000), $25\arcsec$ West and $5\arcsec$ South of the nucleus of NGC 2139. %The pre-discovery detection (magnitude $\sim$17.5) obtained on September 22 \citep{bro98} and the constraints given by spectroscopy lead to estimate the explosion date to be about JD = 2449981$\pm$3 (September 20).
The coordinates of the SN, measured on our astrometrically calibrated images on two different epochs, are $\alpha=06^{h}01^{m}06^{s}.21 \pm0^{s}.05$ and $\delta=-23\degr40\arcmin28\arcsec.90 \pm0\arcsec.05$ (J2000). The object is located in an arm of the host galaxy, $25\arcsec$ West and $5\arcsec$ South of the nucleus of NGC 2139. This position, slightly revised with respect to a previous determination \citep{mcn95}, corresponds to a deprojected distance of $\sim2.8$ kpc from the nucleus.

\begin{figure}
%\vspace{174pt}
\includegraphics[width=\columnwidth]{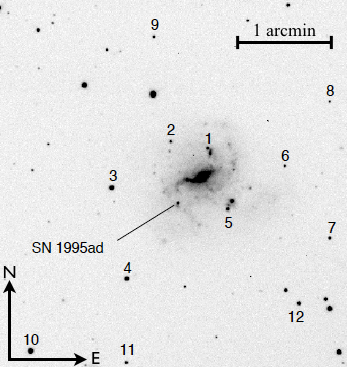}
\caption{R band image of \ad\/ in NGC 2139 obtained with ESO 3.6m+EFOSC1 on December 29th, 1995. The sequence of stars in the field used to calibrate the optical magnitudes of SN 1995ad is indicated.} 
\label{fig:95ad}
\end{figure}

The Galactic reddening has been estimated as  $E_{\rm g}$(B-V)$=0.035$ mag \citep[i.e. A$_{\rm g}$(B)$=0.145$ mag,][]{sc98}. The Na ID interstellar lines associated to the parent galaxy are not visible in the SN spectra. Therefore, hereafter, we consider only the Galactic contribution. 

To obtain the metallicity of the environment of this SN, we analysed the spectra obtained at the ESO1.5m telescope on February 19 and 20, 1996 because of the better resolution than the latest available spectra. We measured the O3N2 and N2 indices \citep{pp04} of an H~{\sc ii} region %extracted
 close to the SN. %along the slit of the spectrograph. 
The average relations then provide $12+log(O/H) = 8.60\pm0.05\pm0.41$, very close to solar.
%corresponding to Z$\sim0.02$.

\item \ww\/ (Fig.~\ref{fig:96w}) was discovered on April 10 UT and confirmed the following night at the Beijing Astronomical Observatory (BAO) \citep{li96} as a type II SN soon after the explosion, showing a blue continuum with strong and broad \Ha\/ ($v\sim14300$ \kms\/) and \Hb.
%Nothing was visible at the SN position in an image obtained on March 1. 
The recession velocity of the host SBc galaxy \citep{hak09a} NGC~4027 corrected for Virgo infall is $v_{\rm Virgo}=1779\pm29$ \kms \citep[][from NED]{mo00}, corresponding to a distance modulus $\mu=31.93\pm0.15$ mag. % (adopting H$_{0}=73$ \kms\/Mpc$^{-1}$).

%The position of the SN has been given by \citet{su96} at $\alpha=11^{h}59^{m}28^{s}.88$ and $\delta=-19^{\degr}15\arcmin21\arcsec.90$ (J2000), $17\arcsec$ West and $34\arcsec$ North of the nucleus of the parent galaxy NGC 4027. 
%Thanks to the spectroscopic information, the explosion is estimated to be happened few days before the discovery and through the thesis we will adopt JD = 2450180$\pm$5 as epoch of explosion.
The coordinates of \ww\/, measured  on our astrometrically calibrated images on two different epochs, are $\alpha=11^{h}59^{m}28^{s}.98 \pm0^{s}.05$ and $\delta=-19\degr15\arcmin21\arcsec.90 \pm0\arcsec.05$ (J2000). The object is located in an arm of the host galaxy, $17\arcsec$ West and $34\arcsec$ North of the nucleus of the parent galaxy NGC 4027. This position, slightly revised with respect to previous determination \citep{su96}, corresponds to a deprojected distance of $\sim$3.0 kpc from the nucleus.

\begin{figure}
%\vspace{174pt}
\includegraphics[width=\columnwidth]{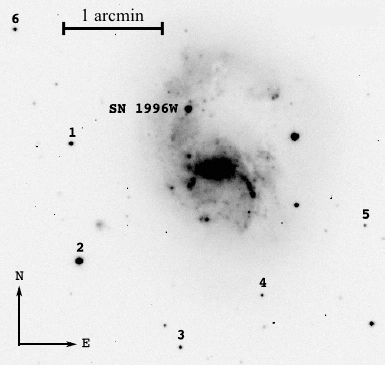}
\caption{R band image of SN~1996W in NGC~4027 obtained with the Dutch 0.9m telescope on May 13th, 1996. The local sequence of stars used to calibrate the optical  magnitude is indicated.} 
\label{fig:96w}
\end{figure}

The Galactic reddening has been estimated as $E_{\rm g}$(B-V)$=0.044$ mag \citep[$A_{\rm g}$(B)$=0.145$ mag][]{sc98}. In the spectra of \ww\/ the absorption features due to interstellar Na ID lines both from our Galaxy and of the host galaxy have been identified, suggesting a reddening $E_{\rm g}$(B-V)$\sim0.05$ and  $E_{\rm i}$(B-V)$=0.187$ mag ($A_{\rm i}(B)=0.77$ mag), respectively. The total extinction $E_{\rm tot}$(B-V)$=0.23$ mag (A$_{\rm tot}$(B)$=0.95$ mag) was adopted.

For the metallicity of the underlying H~{\sc ii} region as representative of the metallicity of the SN, we obtained an oxygen abundance of $12+log(O/H) = 8.60\pm0.06\pm0.41$, also in this case very close to solar.
%equivalent to Z$\sim0.02$. The inferred metallicity is surprisingly similar to those derived for the other objects presented in this paper.

\end{enumerate}

\section{Photometry}\label{sec:phot}

\subsection{Data summary}\label{sec:data}
%\subsection{Optical ground based data}
Optical observations of SNe 2009dd, 2007pk, 2010aj, 1995ad and 1996W were obtained with ground based telescopes and the SWIFT satellite (see Table~\ref{table:cht}).

Observations were reduced following standard procedures in the IRAF\footnote{Image Reduction and Analysis Facility, distributed by the National Optical Astronomy Observatories, which are operated by the Association of Universities for Research in Astronomy, Inc, under contract to the National Science Foundation.} environment. Instrumental magnitudes were measured on the images corrected for overscan, bias and flat field.

Photometric zero points and colour terms were computed for all photometric nights through observations of Landolt standard fields \citep{landolt}. 
The average magnitudes of the local sequence stars were computed and used to calibrate the photometric zero points for the non-photometric nights. 
Magnitudes of the local sequence stars are reported in Tabs.~\ref{table:ls09dd}, \ref{table:ls07pk}, \ref{table:ls10aj}, \ref{table:ls95ad} and  \ref{table:ls96w} along with their r.m.s. (in brackets). 
The calibrated optical magnitudes of the SNe are reported in Tabs.~\ref{table:snm09dd}, \ref{table:snm07pk}, \ref{table:snm10aj}, \ref{table:snm95ad} and \ref{table:snm96w}.

The i-band filter used at the 2.56-m Nordic Optical Telescope (NOT) is an interference filter with central wavelength 7970 \AA, slightly different from the classical Gunn or Cousins I and more similar to Sloan i. In our analysis, however, it was calibrated as Cousins I.
Also, for the LT Sloan photometry we subsequently applied an {\it S}-correction \citep{str02,pig04} to convert the SN magnitudes to the standard Johnson-Cousins photometric system,  finding an average correction of $\Delta U\sim 0.04$, $\Delta R\sim 0.01$ and $\Delta I\sim -0.10$.
 In the case of \dd\/ the discovery magnitude reported in \citet{co09} has been revised and reported as R band in Tab.~\ref{table:snm09dd}.

\begin{table*}
\caption{Instrumental configurations.}
\begin{center}%\tabcolsep=1.0mm
\begin{tabular}{lclclcc}
\hline
\hline
Telescope & Primary mirror & Camera & Array & CCD & pixel scale & field of view \\
 & m & & &  & arcsec/pix & arcmin  \\
 \hline
 Ekar & 1.8 & AFOSC & 1024 x 1024 & TK1024AB & 0.46 & 8.1 \\
          &      & TL512$^\dagger$  & 512 x 512 & TK512& 0.34& 2.9 \\ 
   &  & B\&C$^\dagger$  & 576 x 384 & TH7882 & 0.29 &  2.8\\ 
TNG & 3.6 & DOLORES & 2048 x 2048 & EEV 42-40 & 0.25 & 8.6 \\
 				&  & NICS & 1024 x 1024 & HgCdTe Hawaii & 0.25 & 4.2\\
 LT & 2.0 & RATCam & 2048 x 2048 &  EEV 42-40 & 0.13 & 4.6 \\
 NOT & 2.5 & ALFOSC & 2048 x 2048 & EEV 42-40 & 0.19 & 6.4 \\
 CAHA & 2.2 & CAFOS & 2048 x 2048 & SITe & 0.53 & 16.0 \\
SWIFT& 0.3 & UVOT & 2048 x 2048 &  intensified CCD & 0.48 & 17.0 \\
 NTT  & 3.6 & EFOSC2 & 2048 x 2048 & Loral/Lesser & 0.16 & 5.2 \\
   &  & SOFI & 1024 x 1024 & Hawaii HgCdTe & 0.29 & 4.9 \\
 ESO 1.5 & 1.5 & B\&C & 2048 x 2048 & Loral & 0.82 & 8.0 \\
 MPG-ESO 2.2 & 2.2 & EFOSC2 & 1024 x 1024 & THX31156 & 0.34 & 5.7 \\
 ESO 3.6 & 3.6 & EFOSC1 & 512 x 512 & Tek512 & 0.61 & 3.6 x 5.76 \\
 Dutch & 0.9 & CCD phot. & 512 x 512 & Tek512& 0.48 & 3.8 \\
 Danish & 1.5 & DFOSC & 2052 x 2052 & W11-4 Loral/Lesser & 0.39 & 13.3 \\ 
% CTIO & 1.5 & CCD phot. & 530 x 580 & EEV 42-40 & 0.48 & 2.38 x 3.53 \\
 \hline
\end{tabular}
\end{center}
Ekar = Copernico Telescope (Mt. Ekar, Asiago, Italy); TNG = Telescopio Nazionale Galileo (La Palma, Spain); LT = Liverpool Telescope (La Palma, Spain); NOT = Nordic Optical Telescope (La Palma, Spain); Caha = Calar Alto Observatory 2.2m Telescope (Sierra de los Filabres, Andalucia, Spain); SWIFT by NASA; NTT = New Technology Telescope (La Silla, Chile); ESO 1.5 = ESO 1.5m Telescope (La Silla, Chile); MPG-ESO 2.2 = MPG-ESO 2.2m Telescope (La Silla, Chile); ESO 3.6 = ESO 3.6m Telescope (La SIlla, Chile); Dutch = Dutch 0.9m Telescope (La Silla, Chile); Danish = Danish 1.5m Telescope (La Silla, Chile)\\
$\dagger$ Instruments used until 1998
\label{table:cht}
\end{table*}

\begin{table*}\tiny
\caption{Ultraviolet, optical and infrared photometry of \dd.}
\begin{center}\tabcolsep=0.5mm
\begin{tabular}{cccccccccccccc}
\hline
\hline
Date & JD & uvw2 & uvm2 & uvw1& U & B & V & R & I & J & H & K' & Inst.$\dagger$\\
yy/mm/dd & (+2400000) & & & & & &\\
\hline
09/04/13 & 54935.61 & -& -& -& -& -&  -& 13.88 (.10) & - & - & -&-& 99\\
09/04/14 &  54936.67  & -& -&- &  14.74 (.02)  & 15.07 (.06) &  14.87 (.06)  &  14.60   (.06) & 14.62  (.04)& -& -& -&  1\\
09/04/15 & 54936.63 & -&- & -&- & -& -& -& -& 14.32 (.04)  &  14.25 (.05)	&   14.12   (.06)&2 \\
09/04/15 & 54937.12 &   16.97 (.07)    &   16.88 (.05)   &   15.66 (.05)   &   14.76 (.09) & 15.20  (.08) & 14.85  (.08) & -& -& -& -& &  5 \\
09/04/16 &  54938.40  & -& -&- &  14.86 (.03)  & 15.27 (.05) &  14.83 (.06)  &  14.62   (.06) & 14.64  (.05)& -& -& -&  4\\
09/04/17 & 54939.08   &  17.44 (.13) &  -   &  16.00 (.04)  &  14.96 (.06) & 15.31  (.06) & 14.76  (.06)  & -& -& -& -& -& 5\\
09/04/19 & 54940.87    & 17.25 (.08) & 17.39(.09)   &   -   &  15.19 (.05)   & 15.32 (.06) & 14.83 (.04)  &- & -&- &- & -&   5\\
09/04/21 & 54943.00  & 17.53 (.07)  &  17.51 (.09)  &  16.63 (.08)  &    15.49 (.06)  & 15.44  (.08) & 14.90  (.04)& -& -& -&- &- &5\\
09/04/25 & 54946.62 & 18.10 (.11) &   18.01 (.10) & 16.82 (.09)  &  16.06 (.06) & 15.61 (.05) & 14.90 (.04) & -& -& -& -& -&5  \\ 
09/05/10 & 54961.54 &  19.55 (.08)  &  19.11 (.08)  &  - & 17.18 (.06) & 16.31  (.05) & 15.20 (.04)&- &- & -& -&- &5\\
09/05/11 & 54963.29 & -  &   19.46 (.09)  &  - &  - &   -&- &- &- &- & -& -&5\\
09/05/19  & 54971.58  & - & -& -&17.99   (.10)  & 16.87   (.02)  & 15.48   (.01)   &  14.67    (.04)   &14.57   (.05)  & - &- & -& 2\\
09/07/05 & 55017.50 &  20.17 (.15)  &  20.18 (.12) &  19.29 (.10)  & 19.21 (.06) & 17.83 (.05) & 16.15 (.04) &- & -& -&- & -&5 \\	
09/07/15 & 55028.21  & - & - &  19.52 (.11) &  - & - & -& -& -&- &- &- & 5\\
09/07/16  & 55028.54  & - &-&-&19.72 (.06) & -& -& -&- & -&- & -&5 \\
09/07/19 & 55032.12 & - & - &  19.91 (.13) & -  &  -&  -&- & -&- & -&- &5\\
09/07/20 & 55032.54 & $>$ 20.1&   - & $>$ 20.1  & 19.92 (.13) &-  & - & -& -& -&- & -&5 \\
09/07/20  & 55033.41 & -& -& -& 20.01 (.13) & 18.21   (.09)  & 16.28   (.06)  & 15.45   (.05)  &   15.17  (.04) & -& -& -&2\\
09/07/21  & 55034.31  & -&- &-& -  & 18.20  (.06) &  16.35 (.01)   &  15.47   (.02)  & 15.14   (.02)  &- &- &- & 3 \\
09/07/23  & 55036.39 & -& -&-&  -  & 18.36  (.02) & 16.38  (.01)  &  15.53   (.05) & 15.19  (.07)  & -& -&- &3\\
09/07/26 & 55038.62 &-& $>$ 20.4 & -& -&  -& -& -& -& -& -& -& 5\\
09/07/27 & 55039.62 &- & - & $>$ 20.2 & -&- &- &- &- &- &- & -& 5 \\
09/07/30  & 55042.40 & -& -&-&  -  & 18.81  (.06) & 16.88  (.05)  &  15.95   (.05) & 15.70  (.07)  & -& -&- &4\\
09/08/02 & 55046.00 & $>$ 20.3  &  $>$ 20.3  &  - & - &  -	&-& -& -& -& -&- & 5\\
09/08/03 & 55046.75 &  -  & $>$ 20.2 &  -  &  -&  - &- &- & -&- &- & -&5\\	
09/08/09 & 55052.50  &  $>$ 20.4  & -  &  -&   $>$ 19.2 &-  & -&- &- & -& -&- &5\\
09/08/12  & 55056.50   &- & -&- &  -& 19.63 (.04) &  17.79 (.09)  &  16.74 (.02) & 16.66 (.05) & -&- &- & 1\\
09/08/20  & 55064.36  &- & -& -& - & 19.73  (.20)  & 17.92  (.03)  &  16.84   (.08) & 16.72  (.02) &- & -& -& 3 \\
09/09/04  & 55079.31  &-  &- & -&- &  - & 18.20   (.04)   &  17.04   (.19) &  16.99   (.20)   &- &- & -&4\\
09/09/13 & 55088.50  & - &-   &$>$ 20.3  &-  & -  & -& -& -&- &- &- & 5 \\
09/09/14 & 55089.50 &-&-& -&   $>$ 19.2 &-  & -&- & -&- &- &- &5\\  
09/09/16 & 55091.58 & -& $>$ 20.4 & -&- & -&-  & -&- &- & -&- &5\\
09/09/21 & 55095.46 & -& -& $>$ 20.5 & $>$ 19.3  &-  & -  &- &- & -&- & -&5\\
09/11/19  & 55155.65  & - & -& -&  -&  20.82 (.03) & 18.92 (.10)   &  17.76 (.09) &  17.36 (.06) & -  & -& -& 4\\
09/11/21  & 55157.70  &- &- &- &  - & 20.93  (.04) & 18.98  (.12)  &  17.83   (.10) & 17.44  (.04) & - & -& -& 2\\  
10/01/19  & 55216.20  &- & -&- &   - &  - & 19.73   (.20)   &  18.83   (.20)  &-   &  -&- & -& 3\\
10/05/17  & 55334.43  & -& -& -&-   & $>$ 20.7   & $>$ 20.1   &  $>$ 19.8  & $>$ 19.3   &  -& -& -& 2\\ 
10/10/25  & 55495.67  & -& -& -& -  & $>$ 20.7   & $>$ 20.6    &  $>$ 19.9  & $>$ 20.0  &  -&- & -& 4\\
\hline
\end{tabular}
\end{center}
$\dagger$  1 = NOT,  2 = TNG, 3 = Ekar, 4 = CAHA, 5 = Swift, 99 = CBET 1764 (revised measure) where instruments are coded as in Tab.\ref{table:cht}.
\label{table:snm09dd}
\end{table*}%

\begin{table*}
\caption{Ultraviolet and optical photometry of \pk.}
\begin{center}\tabcolsep=0.9mm
\begin{tabular}{ccccccccccc}
\hline
\hline
Date & JD & uvw2 & uvm2 & uvw1& U & B & V & R & I & Inst.$\dagger$  \\
yy/mm/dd & (+2400000) & & & & & &\\
\hline
07/11/10 & 54415.30  & -&- & -&- &  -   & -&  17.00	( - )	  & -	 &   99 \\
07/11/11 & 54416.40  & -&- &-&  -&16.25 (.02) &	  16.37 (.02) &	 16.37 (.03) &	 16.45 (.03)	&      1 \\
07/11/12 & 54417.50 & -&- &-&  -& 16.00 (.15) &	  16.17 (.17) 	& 16.20 (.18) &	 15.89 (.19)	      &1 \\
07/11/13	&      54417.66  &   	-&      -&    14.50 (.04)&- &- & -& -&- &6 \\
07/11/14& 	  54418.65&  14.54 (.06)	&      14.46 (.06)&     14.64 (.05) & 15.13 (.05) &    16.05 (.06)  &     16.15 (.04)& -&- & 6\\	
07/11/15	&   54420.24&  14.92 (.05)	   &   14.69 (. 05)&  14.78 (.04)&  15.17 (.05)   &  16.07 (.06)  	&   16.09 (.04)	& -& - & 6\\ 		
07/11/16 &  54420.66&  15.14 (.06)	   &   14.84 (.06) &  14.87 (.05) & 15.19 (.05)   &  16.08 (.06) 	&   16.09 (.04)& -& -&  6	\\ 
07/11/16 & 54421.50 &- & -&-& 15.20 (.04)  & 16.14 (.04) &   16.20 (.03) &	 15.88 (.04) 	 &15.99 (.03)	      &2\\ 
07/11/17	 &  54421.68&  15.45  (.06)	 &   15.11 (.06)   &   15.08 (.05) & 15.22 (.06)  &  16.14 (.06) 	&   16.23 (.05) & -& -& 6	 \\
07/11/20	   &    54425.08&  16.03 (.06)	  & 15.73 (.06)   &   15.48 (.05) 	& -& -& -& -&- &  6\\
07/11/25	 &  54429.65 &  16.91 (.07)	 &  16.74 (.07)   &   16.32 (.08) & 15.69 (.06)  &  16.28 (.06)  	&   16.23 (.04)	& -& -& 6\\
07/11/29	&   54433.91&  17.85 (.09)	  & 17.58 (.09) &   17.02 (.08)  & 16.13 (.06)  &  16.42 (.07)	&   16.27 (.05)& -& - & 6	 \\    
07/12/02	 &  54436.96&  18.44 (.10)	  & 18.59 (.14)    &   17.64 (.09)  & 16.45 (.06)  &  16.57 (.06)	&   16.27 (.05)	&- &- & 6 \\
07/12/04 & 54439.39 & -&- &- & - & 16.62 (.24) 	&  16.31 (.24) &	 16.14 (.27) &	 16.01 (.29)	&      1 \\
07/12/06	&   54440.81&  19.01 (.10)	  & -    &   18.07 (.10) & 16.61 (.06)  &   16.69 (.07)  	&   16.32 (.05)	&- &-  & 6   \\ 
07/12/08 & 54443.50 & -& -&-&16.65 (.03)  & 16.86 (.02) &	  16.43 (.02) 	& 16.23 (.02) &	 16.00 (.02)	    &  3 \\
07/12/13 &54448.40  &- & -& -&- & - & 16.51 (.02) 	& 16.38 (.09) &	 16.00 (.04)	     &1	\\
07/12/14 & 54449.50 & -& -&-&17.29 (.03)  & 17.21 (.03) 	 & 16.62 (.03) 	& 16.36 (.03) &	 16.01 (.03)	&     3 \\
07/12/17	&   54451.65 & - & -& -& 17.39 (.06)	&   17.27 (.07)  	&   16.68 (.05)&- & -& 6	 \\
07/12/24 & 54459.50 & -& -&-&18.02 (.10)  & 17.65 (.04) 	&  16.82 (.03) 	& 16.41 (.03) 	& 16.19 (.02)	   &   3 \\
07/12/27	&   54462.01 & - & -&- & 18.20 (.08)  &  17.78 (.08)  & 	  16.95 (.05)	&- & -&  6 \\	
07/12/28 & 54463.29 & -& -& -& - & 17.92 (.12) 	  &16.96 (.10) &	 16.53 (.05) 	& 16.26 (.04)	    &  1 \\
08/01/08 &  54474.09  & -&- &- & 18.52 (.08)   &  18.11 (.10) 	&   17.09 (.06)	& -& -& 6\\
08/01/09  &54475.39  &- &- &-& -& 18.14 (.23) 	&  17.13 (.23) &	 16.63 (.18)	& 	-    &  1\\
08/01/11 & 54477.40 & -&- &-&18.73 (.05) &  18.34 (.03) &	  17.19 (.02) 	& 16.69 (.03) &	 16.40 (.03)	    &  4  \\
08/01/28  &54494.36 & -&- & -& - &18.49 (.93) &	  17.46 (.16) 	& 17.04 (.16) &	 16.69 (.13)	    &  1 \\
08/09/04  &54714.64 &- & -&-& -   &-  &	 $>$ 19.8 	& $>$19.6 &	$>$ 19.4	    &  2 \\
08/09/14  &54723.69 & -&- &-&   -& $>$19.8 &	$>$  20.1	& $>$19.6 & 	-    &  5 \\
\hline
\end{tabular}
\end{center}
$\dagger$  1 = Ekar,  2 = TNG, 3 = LT, 4 = NOT, 5 = CAHA, 6 = Swift, 99 = CBET 1129, where instruments are coded as in Tab.\ref{table:cht}.
\label{table:snm07pk}
\end{table*}

\begin{table*}
\caption{Optical and infrared photometry of \aj.}
\begin{center}\tabcolsep=0.6mm
\begin{tabular}{ccccccccccc}
\hline
\hline
Date & JD &  U & B & V & R & I & J & H & K & Inst.$\dagger$  \\
yy/mm/dd & (+2400000) & & & & & &\\
\hline
10/03/12 & 55268.50  & -  &  - &- &  17.1   ( - )    &	-  & -& -&- & 99\\
10/03/13 &  55269.50  &   -  &-  &   -& 17.0  ( - )  &    -	& -& -& -&  99\\
10/03/24 &  55280.50  &    - &  17.92 (.06)& 17.75 (.03)  &  17.55 (.03) &   17.34 (.05)	  & -& -& -&1 \\
10/03/27 & 55282.95   & -  &  18.30 (.04) & 17.78 (.03)  &  17.57 (.02)   & 17.33 (.04)	& - & -&- & 1 \\
10/04/16 & 55303.50  &   - &  18.72 (.13) & 18.16 (.03)  &  17.82 (.03)   & 17.45 (.06)	& -& -&- &  2 \\
10/04/18 & 55305.50 & -&- &- & -& -& 19.42 (.04)  &  20.19 (.08)	&   20.21   (.05) & 2\\
10/04/24& 55311.52  &   -   &  19.30 (.11) & 18.29 (.04)   & 17.83 (.03)   & 17.54 (.03)	 & -& -& -& 3 \\
10/05/05 & 55321.90   & 20.21 (.17) &	19.55 (.09) & 18.47 (.06)   & 17.99 (.03)   & 17.85 (.03)	 &-  &- &- &1\\
10/05/07 & 55324.00  &  20.27 (.21) &	19.65 (.12) & 18.62 (.07)  &  18.11 (.04)  &  17.81 (.06)	 &- & -&- & 1 \\
10/05/18 & 55334.95   & 20.36 (.25) &	19.90 (.12) & 18.83 (.07)  &  18.24 (.05)  &  18.13 (.08)	 &- & -&- & 1 \\
10/05/22 & 55339.38   & 20.44 (.23) &	19.97 (.10) & 18.93 (.07)  &  18.32 (.05)  &  18.23 (.06)	  &-& -&- & 3 \\
10/05/26 & 55342.92  &  20.81 (.23) &	20.30 (.15) & 19.17 (.07)   & 18.53 (.06) &   18.43 (.06)	  &-& -&- & 1 \\
10/06/09 & 55356.93  &  -&	22.82 (.24) & - &  21.01 (.19)  &  20.74 (.30)	   &-& -& -&1\\
10/06/12 & 55359.89  &  -&	$>$20.7  & 21.91 (.30) &  21.16 (.28)   & 21.03 (.20)	& - & -&- & 1 \\
10/06/16 & 55363.94   &-&	$>$20.7 & 22.28 (.21)   & 21.18 (.11) &   21.11 (.11)	 & -& -&- & 1\\
10/06/17 & 55365.14   & -&	- & $>$20.6 & $>$ 20.2 &  $>$20.0	  &-& -&- & 2   \\
10/06/21 & 55368.91   & -&	$>$20.5& $>$20.5  &  21.25 (.14)&  21.17 (.12)	&-  & -& -& 1 \\
10/06/24 & 55371.91  & - &	$>$20.7& $>$20.7  & 21.58 (.04)&  21.31 (.14)  &-&- &- & 1\\
10/07/08 & 55386.46  &-  &	$>$20.9 & $>$20.6  &  $>$20.1   & $>$20.4	& -  &- &- &3 \\
11/01/01 & 55563.31  &- &	$>$20.5 & $>$20.6  &$>$20.2  &$>$20.0	&- &- &- &  2\\
11/01/25 & 55587.30 & -  &	$>$20.6 & $>$20.6   &$>$20.3 &$>$20.5	 &- & -& -&2 \\
\hline
\end{tabular}
\end{center}
$\dagger$  1 = LT,  2 = NTT, 3 = TNG, 99 = CBET 2201, where instruments are coded as in Tab.\ref{table:cht}.
\label{table:snm10aj}
\end{table*}%

\begin{table*}
\caption{Optical and infrared magnitudes of \ad.}
\begin{center}%\tabcolsep=0.7mm
\begin{tabular}{ccccccccccc}
\hline
\hline
Date & JD & B & V & R & I & J & H & K & Inst.$\dagger$  \\
yy/mm/dd & (+2400000) & & & & & & & & &\\
\hline
95/09/22 & 49983.29  &    -  & - &  15.70 ( - )	  & 	- & -  & -& -&  98\\
95/09/28 & 49989.30  &   -   &  14.25 (.25) & -	  & 	- & - & -&- &   99\\
95/09/29 & 49989.80  &   14.77 (.20)  & 14.73 (.15) &  14.67 (.15)	  & 	- & - & -&- &   1\\
95/10/02 & 49992.88 &15.17 (.01) &	  15.03 (.01) &	 14.85 (.01) &	 14.75 (.01)	&- & -& -&     2 \\
95/10/14 & 50004.90 &- &	  15.29 (.02) 	& 14.98 (.02) &	    -  &-& -& -&2 \\
%95/11/14 & 50036.00 & 16.80 (.25) & 15.70 (.20) &- &- &-& -&- & 7\\
95/11/24 & 50045.53 & 16.90 (.03) &   15.81 (.02) &	 15.30 (.02) 	 &  -  &-& -&- &3\\ 
95/12/26 & 50077.71  &  -	&   -&	 16.92 (.03) &	 -	&   - & -& -&  4 \\
95/12/26 & 50077.73 & 19.79 (.01) &	  17.96 (.06) 	& 16.93 (.03) &	 16.37 (.02)	    & -& -& -& 4 \\
96/12/29 & 50080.74  &  -& 18.00 (.09) 	& 17.07 (.07) &	     - &- & -&- &2	\\
96/01/17 & 50100.42  & 19.83 (.30) 	 & 18.32 (.19) 	& 17.35 (.12) &	-	&  - &- & -&   3 \\
96/01/19 & 50102.45 & 19.91 (.25) 	&  18.35 (.12) 	& 17.38 (.10) 	& 16.89 (.10)	   & -& -& -&  5 \\
96/02/18 & 50131.63   & 19.94 (.10) 	  &18.59 (.05) &	 -&  -& - & -& -&4 \\
96/02/22  &50135.65  & 20.08 (.30) 	&  18.51 (.15) &	 17.61 (.10)	& 	17.15 (.10)    & -& -& -& 5\\
96/02/23 & 50136.65 &  19.96 (.30) &	  18.53 (.15) 	& 17.61 (.10) &	 17.18 (.10)	    &- & -&- & 5  \\
96/04/20  &50193.52   &20.16 (.35) &	  19.14 (.20) 	& 18.20 (.15) &	 17.76 (.15)	    &- &- & -& 5 \\
96/04/21  &50194.55  &  -&	  19.07 (.20) 	& 18.19 (.15) &	 17.75 (.15)	    & - & -& -&5 \\
96/04/29 & 50203.50 & -& -& -&- & 17.69 (.20)  &  17.67 (.30)	&   17.13   (.50) & 2\\
96/05/14  &50217.50  & $>$20.5 &	  19.54 (.25) 	& 18.58 (.20) & 	  -  &  -&- & -&5 \\
96/10/02  &50358.87 & $>$23.0 &	  22.42 (.55) 	& 20.21 (.15) & 	  -  & - & -& -&6 \\
96/11/19  &50406.85 &-  &	   -	& 20.79 (.40) & 	   - & -& -& -& 5 \\
97/02/19  &50489.53 & - &	 -	& $>$21.5  & 	 -   & -& -& -& 2 \\
\hline
\end{tabular}
\end{center}
$\dagger$ 
1 = ESO 1.5,  2 = ESO 3.6 , 3 = Ekar, 4 = MPG-ESO 2.2, 5 = Dutch, 6 = Danish, 98 = IAUC 6852, 99 = IAUC 6239, where instruments are coded as in Tab.\ref{table:cht}. \\
\label{table:snm95ad}
\end{table*}%

\begin{table*}
\caption{Optical magnitudes of \ww.}
\begin{center}%\tabcolsep=0.7mm
\begin{tabular}{cccccccc}
\hline
\hline
Date & JD & U & B & V & R & I & Inst.$\dagger$  \\
yy/mm/dd & (+2400000) & & & & & &\\
\hline
96/04/10 & 50183.50  & -&  -  & 16.00 ( - ) &  	-  & -	 &     6 \\
96/04/11 & 50184.50  &- &  -  & 16.00 ( - ) &  	-  & -	 &     6 \\
96/04/13 & 50186.60  & -&  -  & 15.10 ( - ) &  	-  & 	- &     6 \\
96/04/16 & 50190.40  & -&   15.53 (.02)  & 15.18 (.02) &  14.77 (.04)	  & 	- &     1 \\
96/04/18 & 50192.40  & - &15.51 (.02) &	  15.26 (.01) &	 14.79 (.01) &	 	-&      1 \\
96/04/18 & 50192.41 &  -& -&	  15.26 (.01) 	& - &	-      &1 \\
96/04/19 & 50192.50 &-& 15.45 (.03)&	  15.15 (.01) 	& 14.77 (.01) &	14.60 (.02)      &2 \\
96/04/20 & 50193.50 & - & 15.51 (.03) &   15.12 (.01) &	 14.75 (.01) 	 &  14.59 (.02)  &2\\ 
96/04/25 & 50198.54   &  15.21 (.30) &  15.53 (.15)	& 15.20 (.10)  &	 14.79 (.10) & 14.62 (.15)	 	&      3 \\
96/05/09 & 50212.50 & -& 15.89 (.20) &	  15.22 (.15) 	& 14.74 (.10) &	 14.49 (.15)	    &  3 \\
96/05/11 & 50215.63  &- &  15.95 (.20) & 15.24 (.15) 	& 14.76 (.10) &	14.49 (.15)      & 3	\\
96/05/13 & 50217.50 & 16.75 (.05)  & 16.12 (.03) 	 & 15.25 (.03) 	& 14.79 (.03) &	14.49 (.03)	&      2 \\
96/05/14 & 50218.50 & 16.82 (.07) & 16.15 (.03) 	&  15.26 (.03) 	& 14.79 (.03) 	& 14.50 (.04)	   &   2 \\
96/05/19 & 50222.54  & 16.90 (.03) & 16.32 (.01) 	  &15.31 (.01) &	14.79 (.01) & 14.49 (.02) &  4 \\
96/12/15  &50432.81  &- & 19.24 (.13) 	&  18.34 (.12) &	 17.38 (.06)	& 	17.02 (.05)    &  4\\
97/01/30  &50478.80  &- & 19.33 (.11) 	&  18.55 (.08) &	 17.76 (.09)	& 	17.41 (.05)    &  5\\
97/02/12 & 50491.50 & 18.64 (.20) &  19.40 (.20) &	  18.74 (.20) 	& 17.96 (.10) &	 17.65 (.05)	    &  4  \\
97/03/31  &50538.50  & - &19.77 (.20) &	  19.25 (.20) 	& 18.34 (.20) &	 17.91 (.10)	    &  2 \\
\hline
\end{tabular}
\end{center}
$\dagger$  1 = Ekar,  2 = Dutch, 3 = ESO 1.5, 4 = MPG-ESO 2.2, 5 = Danish,  6 = IAUC 6379, where instruments are coded as in Tab.\ref{table:cht}. \\
\label{table:snm96w}
\end{table*}%

Swift U, B, V aperture magnitudes were transformed to Johnson system through the colour transformations by \citet{li06}. After comparison with optical ground-based data, offsets were applied when necessary (cfr. Sect. \ref{sec:pe}).
The magnitudes of the SNe were obtained through a point spread function (PSF) fitting 
%realized with the "SNOOPY"\footnote{SNOOPY is a custom package implemented in IRAF by E.Cappellaro, based on DAOPHOT.} package; 
sometimes applied after template subtraction, depending on the background complexity and the availability of suitable template images.
The uncertainties reported for each optical band in Tabs.~\ref{table:snm09dd}, \ref{table:snm07pk}, \ref{table:snm10aj}, \ref{table:snm95ad} and \ref{table:snm96w} were estimated by combining in quadrature the errors of photometric calibration and those on the instrumental magnitudes. The latter were obtained through artificial star experiments. When the object was not detected,
limiting magnitudes were estimated by placing artificial stars of different magnitudes at the expected SN position. 
Only significant limits are presented in  Figs.~\ref{fig:sn_lc09dd}, \ref{fig:sn_lc07pk}, \ref{fig:sn_lc10aj}, \ref{fig:sn_lc95ad} and \ref{fig:sn_lc96w}.

Ultraviolet \citep[uvw2, uvm2, uvw1; see][]{swift} observations, obtained by UVOT on board of the SWIFT satellite are available for twenty-four epochs in a period of 160d in the case of \dd\/ and for ten epochs in 23d for \pk. % with spatial resolution FWHM$\sim2\arcsec$.
We reduced these data using the HEASARC\footnote{NASA's High Energy Astrophysics Science Archive Research Center} software. For each epoch all images were co-added, and then reduced following the guidelines presented by \citet{swift}. 

The NIR images of the SN fields were obtained combining several sky-subtracted, dithered exposures. Photometric calibration was achieved relative to the 2MASS photometry of the same local sequence stars as used for the optical calibration. 
NIR photometry was obtained just at a single epoch for SNe 2009dd, 2010aj and 1995ad.
In the case of \dd\/ the K' filter was used but was calibrated as K band. % and as such reported in Fig~\ref{fig:sn_lc09dd}.

The follow-up coverage of individual SNe is not exceptional. However, these objects taken together provide a fairly complete picture of the photometric evolutions of luminous type II SNe. 

\subsection{Light curves}\label{sec:pe}
In this Section we present the photometric information for the full SN sample. The main data are reported in Tab.~\ref{table:main}. 

\subsubsection{\dd}\label{ss:09dd}
The optical monitoring of SN~2009dd started on April 14, 2009, the day after the discovery, and continued until October 2010. 
%We attempted to recover the object later but the object was already fainter than the instrumental detection limit. 
Because of the location of the SN very close to the galaxy nucleus, the optical photometric measurements of \dd\/ were performed using the template subtraction technique (Tab.~\ref{table:snm09dd}).

By comparing space and ground-based UBV magnitudes, computed interpolating the light curves with low-order polynomials at corresponding epochs, we found average differences (ground--space) $\Delta U \sim 0.20\pm0.03$, $\Delta B \sim 0.06\pm0.03$, $\Delta V \sim 0.10\pm0.03$.  These corrections have been applied to all UVOT magnitudes and the resulting values are reported in Tab. \ref{table:snm09dd}.

\begin{figure}
\includegraphics[width=\columnwidth]{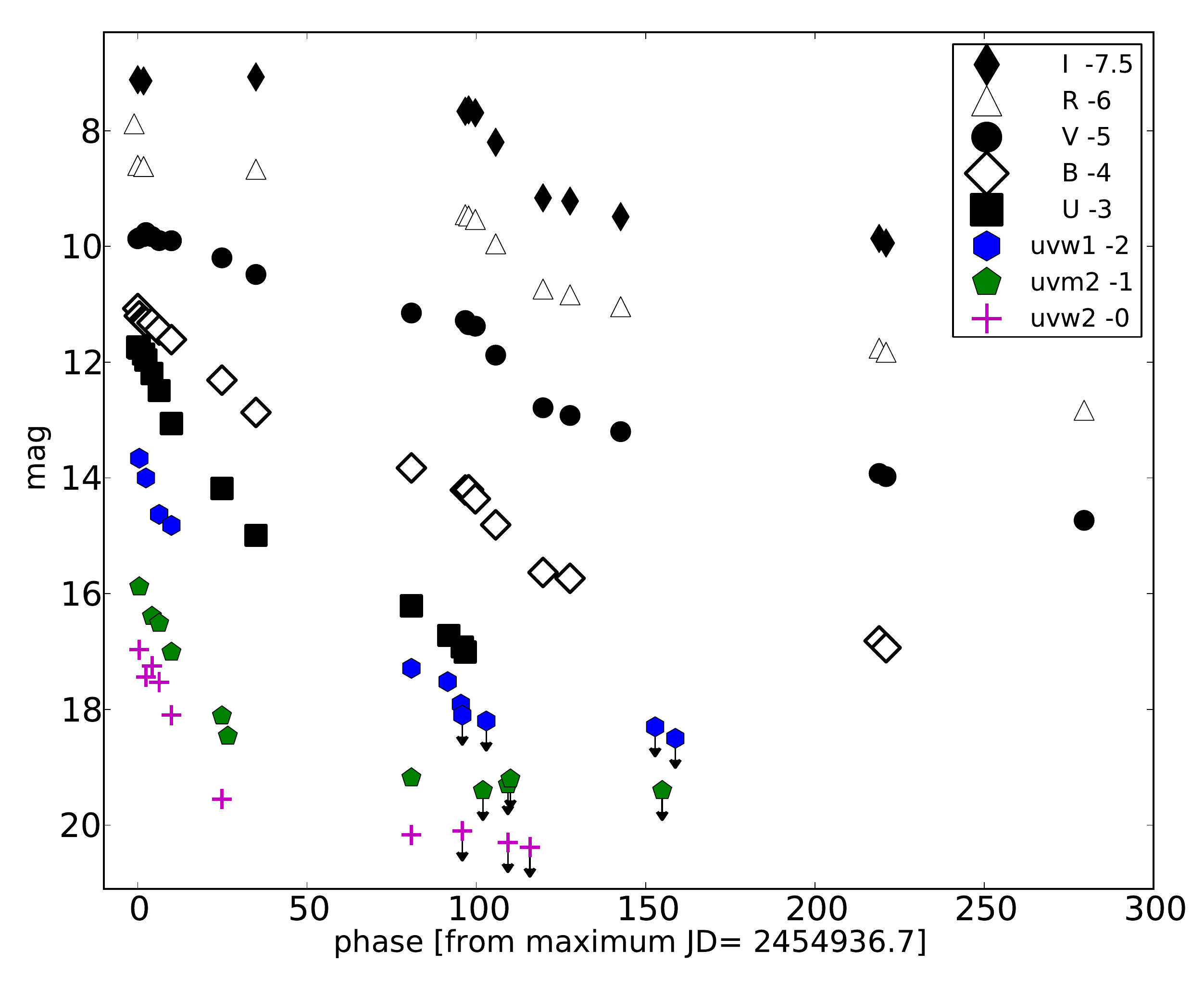}
\caption{Synoptic view of the light curves of \dd\/  in uv and optical bands. 
Magnitude shifts for the different bands are in the legend.}
\label{fig:sn_lc09dd}
\end{figure}

In Fig.~\ref{fig:sn_lc09dd} the uvw2, uvm2, uvw1, U, B, V, R, I light curves of \dd\/ are plotted. 
NIR magnitudes are not shown because available at a single epoch.
The light curves do not constrain the explosion epoch, that was determined through a comparison of the early spectra with a library of SN spectra performed with the {\it GELATO} code \citep{ha08}, and was found to be JD 2454925.5$\pm$5.0 (April 04 UT).

Assuming the distance and extinction discussed in Sect.~\ref{sec:sne}, we find $M^{\rm max}_{U} \leq -18.22 \pm 0.25$, $M^{\rm max}_{B} \leq -17.61 \pm 0.27$, $M^{\rm max}_{V} \leq  -17.46 \pm 0.23$, $M^{\rm max}_{R} \leq  -17.38 \pm 0.21$ and $M^{\rm max}_{I} \leq  -17.09 \pm 0.18$, where the reported errors include both measurement errors and the uncertainties on the distance modulus.

An initial rapid decline is visible from U to V during the first 30 days ($\Delta m(B)\sim1.2$ mag).
It was followed by a plateau lasting about 50--70 days, more clearly visible in the R and I bands.
The luminous peak, m$_{V}$$\sim$15.5 corresponding to M$_{V}\sim -16.6$ is slightly higher than the average for SNIIP \citep[M$_{V}\sim -15.9$,][]{pa94,li11} and fainter than the luminous IIP 1992H \citep[M$_{V}\sim-17.3$,][]{cl96},  2007od \citep[M$_{V}\sim-17.4$,][]{07od} or 2009bw \citep[M$_{V}\sim-17.2$,][]{09bw}.
The rapid post plateau decline occurring at about 100d signals the onset of the nebular phase. The drop in magnitude between the photospheric and the nebular phase is $\Delta m(V)\sim$1.4 mag in $\sim$20d  in the V band, somewhat lower than the value of 2 mag in normal SNe IIP. The late decline rates in the various bands are similar. During the 120--280 d interval the V band decline rate is 1.15 \mcento\/, marginally higher than the decline rate expected from  \co\/ to \fe\/  decay (0.98 \mcento) in case of complete $\gamma$-ray trapping.

\subsubsection{\pk}\label{ss:07pk}
The optical photometric monitoring of \pk\/ started on November 11th, 2007 and continued until January 2008.
%Unfortunately the attempts to recover the 
SN observations during the nebular phase turned out with only upper limits. %, preventing the determination of the \ni\/ mass.
Because of the proximity to the nucleus of the host galaxy and a nearby H~{\sc ii} region, also in this case the SN optical photometry was measured using the template subtraction technique.
The Swift data have been treated as for \dd. The comparison of space and ground-based photometry at corresponding epochs pointed out systematic average differences (ground--space) $\Delta(U)\sim0.20\pm0.05$, $\Delta(B)\sim0.06\pm0.02$, $\Delta(V)\sim0.07\pm0.03$.  As before, Tab.~\ref{table:snm07pk} reports the corrected magnitudes. Our estimates
of the Swift ultraviolet (uvw2, uvm2, uvw1) and optical (u, b, v, before transformation to the Johnson--Cousins system)  photometry are in agreement with those presented by \citet{pri12}.

\begin{figure}
\includegraphics[width=\columnwidth]{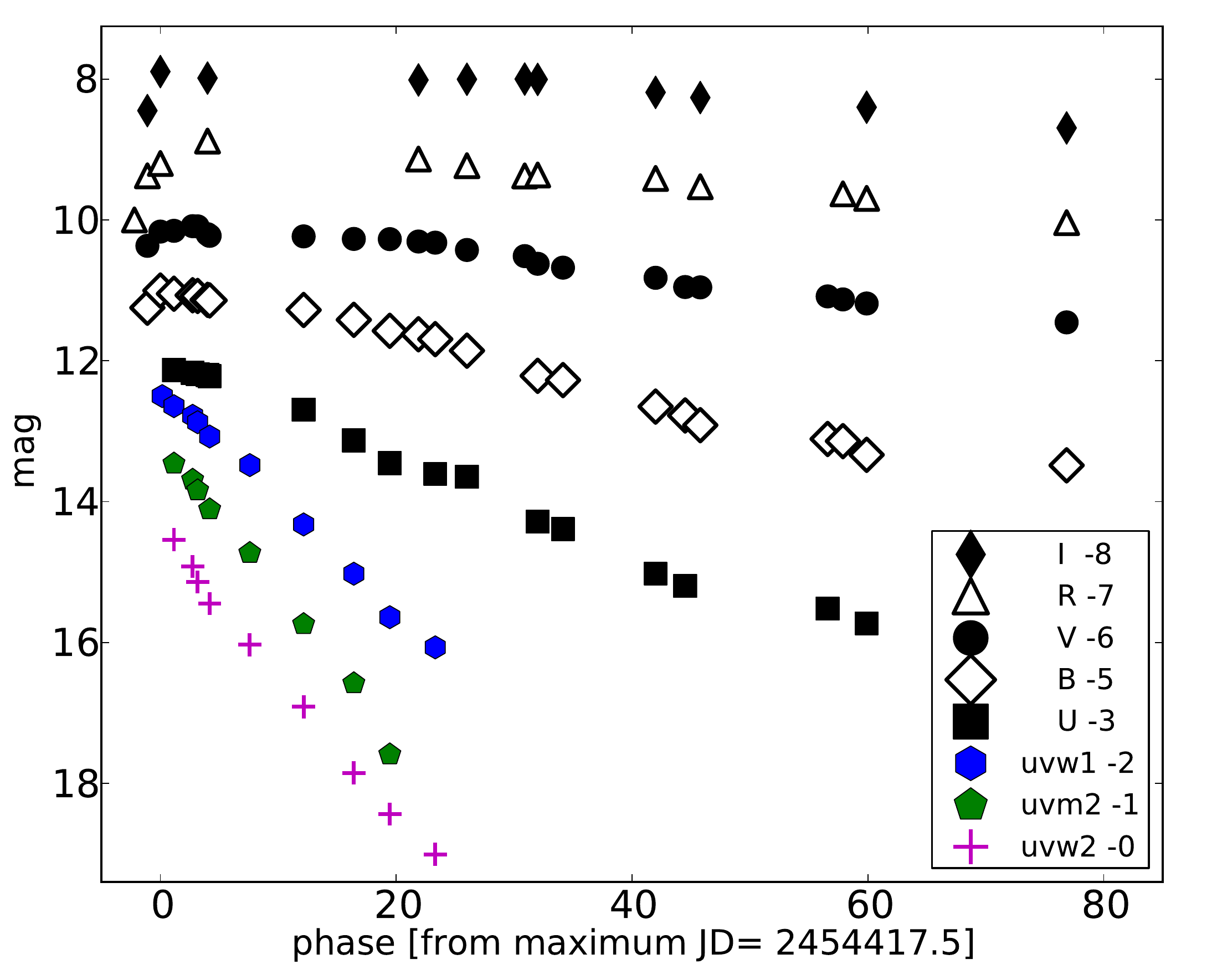}
\caption{Synoptic view of the light curves of \pk\/  in all available bands. The magnitude shifts with respect to the values reported in Tab.~\ref{table:snm07pk} are listed in the insert.}
\label{fig:sn_lc07pk}
\end{figure}

The light curves are plotted in Fig.~\ref{fig:sn_lc07pk}. 
The very early observations show a rise to maximum.
The somewhat slower rise and late peak in the R band might be an effect of the large errors affecting the measurements of Nov. 12, 2007.
The B-band peak (JD 2454417.5$\pm$1.0) is consistent with the phases derived 
from the spectral comparison performed with $GELATO$  \citep[][]{ha08}.  Therefore, hereafter
we will adopt JD 2454412.0$\pm$5 (November 7.5 UT) as the best estimate for the explosion epoch.

%Afterward the light curves resemble those of normal SNe II, flatter with a more pronounced plateau at longer wavelengths than at the shorter ones.
% slight, slow rise to the peak shown by R band light curve supports the classification of this object as type IIn, even if after a period of 10 days post maximum in B band, , the light curves resemble those of a normal type II.
%Despite the classification of type IIn and the similarity with early spectra of \s\/, the light curves show a non slow rise to the peak, especially in R where there are available more nights respect the other bands, estimated around JD 2454417.5$\pm$1.0 in B band.

Having adopted the distance and the extinction (Sect.~\ref{sec:sne}), we can determine the absolute magnitudes at maximum: $M^{\rm max}_{B}=-18.70\pm0.23$, $M^{\rm max}_{V}=-18.44\pm0.24$, $M^{\rm max}_{R}= -18.64\pm0.25$ and $M^{\rm max}_{I}=-18.52\pm0.26$, where the associated errors include the uncertainty on the distance modulus and measurement errors. Therefore, \pk\/ is a bright SN~II according to the criteria of \citet{pa94} and \citet{li11}.
%{\bf MT: sono molto perplesso. Sembra IIP. E' piatta in 3 bande ed il Delta(m) sono molto simili a quelli della 2009dd. il Beta100(B) e' misurato solo sui primi 50gg ossia prima che si instauri un possibile plateau. Cmq beta100=3.7 e' al limite tra IIP e IIL.  Al momento non mi spiego un'evoluzione come quella indicata in Fig.12 }
%After the peak, the decline of the light curves is similar to those of type IIL SNe, especially in the B band, but with a very slow decline in the red bands.  The behaviour of the light curves may resemble that of type IIL SNe such as SN 1979C rather than that of the SNe IIP selected as comparison objects in Sect.~\ref{sec:bol}. 
The average decline rate in the first 100 days post maximum is $\beta_{100}^{B}$(07pk)$\sim3.6$ \mcento, closer to those of type IIL than to those of type IIP SNe \citep{pa94}.
On the contrary, $\beta_{100}^{V}$(07pk)$\sim1.8$ \mcento is typical of type IIP, making \pk\/ a transitional object between the two subclasses.

\subsubsection{\aj}\label{ss:10aj}
\begin{figure}
\includegraphics[width=\columnwidth]{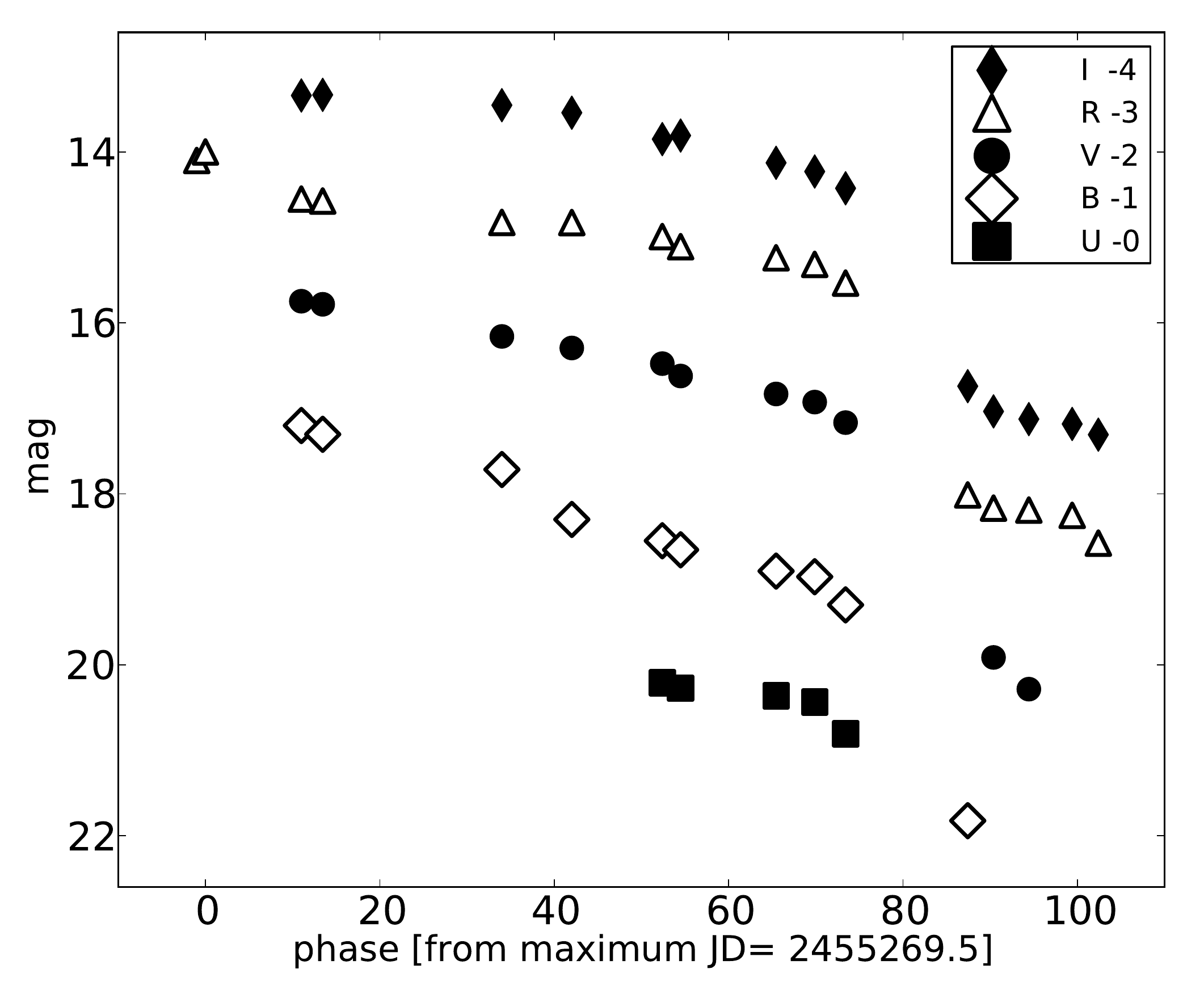}
\caption{
Light curves of \aj\/  in optical bands. The magnitude shifts with respect to the values reported in Tab.~\ref{table:snm10aj} are in the legend.} 
\label{fig:sn_lc10aj}
\end{figure}

Our observations (cfr. Tab.~\ref{table:snm10aj} and Fig.~\ref{fig:sn_lc10aj}) cover a period of almost one year, although the SN has been detected only until $\sim$100 days after discovery.
Because of the complex background, the late time photometry was performed using the template subtraction.
The large errors, estimated with the artificial stars method, are due to non ideal sky conditions.

%U, B, V, R, I, J, H, K light curves of \aj\/ are plotted on Fig.~\ref{fig:sn_lc10aj}. 
The early magnitudes reported in the CBET, give a weak indication that the  R band peak  occurred around JD 2455269.5 (March 13 UT), in agreement with the spectral age reported by \citet{ce10} and with the epoch provided by the GELATO comparisons. Then we adopted JD 2455265.5$\pm$4.0 (March 10 UT) as epoch of the explosion.

The maximum absolute magnitudes are: 
$M^{\rm max}_{B} \leq -16.95 \pm 0.18$,
$M^{\rm max}_{V} \leq -17.08 \pm 0.17$, 
$M^{\rm max}_{R} = -17.80\pm 0.16$ and
$M^{\rm max}_{I} \leq -17.44 \pm 0.16$, 
where the reported errors include the uncertainties in our photometry, in the adopted distance modulus, and in the
interstellar reddening. 
Note that only in the R band the reported value is the absolute magnitude at maximum, while in the other bands they are the brightest measured magnitudes.

The early post maximum decline in the R band is about 0.55 mag in 11d. 
After about 15 days,  the V, R and I light curves settle in a long slanted plateau with average magnitudes V$\sim18.2$,  R$\sim17.8$ and I$\sim17.5$ (M$_{V}\sim-16.6$ , M$_{R}\sim-17.0$ and M$_{I}\sim-17.3$), while the B band shows a monotonic decline. 
The plateau of \aj\/ is, therefore, relatively
luminous when compared with those of more typical SNe IIP and similar to those of SNe 2009bw \citep[M$_{V}\sim-17.2$,][]{09bw} and 2009dd  (M$_{V}\sim-16.6$, cfr.~\ref{ss:09dd}).
The tail has been observed in the V, R, I bands and, with a single data point, in the B band. However, because of the early phase and short time baseline the measured decline rates, $\gamma_R\sim3.0$ and $\gamma_I\sim2.1$ \mcento,  are not indicative of the true decline in the tail. 
The drop between the photospheric and the nebular phase is $\Delta (R)\sim2.5$ mag in $\sim$16d, similar to that of~\bw\/ ($\sim$2.2 mag in $\sim$13d).
%The V band decline rate, calculated with the only two values available in the tail, is 5.0$\pm$0.3 \mcento\/ extremely different from 0.98 \mcento\/ that is the decay rate of \co\/ to \fe\/.

\subsubsection{\ad}\label{ss:95ad}

\begin{figure}
\includegraphics[width=\columnwidth]{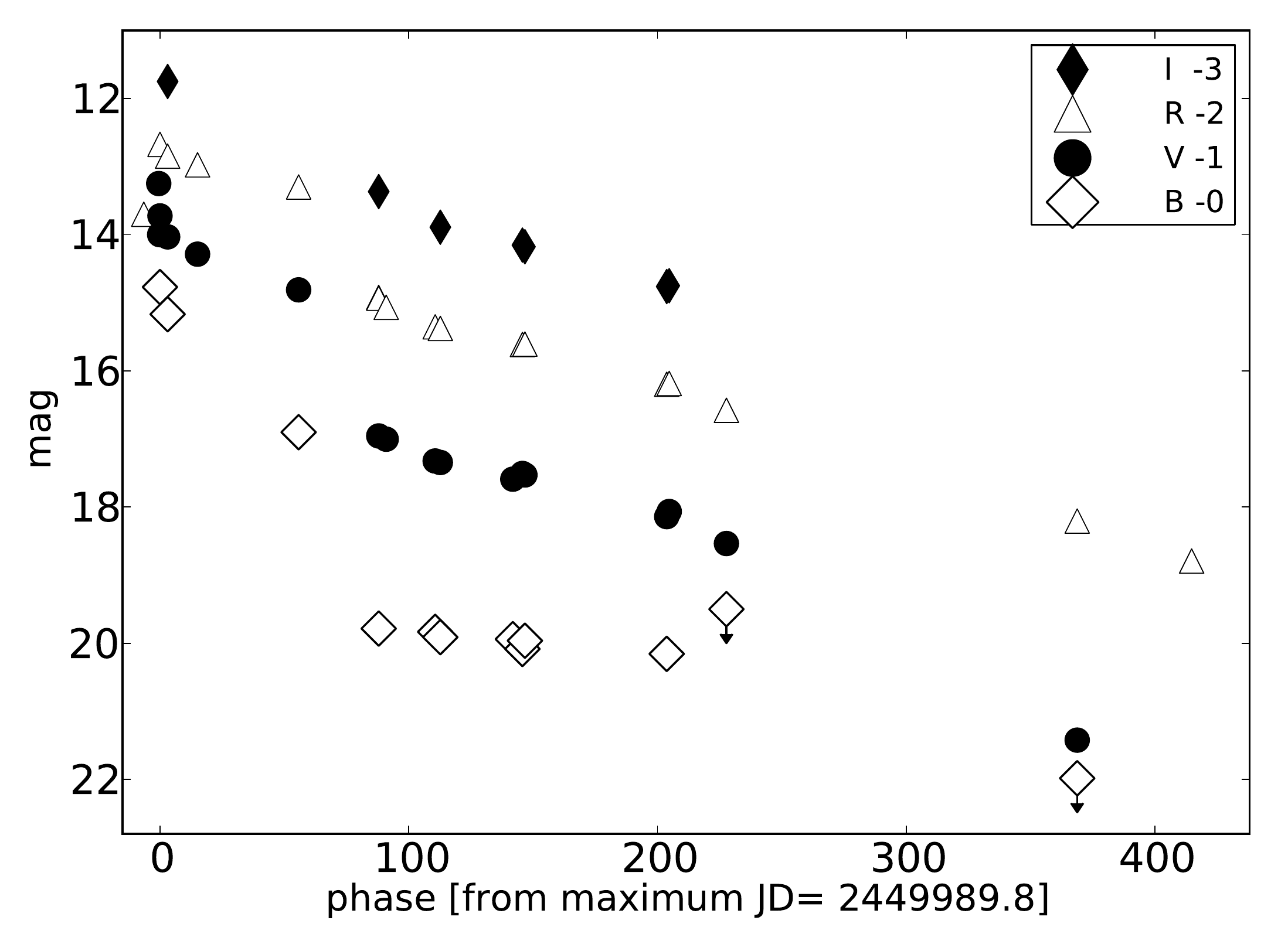}
\caption{
Light curves of \ad\/  in uv and optical bands. The magnitude shifts with respect to the original value reported in Tab.~\ref{table:snm95ad} are in the insert.} 
\label{fig:sn_lc95ad}
\end{figure}

Observations of \ad\/ started the day after the discovery by \citet[$V\sim14.25$,]{ev95}, about one week after the first (pre-discovery) detection  \citep[$R\sim 15.7$,][]{bro98}, and span a period of more than 400 days.
SN magnitudes are reported in Tab.~\ref{table:snm95ad} and shown in Fig.~\ref{fig:sn_lc95ad}.
The pre-discovery detection together with spectroscopic constraints allow us to estimate the explosion epoch to about JD $2449981.0\pm3$ (on September 20).

Considering the adopted extinction and distance (Sect.~\ref{sec:sne}), the SN reached at maximum $M^{\rm max}_{B} = -17.18 \pm 0.26$,
$M^{\rm max}_{V} = -17.67 \pm 0.22$, 
$M^{\rm max}_{R} = -17.22\pm 0.22$ and
$M^{\rm max}_{I} = -17.12 \pm 0.15$.

The light curves show a steep decline from the maximum to the plateau, 
%($\gamma_{B}\sim12.96$, $\gamma_{V}\sim4.14$ and $\gamma_{R}\sim1.73$ {\bf IGNOREREI QUESTE STIME ANCHE PERCHE' CI SONO DATI DA CIRCOLARE})
the latter lasting about 50 days (with an average M$_{V}\sim-16.6$) resembling those of \od\/, \h\/ and \aj. 
Between day 60 and 90 the SN abruptly faded reaching the radioactive tail, marked by a decline very close to the \co\/ to \fe\/ decay rate ($\gamma_{V}\sim0.93$ \mcento\/ between 95-220d).
The available very late time photometry (t$>$300d) shows an increase of the luminosity decline, being larger in V than in the R band.
%  $\gamma_{V}\sim2.02$ \mcento\/ between  21 and 425d. 
This may be due to dust formation, as suggested by the detection of CO emission  in the late-time NIR spectra \citep{sp96}.

\subsubsection{\ww}\label{ss:96w}

\begin{figure}
\includegraphics[width=\columnwidth]{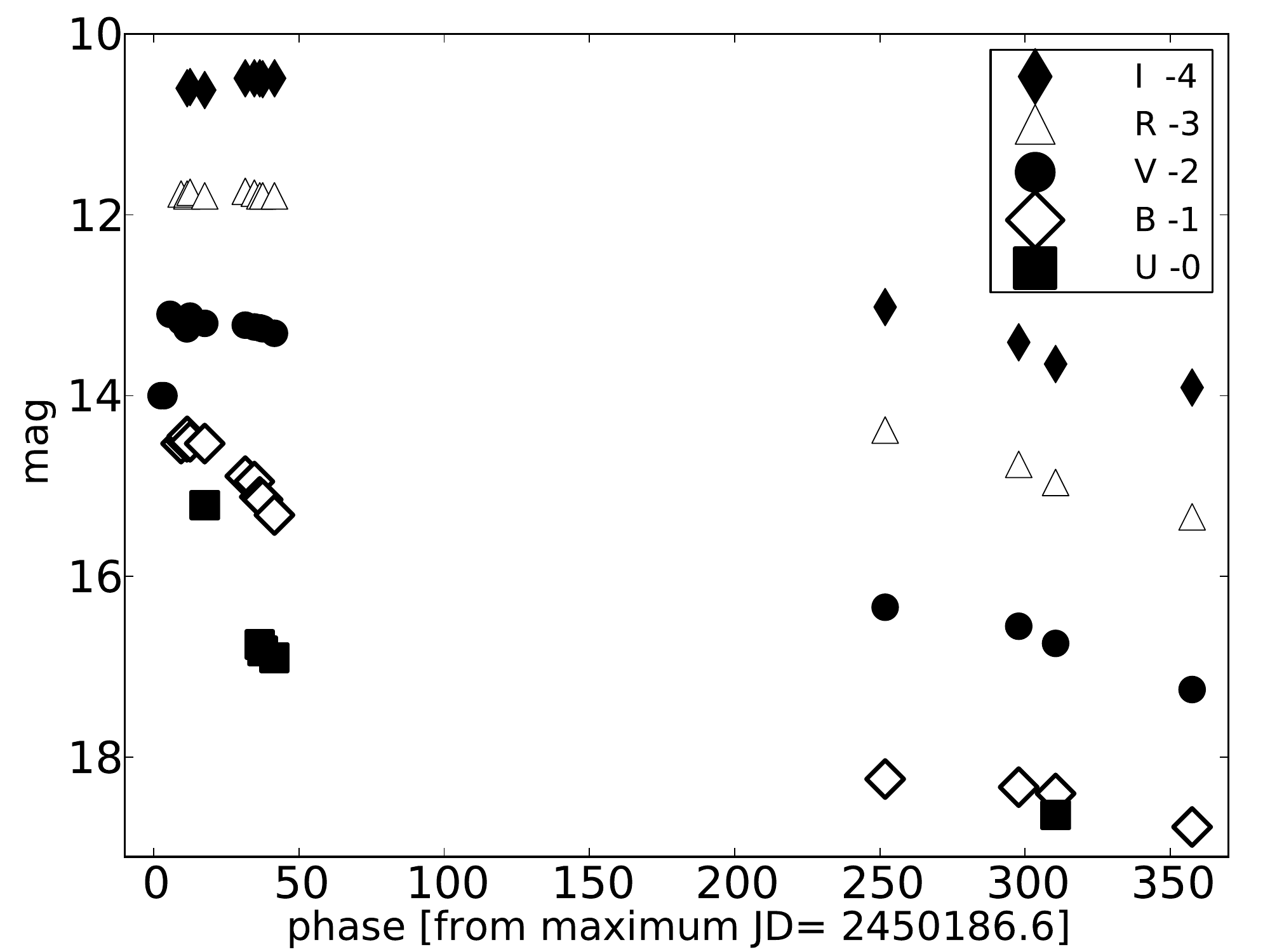}
\caption{Light curves of \ww\/  in all available bands. Shifts compared to the original values reported in Tab.~\ref{table:snm96w} are in the legend.} 
\label{fig:sn_lc96w}
\end{figure}

Our photometric observations started a few days after the discovery. The SN magnitudes are reported in Tab~\ref{table:snm96w} along with photometry reported in the IAUC; the light curves are shown in Fig.~\ref{fig:sn_lc96w}.
The early discovery magnitudes and the spectral comparison with GELATO agree in dating the explosion shortly before the discovery. 
Thus we  adopt  as epoch of the explosion JD $2450180.0\pm3$. 

Assuming the distance and extinction values discussed in Sect.~\ref{sec:sne}, we find
$M^{\rm max}_{B} = -17.59 \pm 0.26$,
$M^{\rm max}_{V} = -17.51 \pm 0.22$, 
$M^{\rm max}_{R} = -17.77\pm 0.20$ and
$M^{\rm max}_{I} \leq -17.78 \pm 0.18$.

The available observations show a flat and bright (M$_{V}\sim-17.5$)
plateau in the VRI bands during the first $\sim$40 days and instead a linear decline in the U and B bands until the SN disappeared behind the Sun. 
The SN was recovered after day 250. The late time decline rates in the various bands are $\gamma_{B}\sim0.50$, $\gamma_{V}\sim0.86$, $\gamma_{R}\sim0.92$ and $\gamma_{I}\sim0.86$ \mcento, not dissimilar from that of the  \co\/ to \fe.

\begin{table}
  \caption{Main parameters of the SNe~II used as reference.}
  \begin{center}
  \begin{tabular}{cccccc}
  \hline
  \hline
  SN & $\mu^{*}$ & \ebv & M$^{\rm pl}_V$$^{*}$ &Parent Galaxy & References  \\
 \hline
 1979C & 31.16 & 0.01 & - & NGC4321 & 1\\
  1987A & 18.49 & 0.19 & -& LMC & 2 \\
 1992H & 32.38  & 0.03 &-17.3 &NGC 5377& 3 \\
 1998S & 31.08 & 0.23 & -&NGC 3877&4 \\
 1999em & 29.47 & 0.10 &-15.6&NGC 1637& 5,6 \\
 2004et & 28.85 & 0.41 & -17.0 &NGC 6946 & 7 \\
 2005cs & 29.62 & 0.05 & -15.0 &M 51& 8 \\
 2005gl & 34.03 & 0.30 & -17.7$^{\ddagger}$ & NGC 266 &9\\
 %2006bp & 31.44 & 0.03 & -16.5$^{\dagger}$ &NGC 3953 & 10\\
 2007od & 32.05 & 0.04 & -17.4 &UGC 12846 & 10\\
 2009bw & 31.53 & 0.31 & -17.2 &UGC 2890 & 11\\
 2009kf & 39.69 & 0.31 & -18.3 & SDSS J16& 12\\
  \hline
\end{tabular}
\end{center}
$^{*}$ Reported to a H$_{0}=73$ \kms\/ Mpc$^{-1}$ distance scale\\
$^{\ddagger}$ unfiltered\\
%$^{\dagger}$ R band\\
REFERENCES: 1 - \citet{bal80}, 2 - \citet{ar89}, 3 - \citet{cl96}, %3 - \citet{94w}, 
%3 - \citet{pphdt}, 
4 - \citet{fa01}, 5 - \citet{elm03}, 6 - \citet{bar00}, 7 - \citet{mag10}, %6 - \citet{05cs}, 
8 - \citet{pa09}, 9 - \citet{gy07}, %10 - \citet{qu07}, 
10 - \citet{07od}, 11 - \citet{09bw}, 12 - \citet{bot10}.
\label{table:snc}
\end{table}

\subsection{Colour evolution}\label{sec:col}

%{\bf MT: quanto segue, in italico, mi sembra fuori luogo. Cosa vuoi dire? che relazione ha con color curves ?}
%{\it In order to estimate the intrinsic luminosity of the objects of our sample, it is essential to establish their distances, the phases and the extinctions to which they are subject. All this informations have been presented in the previous Sect.~\ref{sec:sne} \& \ref{sec:pe}. Only for the \aj\/ is missing information about a possible extinction related to the SN position in the parent galaxy. 
%{\bf MT: non mi sembra. Il fatto che non ci siano assorbimenti e' (!!) informazione che reddening e' basso. Lo stesso dicasi x 95ad. }\\
%As we will show in Sec.~\ref{sec:spec}, the SNe, presented as new set of luminous objects, show early spectra with blue continua (except \aj\/ for which the first spectrum is at $\sim$22d post explosion) and broad P-Cygni of H (Balmer series) and HeI  $\lambda$5876 lines, typical of young type II. As reported in the previous Sec.~\ref{sec:sne} both spectroscopic and photometric evidences lead us to conclude that these SNe have been discovered close to explosion. More uncertain is the explosion epoch of \aj\/, for which we have assumed that it exploded 3 days before discovery (cfr. Sect.~\ref{ss:10aj}). }

 \begin{figure}
%\vspace{174pt}
%\resizebox{\hsize}{!}{\includegraphics{col_sample.pdf}}
\includegraphics[width=\columnwidth]{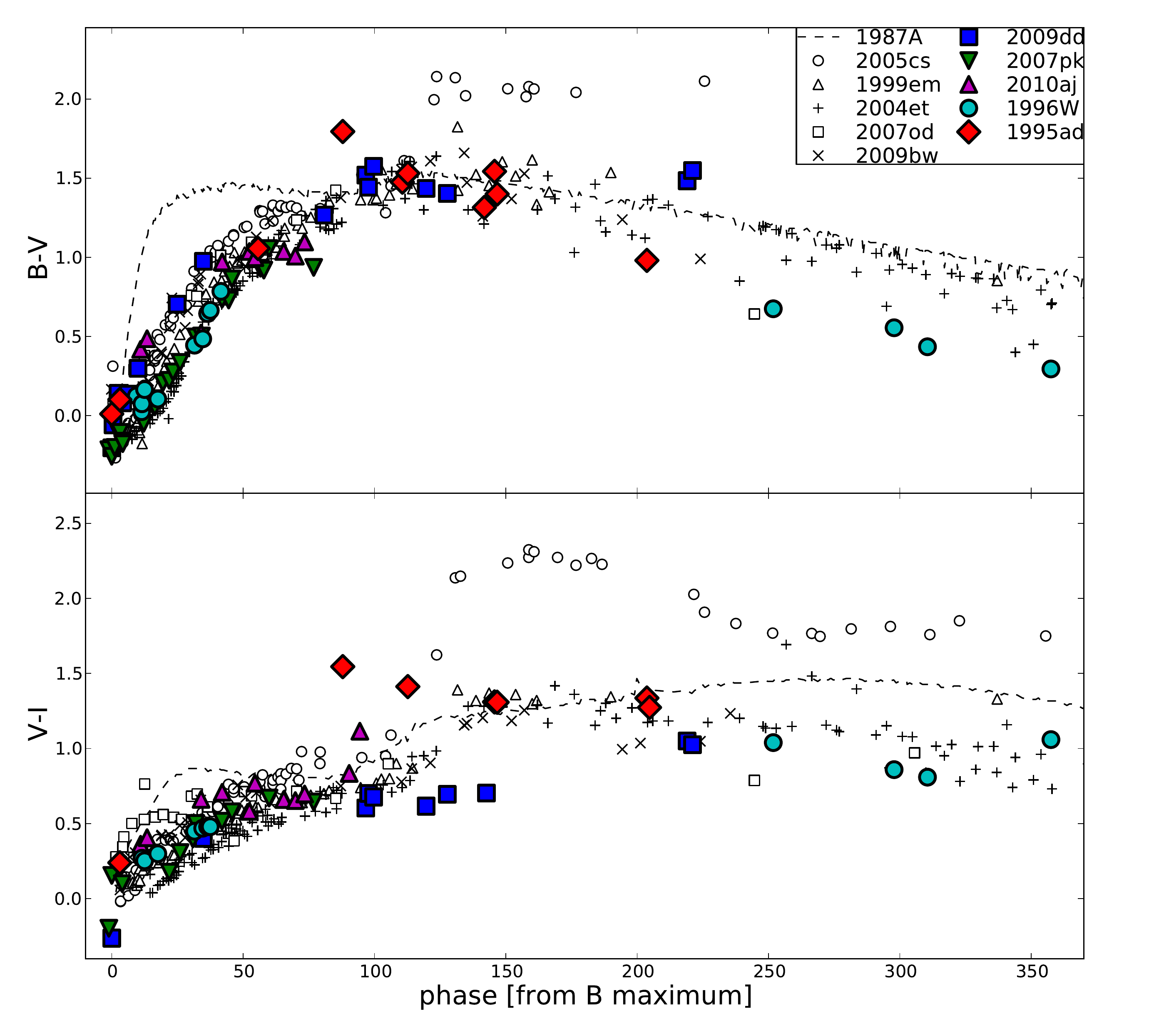}
\caption{
Comparison of the dereddened colours of our sample of SNe and those of  SNe 1987A, 2005cs, 1999em, 2004et, 2007od and 2009bw. The phase of SN 1987A is with respect to the explosion date.}
\label{fig:col}
\end{figure} 

The colour curves of the five SNe, valuable to test their degree of homogeneity, are reported in Fig.~\ref{fig:col} along with those of \a, the faint \cs\/, the normal \em\/ and the luminous SNe 2004et, 2007od and 2009bw (Tab.~\ref{table:snc}). 
All these SNe IIP show quite similar colour evolutions with a rapid increase of the (B--V) colour as the SN
envelope expands and cools down. After about 40 days the colour varies more
slowly as the cooling rate decreases, reaching a value of $\sim$1.5 mag at $\sim$100d. 
The remarkable exceptions to this uniform trend are SN 1987A, faster at early times, and SN 2005cs having a red
spike at about 120d, which seems to be a common feature in low-luminosity SNe IIP \citep{pa04}. 
After 150 days all SNe show a slow turn to bluer colours. Among the SNe presented in this paper, the (B--V) colour curve of \dd\/ seems redder although this might be related to an underestimate of the colour excess.

The (V--I) colour increases for all SNe II during the first days past explosion; it remains roughly constant during the plateau phase ($V-I\sim0.7$) and has a further increase during the post-plateau drop. Then it is fairly constant during the nebular phase (Fig.~\ref{fig:col} bottom). 
The objects of our sample have similar colours at all epochs as other type II SNe.
%{\bf Only the V--I color of \dd\/ remains always roughly constant after 50 days. NON ENFATIZZEREI LA COSA, MI PARE IRRILEVANTE}.
%The comparison with SNe 1999em and 2004et highlights as \ad\/ begins to be red $\sim$30 days before the others.
%The SNe with nebular points in V--I (SNe 2009dd, 1995ad and 1996W) have the same behaviour of SN 2004et and SN 1999em.

\subsection{Bolometric light curves and Ni masses}\label{sec:bol}

\begin{figure*}
%\vspace{174pt}
%\includegraphics[width=\columnwidth]{cfr_bol_all_n2.pdf}
\includegraphics[width=18cm]{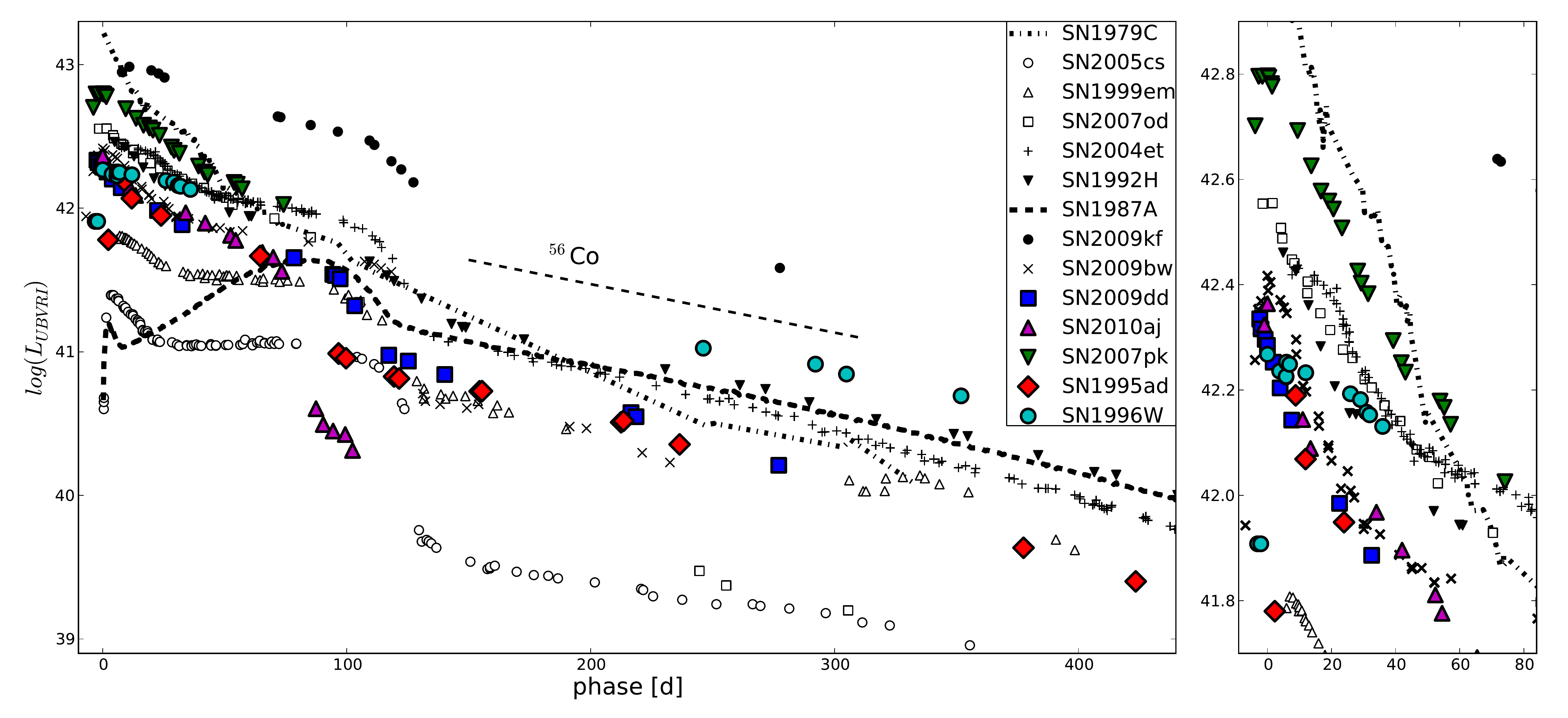}
\caption{Comparison of quasi-bolometric light curves of our sample (UBVRI-integrated for \dd, \pk, \aj\/ and \ww\/ and BVRI-integrated for \ad\/)  with those of other type II SNe; most of them are bright and thus different from the general distribution in luminosity of type II SNe. The phase is with respect to the maximum, only for SN 1987A is with respect to the explosion epoch.
Minor misalignments in the epoch of maxima are due to the different epochs adopted for the maxima of the reference band light curve and the quasi-bolometric curve. On the right a blow up of the SNe of our sample in the first 80d.}
\label{fig:bol}
\end{figure*}

Because of the incomplete UV-optical-NIR coverage, it is impossible to obtain true bolometric light curves. We computed the quasi-bolometric light curves by integrating the fluxes of the available UBVRI photometry (B to I for \ad). % and refer hereafter to these simply as $bolometric$.
Broad band magnitudes were converted into fluxes at the
effective wavelengths, then were corrected for the adopted extinctions (cfr. Sect.~\ref{sec:sne}),
and finally the resulting Spectral Energy Distributions (SED) were integrated
over wavelengths, assuming zero flux at the integration limits.
Fluxes were then converted to luminosities using the distances adopted in
Sect.~\ref{sec:sne}. The emitted fluxes were computed at phases in which R or V
observations were available. When observations in other bands were
unavailable in a given night, the magnitudes were obtained by
interpolating the light curves using low-order polynomials, or were extrapolated using constant colours.
%In the \pk\/, \aj\/, \ad\/ and \ww\/ cases, 
Pre-maximum estimates are based mainly on single R- or V-band observations and
should be regarded as most uncertain.
%, especially for the \aj\/ because of the reasons explained in Sec.~\ref{ss:10aj}. 
The quasi-bolometric light curves of our sample are displayed in Fig.~\ref{fig:bol} along with those of the reference SNe of Tab.~\ref{table:snc}. 
We notice that the possible contribution to the total flux in  the nebular phase from the NIR bands \citep[likely $15\%-20\%$,][]{07od}  was neglected.

The quasi-bolometric peaks of \pk\/, \aj\/ and \ad\/  are reached close to the R maximum and those of \dd\/ and \ww\/ closer to the V maximum.
The quasi-bolometric luminosities at maximum light are reported in Tab.~\ref{table:main}. They range between 1.5 and $6.3 \times 10^{42}$ erg s$^{-1}$.
%L$_{bol}$ = 2.16 x 10$^{42}$ erg s$^{-1}$,  L$_{bol}$ = 6.26 x 10$^{42}$ erg s$^{-1}$, L$_{bol}$ = 2.68 x 10$^{42}$ erg s$^{-1}$, 
%L$_{bol}$ =  1.55 $\times$ 10$^{42}$ erg s$^{-1}$ and L$_{bol}$ =  2.31 $\times$ 10$^{42}$ erg s$^{-1}$ respectively.
The peak luminosities for all objects are moderately bright, only slightly fainter than those of the luminous SNe 2007od, 2004et, 1992H.  
%The luminosities are significantly lower than that of SN 2009kf \citep[][]{bot10}, the brightest %known type IIP, suggesting a different explosion mechanism and progenitor for this event. 
Four out of five objects have long and bright plateaus, comparable in luminosity to that of \bw. Fig.~\ref{fig:bol} points out the early transition ($\sim$80d--100d) to the radioactive tail of \aj\/, \dd\/ and \ad~ (the end of the plateau of \ww\/ was missed).
%and \aj, but the drop in magnitude that leads to the nebular phase occurs slightly before the other SNe of the comparison (except \aj\/) and it is similar in phase to that of \od\/.
Instead \pk\/ shows a peak luminosity higher than other objects of our sample and a linear decline during the photospheric phase though with a rate that is smaller than for the prototypical type IIL SN 1979C.

%As written before, the peak luminosity of \aj\/ is similar to that of \dd\/ and \bw\/. It shows a dramatic drop from plateau similar to that of \od\/. This can suggest a progenitor with a small H layers and/or early dust formation as for \od\/.
%The comparison shows as also \ad\/ and \ww\/ have peak luminosity similar to that of \bw\/ and \dd\/.  \ad\/ is rather comparable to \aj\/ and presents a behaviour also similar to \od\/. 

The tails of the quasi-bolometric light curves of \dd,  \ad\/ and \ww\/ 
have slopes close to that of the decay of \co\/ to \fe\/ (cfr. Tab.~\ref{table:main}) allowing the determination of the ejected \ni\/ mass.
The observations of \aj\/ ended just at the beginning of the radioactive tail, thus we feel confident to provide only an upper limit to the \ni\/ mass, while not even this is available for  \pk.

The $^{56}$Ni mass has been derived by comparing the late quasi-bolometric light curves integrated over the same wavelength range as  \a, assuming a similar $\gamma$-ray deposition fraction
\begin{equation}\label{eq:1}
M(^{56}Ni)_{\rm SN}= M(^{56}Ni)_{\rm 87A}\times\frac{L_{\rm SN}}{L_{\rm 87A}} M_{\odot} 
\end{equation}
where the mass of \ni\/ ejected by \a\/ is M(\ni)$_{\rm 87A}=0.075\pm0.005$ \M\/ \citep{ar96}, and L$_{\rm 87A}$ is the
quasi-bolometric luminosity at comparable epoch. The comparisons give M($^{56}$Ni)$_{\rm 09dd}\sim0.029$ M$_{\odot}$, M(\ni)$_{\rm 95ad}\sim0.028$ \M\/ and  M(\ni)$_{\rm 96W}\sim0.14$ \M\/. 
With the same assumption on the full thermalization of the $\gamma-$rays, we cross-checked these results with the formula
\begin{equation}\label{eq:2}
\textstyle M(^{56}Ni)_{\rm SN} = \left(7.866\times10^{-44}\right) L_{\rm t}exp\left[\frac{(t-t_{0})/(1+z)-6.1}{111.26}\right] M_{\odot}
\end{equation}
from \citet{h03a}, where t$_{\rm o}$ is the explosion epoch, 6.1d is
the half-life of $^{56}$Ni, and 111.26d is the \textit{e}-folding time
of the $^{56}$Co decay
%, which releases 1.71 MeV and 3.57 MeV respectively as $\gamma$-rays 
\citep{ca97,wo89}. This method
yields M(\ni)$_{\rm 09dd}\sim0.027$ \M\/, M(\ni)$_{\rm 95ad}\sim0.025$ \M\/ and M(\ni)$_{\rm 96W}\sim0.13$ \M, in agreement with the above determinations. 
%and in graphic confirmation with Fig.~\ref{fig:bol}.
For \aj\/ we estimated  an upper limit M(\ni)$_{\rm 10aj}<0.007$ \M\/, based on the last epoch in which the SN was detected.

\section{Spectroscopy}\label{sec:spec}

In this Section we present and discuss the spectral evolution of the five SNe of our sample (Tab.~\ref{table:sp}).

The data were reduced (trimming, overscan, bias correction and flat-fielding) using standard IRAF routines. Optimal extraction of the spectra was adopted to improve the signal-to-noise (S/N) ratio. Wavelength calibration was performed using spectra of comparison lamps acquired with the same configurations as the SN observations. Atmospheric extinction correction was based on tabulated extinction coefficients for each telescope site. Flux calibration was done using spectro-photometric standard stars observed in the same nights with the same set-up as the SNe. The flux calibration was checked against the photometry, integrating the spectral flux transmitted by standard {\it BVRI} filters and adjusted by a multiplicative factor when necessary. 
The resulting flux calibration is accurate to within 0.1 mag. The spectral resolutions in Tab.~\ref{table:sp} were estimated from the full widths at half maximum (FWHM) of the night sky lines.  Whenever possible we used the spectra of standard stars to remove telluric features in the SN spectra. %The regions of the strongest atmosphere features are marked in Fig.~\ref{fig:spec_ev09dd}, \ref{fig:spec_ev07pk}, \ref{fig:spec_ev10aj}, \ref{fig:spec_ev95ad} \& \ref{fig:spec_ev96w} (the NIR spectrum in Fig.~\ref{fig:}.

\subsection{Individual properties} \label{sec:indiv}

In the following we will discuss the spectroscopic properties of the individual SNe showing relatively normal behaviors for four objects, with some evidences of  weak ejecta-CSM interaction, and more marked signatures of interaction  during the first months for \pk\/.

\begin{figure*}
%\vspace{174pt}
\includegraphics[width=18cm]{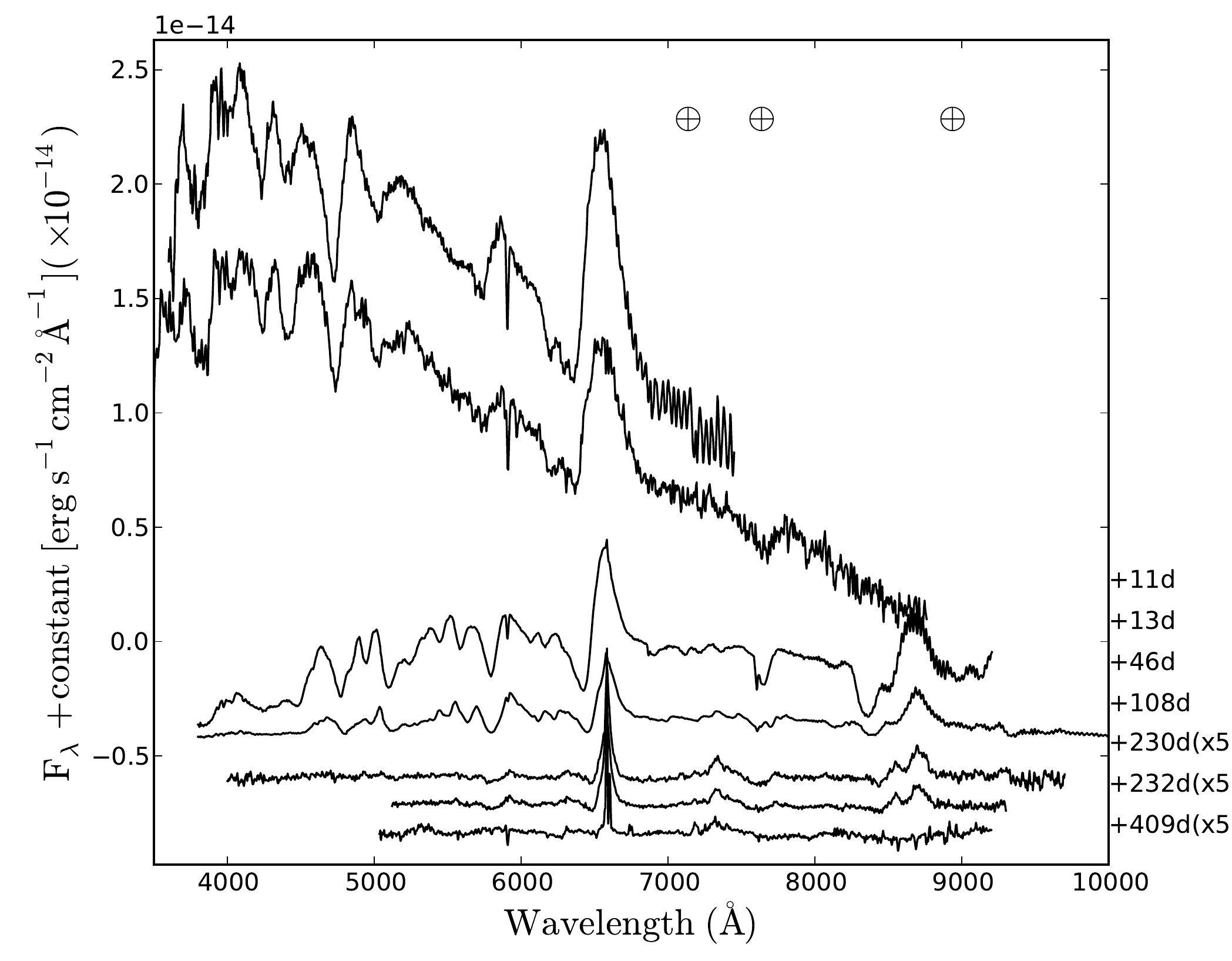}
\caption{The complete spectral evolution of \dd. Wavelengths are
  in the observer's rest-frame. The phase for each spectrum
  is relative to the adopted explosion date (JD 2454925.5), late spectra
  have been multiplied by a factor 5 to emphasise the nebular lines. The $\oplus$ symbols
  mark the positions of the strongest telluric absorptions. The
  ordinate refers to the top spectrum; the other spectra are shifted
  downwards by $4\times 10^{-15}$ (second and third)
  and $1.4\times 10^{-15}$ (others) erg s$^{-1}$ cm$^{-2}$ \AA$^{-1}$.}
\label{fig:spec_ev09dd}
\end{figure*}

\subsubsection{\dd}\label{sssp:09dd}
Seven spectra are available for this object tracing the evolution from about 11d post explosion to about 14 months (Fig.~\ref{fig:spec_ev09dd}). 
%In the last spectrum however it is difficult to recognize the SN spectrum from the galaxy background.
The first spectrum shows a blue continuum comparable to that of other young SNe II. It is characterised by the P-Cygni features of the H Balmer series, He~{\sc i} $\lambda$5876, Fe~{\sc ii} multiplets (e.g. $\lambda\lambda$4924, 5018, 5169) and H\&K of Ca~{\sc ii}. 
On the blue side of the \Ha\/ emission, at about  6200\AA, three absorption features are present.
The bluest absorption, at $\sim$6174\AA\/ is possibly identified as Si~{\sc ii}, with an expansion velocity comparable to those of the other metal ions.
This line was also identified in other type II SNe \citep[e.g. \h\/, \od\/, \bw\/;][and references therein]{07od,09bw}.
The middle feature is probably Fe~{\sc ii} $\lambda$6456  but the line strength would indicate a much higher optical depth than usual for metal lines at this stage. %Similarly also the Fe~{\sc ii} multiplet $\lambda$5169 %has a strength uncommon for this phase presents a deeper absorption. 
Alternatively, this line might be a high velocity component (HV) of  \Ha. 
%However, no other H lines related to other Balmer lines are clearly visible. 
Actually a tiny absorption blue ward of \Hb\/  is visible at $\sim$4620\AA\/ but 
the resulting expansion velocity is larger than the putative \Ha\/ component (HV(\Ha)$=13800$ \kms\/ vs. HV(\Hb)$=14800$ \kms).  
The presence of a HV feature could be a signature of early interaction of the ejecta with the CSM \citep{chu07}, consistent with the X-ray detection (cfr. Sect. \ref{sec:sne}).
Unfortunately, we do not have other spectra at similar epochs to confirm this line identification.
The reddest observed component is the \Ha\/ P-Cygni absorption, indicating an expansion velocity of $\sim$ 11000 \kms.

The third spectrum was taken during the plateau phase. It shows well-developed P-Cygni profiles of the Balmer series and a number of metal lines, including Fe~{\sc ii} at $\sim$4500\AA\/ and Sc~{\sc ii} $\lambda$5031 on the red side of \Hb\/. Fe~{\sc i} and Sc~{\sc ii} are visible at about 5500\AA\/, while Na~{\sc iD} has now replaced He~{\sc i} $\lambda$5876. Other prominent metal lines  are Ba~{\sc ii} $\lambda$6142 and Sc~{\sc ii} $\lambda$6245. In the red  ($\lambda>7000$ \AA\/) the Ca~{\sc ii} near-IR triplet is one of the strongest spectral features. In addition, O~{\sc i} at $\sim$7774\AA\/ and N~{\sc ii} at $\sim$8100\AA\/ are possibly detected. The same features are visible in the fourth (late-photospheric) spectrum (+108d).

The series of three nebular spectra (230d-409d) show emissions of Na~{\sc iD}, [O~{\sc i}] $\lambda\lambda6300, 6364$, \Ha,  [Fe~{\sc ii}] $\lambda7155$, [Ca~{\sc ii}] $\lambda\lambda7291$, 7324 doublet and the Ca IR triplet.  In the latest spectrum (409d) SN emission features are still visible (namely [O~{\sc i}], [Fe~{\sc ii}] and [Ca~{\sc ii}] features) though heavily contaminated by narrow emission lines of an underlying H~{\sc ii} region (\Ha, [N~{\sc ii}], [S~{\sc ii}]).

%%%%%%%%%%%%%%%%%%%%%%%%%%%%%%%%%%%
\subsubsection{\pk}\label{sssp:07pk}
\begin{figure*}[]
\centering
%\vspace{174pt}
\includegraphics[width=14cm]{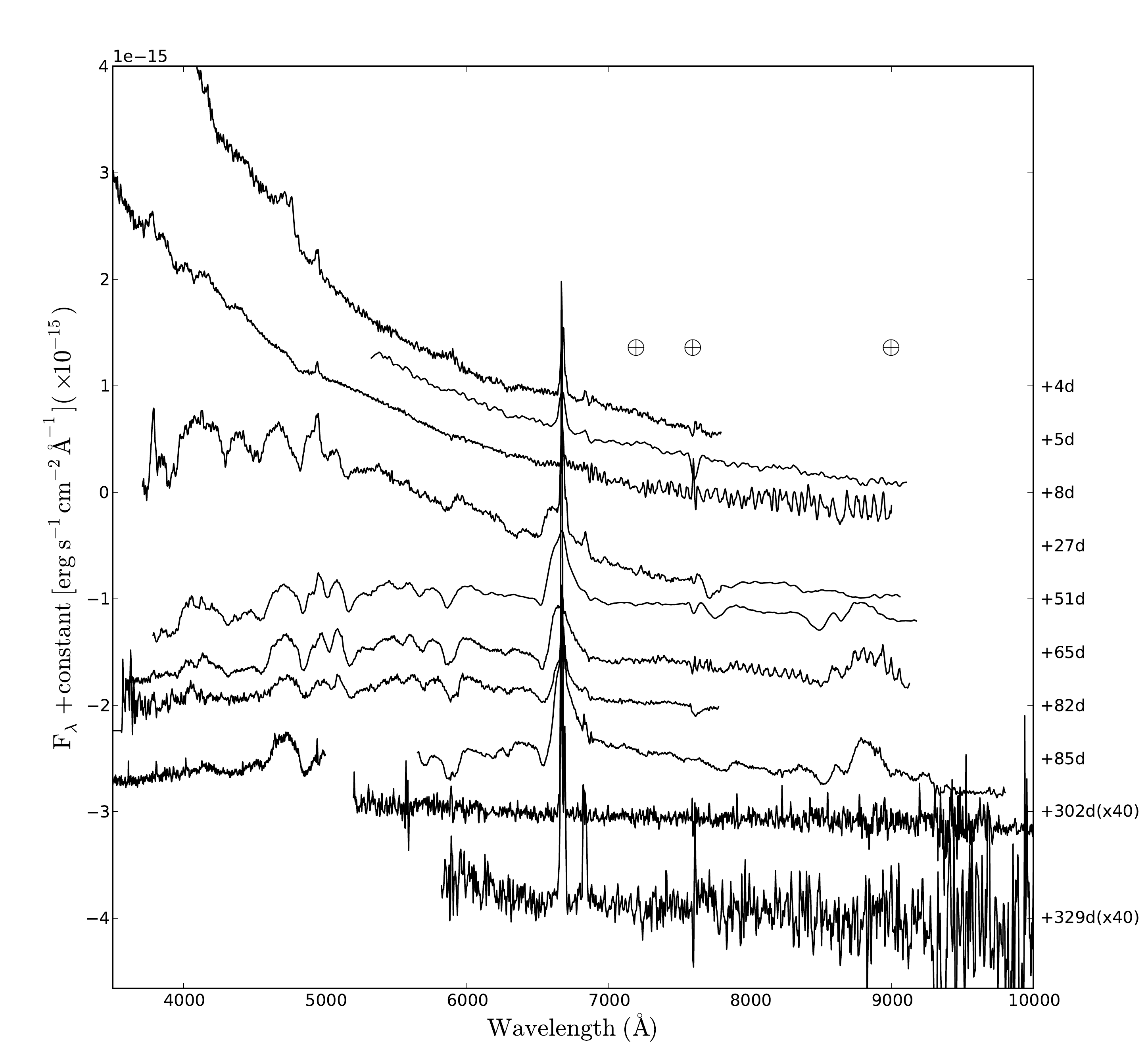}
\caption{The spectral evolution of \pk\/. Wavelengths are
  in the observer's rest frame. The phase reported for each spectrum
  is relative to the explosion date (JD 2454412), late spectra
  have been multiplied by a factor 40 to emphasise the nebular lines. The $\oplus$ symbols
  mark the positions of the strongest telluric absorptions. The
  ordinate refers to the top spectrum; the other spectra are shifted
  downwards with respect to the previous one by $7\times 10^{-16}$,
  except the third, which is shifted by $1\times 10^{-15}$  erg s$^{-1}$ cm$^{-2}$ \AA$^{-1}$}
\label{fig:spec_ev07pk}
\end{figure*}

Fig.~\ref{fig:spec_ev07pk} shows the photospheric evolution from a few days post explosion ($\sim$4d) to about 3 months. Two spectra at 10 and 11 months were also obtained when the SN was barely visible and narrow lines from the underlying H~{\sc ii} region are mostly detected.

The first spectra (4d-8d) show a blue continuum  with \Ha\/ and \Hb\/  emission and no prominent absorption features.
As reported in \citet{par07}, these are the characteristics of type IIn SN spectra.
The prominent feature bluer of \Hb\/ is identified as He~{\sc ii} $\lambda$4686 possibly blended with C~{\sc iii}/N~{\sc iii} like in the case of \s\/ \citep{fa01}. Broad Balmer absorptions begin to emerge on day 8 (although they are still weak). %From the third spectrum (8d) the absorption components of the Balmer series are roughly visible but the emission components are still predominant. 
%This suggests that SN 2007pk is an interacting core-collapse SN.

\begin{figure}[!hb]
\begin{center}
\includegraphics[width=\columnwidth]{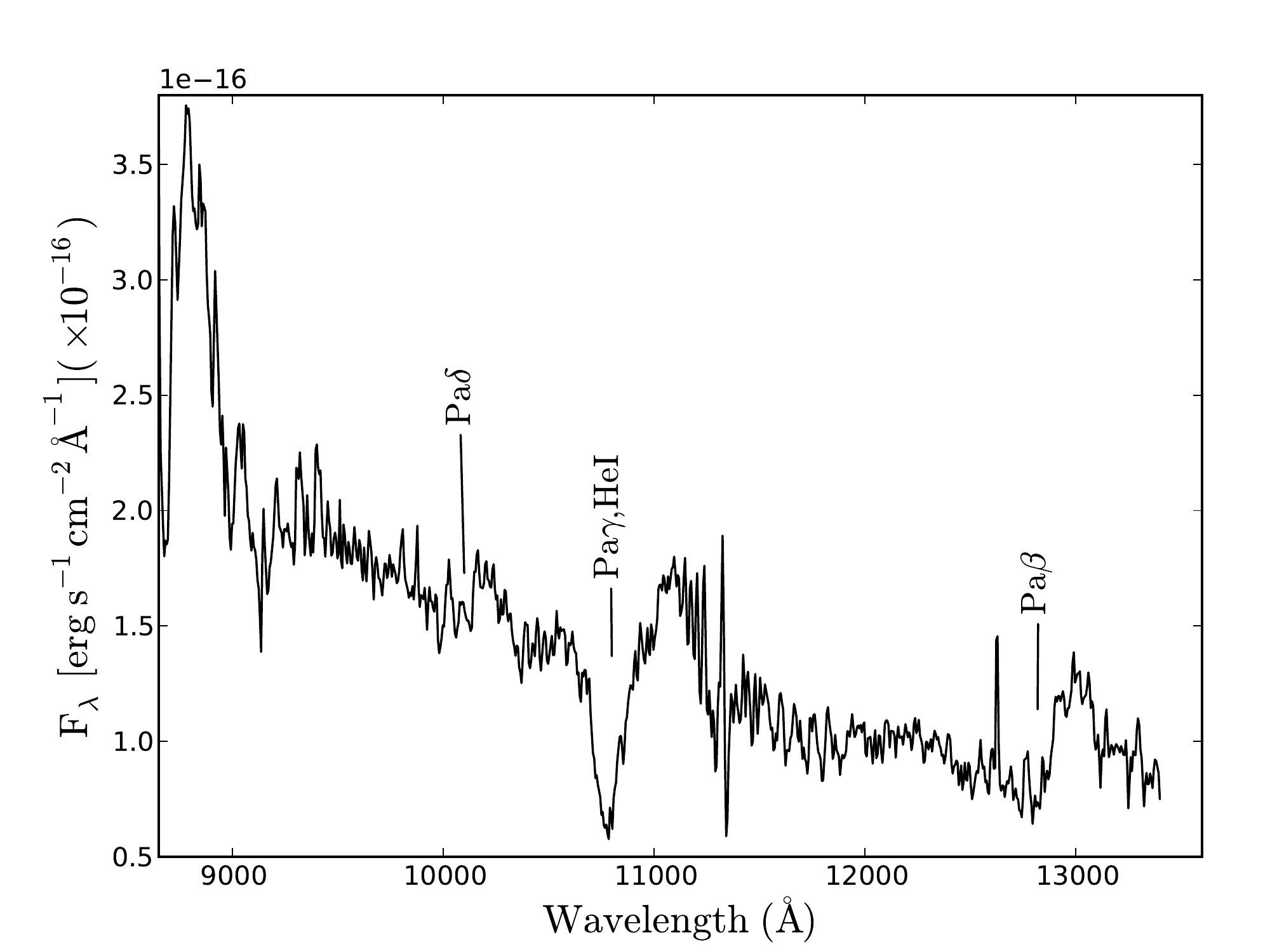}
\caption{The NIR spectrum of SN 2007pk at $\sim66$d after the explosion date (JD 2454412). Wavelengths are in the observer's rest-frame.} 
\label{fig:spec_nir07pk}
\end{center}
\end{figure}

The spectrum at 27d shows well-developed P-Cygni lines of metal elements such as Fe~{\sc ii} at $\sim$4500\AA\/ and Sc~{\sc ii} $\lambda$503. Also the Fe~{\sc ii} multiplet $\lambda\lambda$4924, 5018, 5169 and Sc~{\sc ii} $\lambda$6245 are visible. The Na~{\sc iD} feature is blended with He~{\sc i} $\lambda$5876. The \Ha\/ and \Hb\/ still show a dominant emission component with an absorption component comparable with those of other SNe II at the same phase. Nevertheless, the absorption profiles display flat, blue shoulders, possibly an evidence of a residual interaction of the SN ejecta with the CSM \citep[see][]{ch82,ch94}. 
For this reason, the 27d spectrum may  represent the rare snapshot of the spectral transition from a type IIn to a normal, non-interacting type II SN. 
%{\bf MT:  non condivido questa frase. Mi pare la transizione da una very early SNII a una normal SNII. Certo, la blue shoulder e' particular ed hai fatto bene a menzionarla. CI: ovviamente non sono d'accordo e confermo la mia idea. AP: diciamo che la veritˆ' sta nel mezzo, e ho risistemato la frase. Lo spettro early e' di una type IIn sono d'accordo; qualcosa ionizza un denso CSM, c'e' poi chiara interazione dai profili boxy, e una graduale transizione verso una type II. chiaramente la sparizione di ogni traccia della SN negli spettri late e' assai sorprendenteÉ Non certo tipica di una type IIP.}.

The last set of photospheric spectra (51d-85d) show the evolution of a  canonical type II during the H recombination phase. Fe~{\sc i} $\lambda$4541, Sc~{\sc ii} $\lambda$5527, Ba~{\sc ii} $\lambda$6142, Na~{\sc iD}, O~{\sc i} $\lambda$7774 are visible as well as the Ca~{\sc ii} near-IR triplet ($\lambda\lambda$8498, 8542, 8662).
P-Cygni line profiles of the Paschen series, in particular Pa$\beta$, Pa$\gamma$ and Pa$\delta$ are visible in the day 66 NIR spectrum (Fig.~\ref{fig:spec_nir07pk}). Pa$\gamma$ is blended with He~{\sc i} $\lambda$10830.

The two late-time spectra (302d to 330d) do not show unequivocal  SN features. The narrow unresolved emissions, i.e. \Ha\/ ($\Delta$v$\sim$800 \kms\/), [N~{\sc ii}] $\lambda$6583 and the [S~{\sc ii}] $\lambda\lambda$6717, 6731 doublet, are probably related to the underlying H~{\sc ii} region. 
%{\bf SIAMO SICURI CHE SIANO NON RISOLTE? SE SONO NON RISOLTE, SONO HII REGION, SE SONO RISOLTE A VELOCITA' 800 KM/S SONO PROBABILMENTE CSM IN ESPANSIONE! QUESTA E' UNA DIFFERENZA IMPORTANTE! CI: sono non risolte}. 

\begin{figure*}
%\vspace{174pt}
\includegraphics[width=18cm]{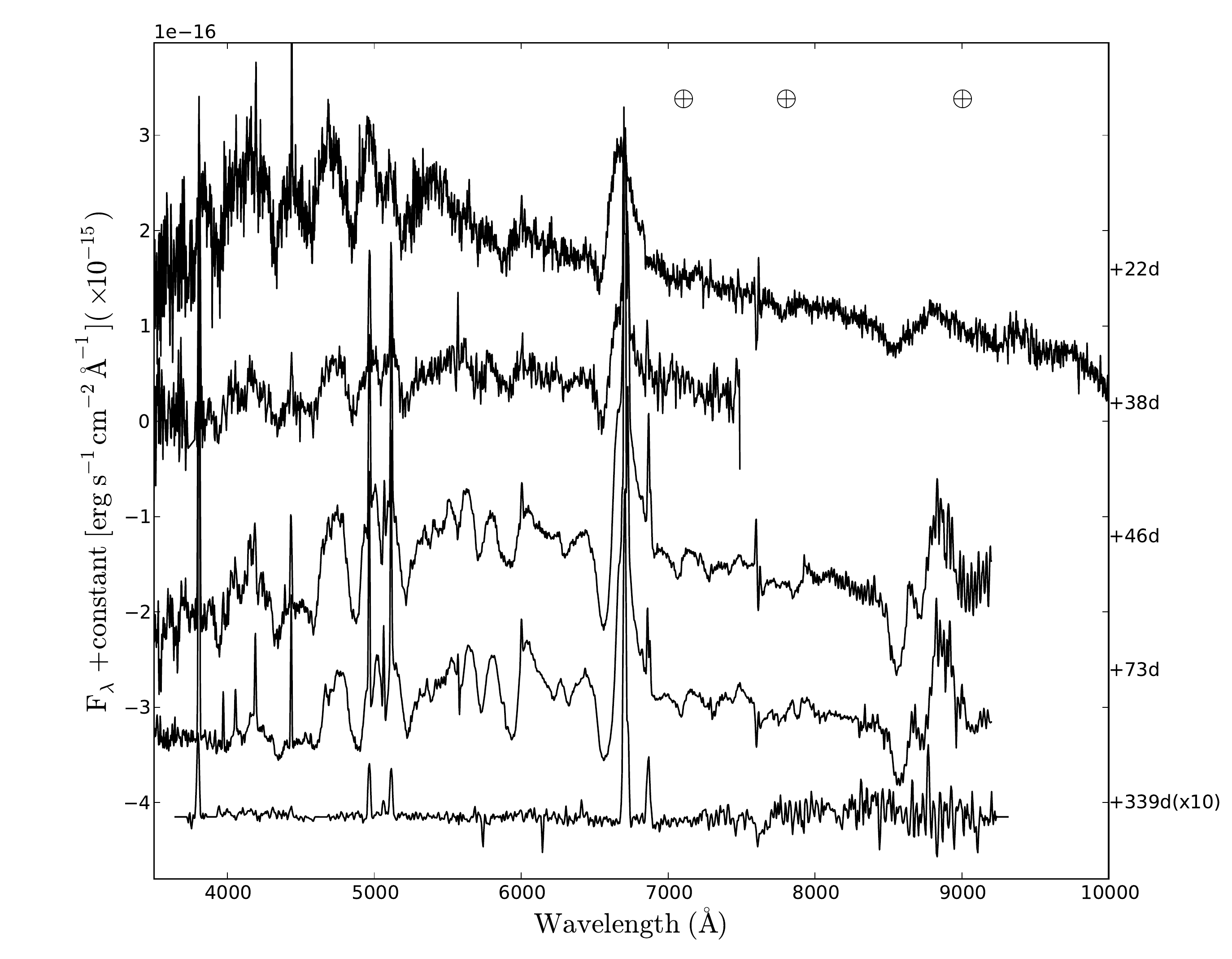}
\caption{The spectral evolution of \aj\/. Wavelengths are
  in the observer's rest frame. The phases reported to the right
  are relative to the explosion epoch (JD 2455265.5), The late spectrum
  has been multiplied by a factor 4 to emphasise the features. The $\oplus$ symbols
  mark the positions of the strongest telluric absorptions. The
  ordinate refers to the top spectrum; other spectra are shifted
  downwards with respect to the previous one by $2\times 10^{-16}$
  and $6\times 10^{-16}$ (only the third) erg s$^{-1}$ cm$^{-2}$ \AA$^{-1}$.}
\label{fig:spec_ev10aj}
\end{figure*}

%%%%%%%%%%%%%%%%%%%%%%%%%%%%%%%%%
\subsubsection{\aj}\label{sssp:10aj}
% As for \dd\/, also \aj\/ has a few and sparse spectra.
The spectral evolution of SN~2010aj  shown in Fig.~\ref{fig:spec_ev10aj} spans $\sim$2 months during the photospheric phase, complemented by an additional observation in the nebular phase, about one year after the explosion.

The first spectrum shows the H Balmer lines (the P-Cygni profile has a weak absorption component), He~{\sc i} $\lambda$5876, Ca~{\sc ii} H\&K, Fe~{\sc ii} lines ($\lambda\lambda$4924, 5018, 5169), and possibly Si~{\sc ii} $\lambda$6355, O~{\sc i} $\lambda$7774 and the Ca~{\sc ii} IR triplet.
In the following three spectra (38d - 73d) other metal lines become prominent, including Ti~{\sc ii}  around 4100\AA\/, while Fe~{\sc ii} and Ba~{\sc ii} contribute to the feature at $\sim$4930\AA. Fe~{\sc i} and Sc~{\sc ii} lines are also clearly detected at about 5500\AA\/, as well as Sc~{\sc ii} $\lambda$6245. The Na~{\sc iD} feature replaces He~{\sc i} $\lambda$5876. The H Balmer lines develop well-designed P-Cygni profiles, always contaminated by the strong narrow components due to an underlying H~{\sc ii} region.
We cannot perform a more detailed analysis because of the sparse temporal sampling.

The late-time spectrum does not show evidence of broad lines due to the SN. 
The \Ha, \Hb, [O~{\sc ii}] and [S~{\sc ii}] unresolved ($\Delta\lambda=17\AA$) lines are related to the host galaxy H~{\sc ii} region.
%Because of the modest resolution of the last spectrum, we can not exclude that \Ha\/ is blended with [N~II].
%Also the blue continuum is probably due to the contamination from a nearby cluster of young hot stars.

%%%%%%%%%%%%%%%%%%%%%%%%%%%%%%%%%%%%%%%
\subsubsection{\ad}\label{sssp:95ad}

\begin{figure*}
\centering
%\vspace{174pt}
\includegraphics[width=15cm]{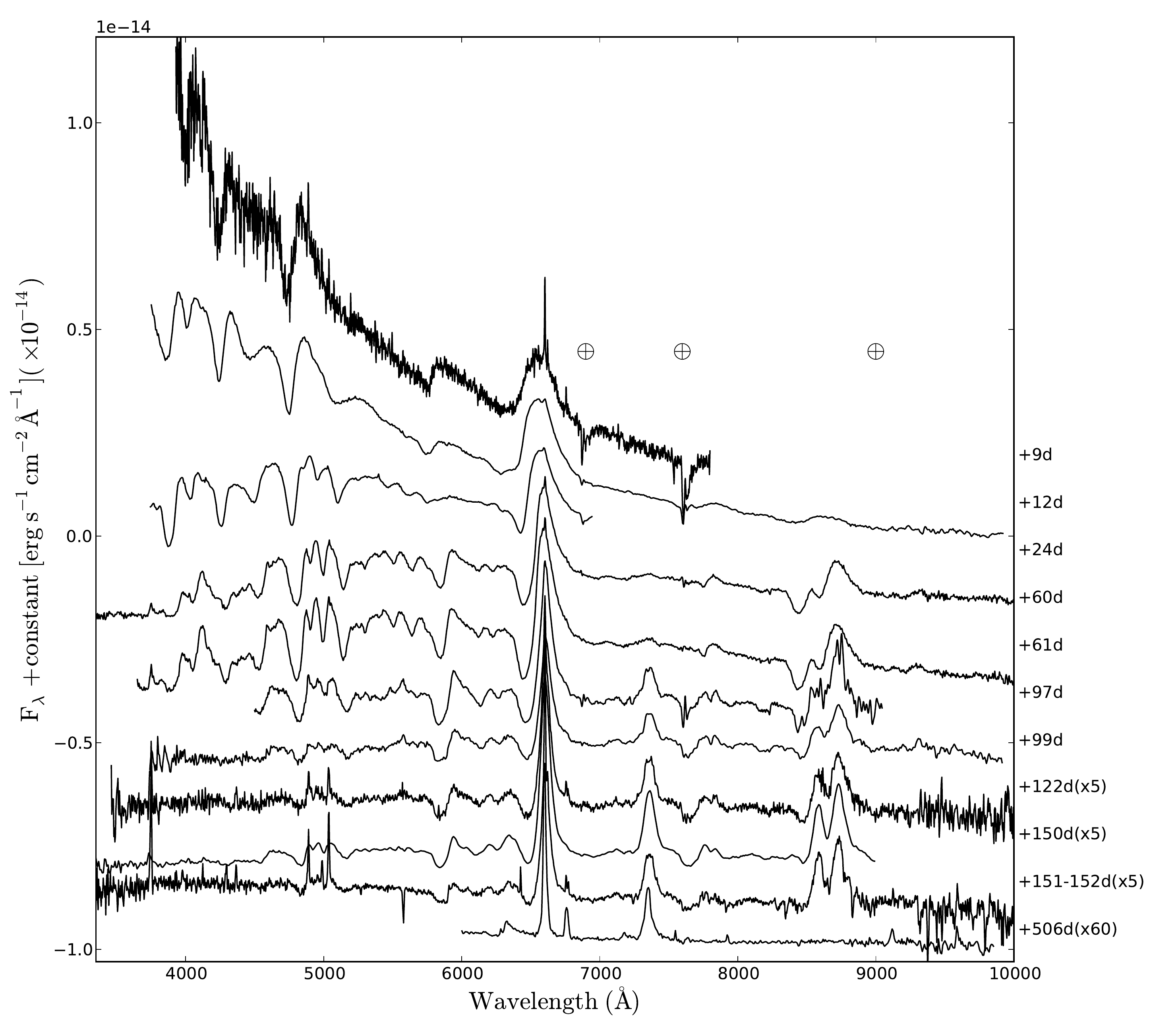}
\caption{Spectral evolution of \ad\/. Wavelengths are in the observer's rest frame. The phase reported for each spectrum is relative to the explosion date (JD 2449981), late spectra
  have been multiplied by a factor 5 and 60 to emphasise the features. The $\oplus$ symbol marks the positions of the most significant telluric absorptions. The second and third spectra are shifted upwards by $1\times 10^{-15}$; the other ones are shifted downwards with respect to the previous by $0.7\times 10^{-15}$  erg s$^{-1}$ cm$^{-2}$ \AA$^{-1}$.} 
\label{fig:spec_ev95ad}
\end{figure*}

The available spectra cover the period from one week to over 500 days after the explosion (see Fig.~\ref{fig:spec_ev95ad}). 
The continuum is very blue at the first epoch and progressively becomes redder and
the most prominent features are those usually detected in type II SNe during the photospheric phase (H Balmer, He~{\sc i}/Na~ID, Fe~{\sc ii}, Ca~{\sc ii}, Sc~{\sc ii} and other metal lines).
Also the line profiles evolve in standard fashion,  from broad P-Cygni profiles at the early epochs to narrow emissions during the nebular phase.
The \Ha\/ profile in the early phase (9d-24d) is reminiscent of that of \od\/ at a similar stage with flat-topped emission and significant absorption only after day 24.
Though one may argue that the flat profile is due to the CSM-ejecta interaction similar to \od\/ \citep{phoenix07od,07od}, most likely the peculiar profile is the result of blending of the blue-shifted SN emission with the narrow emissions from the H II region.
Searching for possible signatures of interaction, we have investigated several high signal-to-noise spectra for the possible presence of  HV features of \Ha\/ and \Hb\/ with no conclusive results.

\begin{figure}
\begin{center}
\includegraphics[width=\columnwidth]{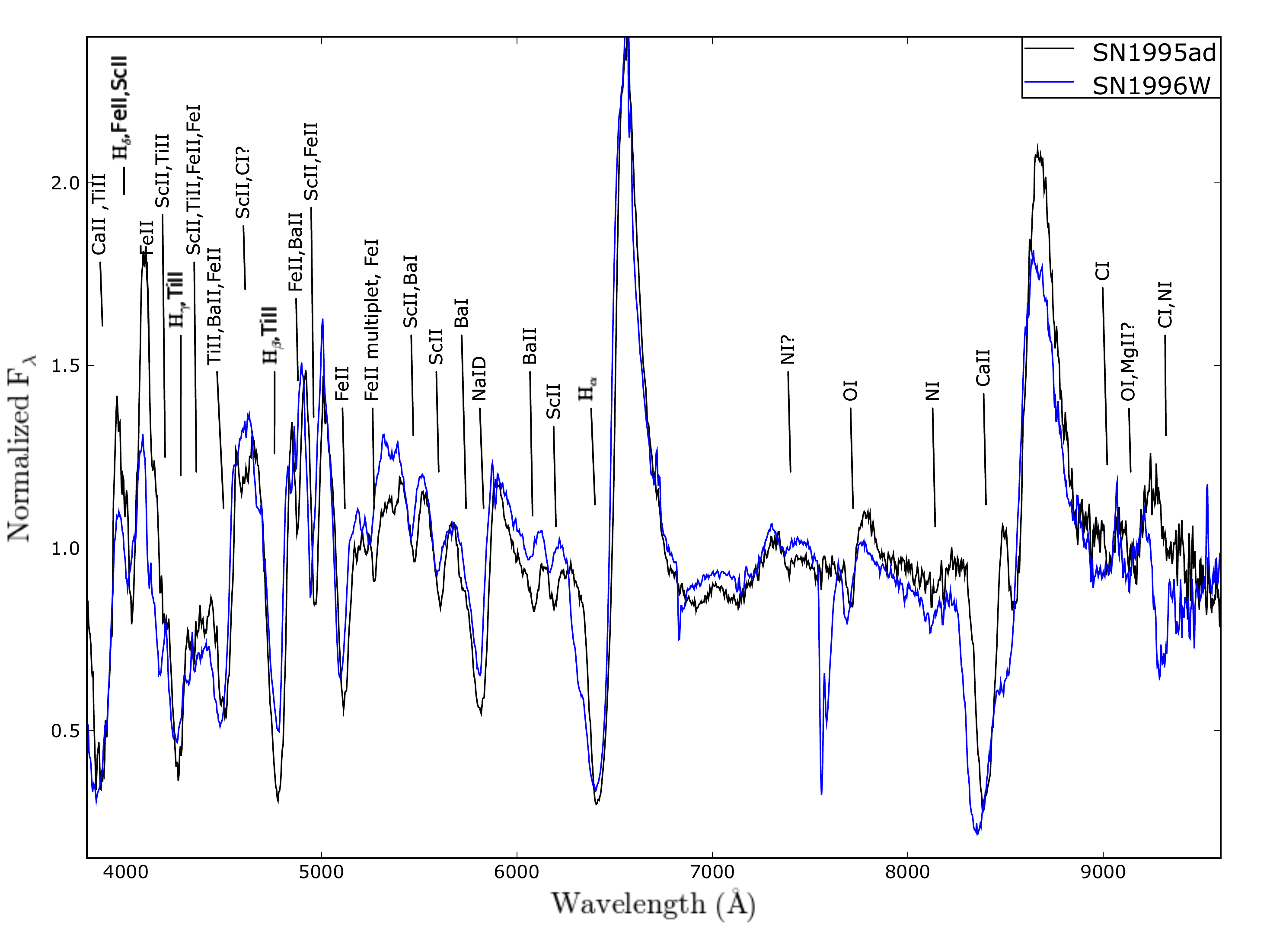}
\caption{Spectra of SN 1995ad at $\sim$61d and SN 1996W at $\sim$67d post explosion. The spectra were corrected for extinction and reported to the SN rest frame. The most prominent absorptions are labelled.}
\label{fig:lineid}
\end{center}
\end{figure}

The line identification during the plateau is shown in Fig.~\ref{fig:lineid} for both \ad\/ and \ww, using spectra of high signal-to-noise and good resolution (FWHM$\sim10$ \AA).
The spectra of \ad\/ show the possible contribution of Ti~{\sc ii} lines in the bluer part  ($<$4500 \AA\/) of the spectrum. Also prominent are the lines of Ba~{\sc ii}, Fe~{\sc ii} and Sc~{\sc ii} around 5000 \AA\/, the strong Sc~{\sc ii} $\lambda$5527 (possibly blended with Ba~{\sc i} $\lambda$5535) and $\lambda$6245, as well as the Ba~{\sc ii} $\lambda$5854 and $\lambda$6142. 
In the red wavelength, in addition to the evident Ca~{\sc ii} NIR triplet, we identify lines of O~{\sc i} $\lambda$7774, N~{\sc i} at $\sim$8120 \AA\/ and O~{\sc i} $\lambda$9260. By comparison with \bw\/ the absorption feature at 9030 \AA\/ may be due to C~{\sc i}.
The presence of O lines becomes more evident in the subsequent spectra.  
Finally we note that [O~{\sc i}] and [Ca~{\sc ii}] emission of the SN are clearly visible in the nebular spectra despite the contamination of nebular lines from the underlying H~{\sc ii} region. %{\it SB: a me sembra che le righe dell'[OI] siano particolarmente deboli a fase +506d, specie in confront a quell del [CaII]!!!}
%In the {\bf two} spectra (151-506d) the [S~II] $\lambda\lambda$6717, 6731 {\bf are associated to an} underlying H~II region.

\begin{figure*}
\centering
%\vspace{174pt}
\includegraphics[width=15cm]{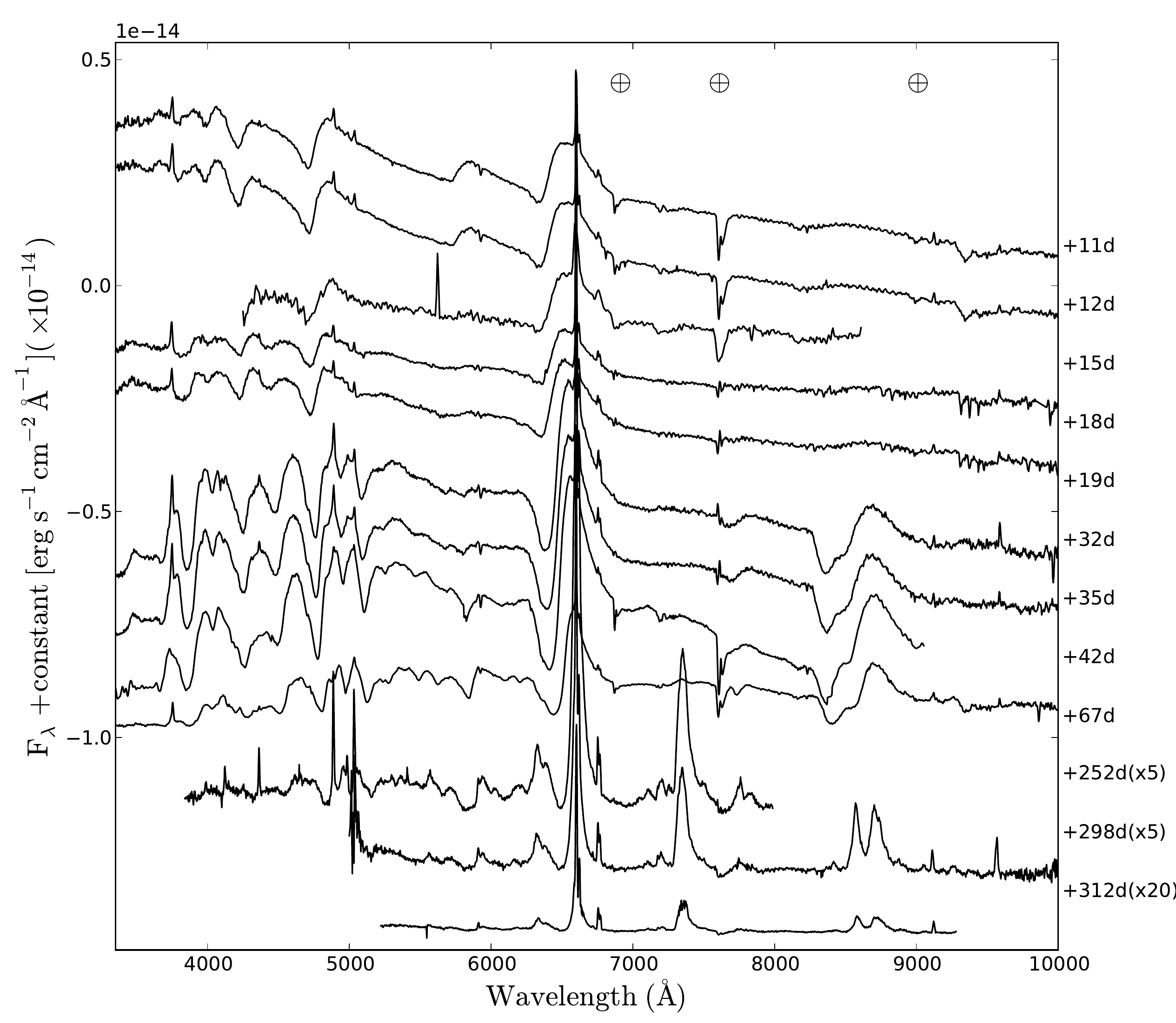}
\caption{The overall spectral evolution of \ww\/. Wavelengths are in the observer's rest frame. The phase reported for each spectrum is relative to the explosion date (JD 2450180), late spectra
  have been multiplied by a factor f5 and 20 to emphasise the features. The $\oplus$ symbol marks the positions of the most important telluric absorptions. 
The spectra are shifted with respect to the first one.}
%The second is shifted upwards by $2\times 10^{-15}$; the other spectra are shifted downwards with respect to the previous by $2.7\times 10^{-15}$ (third spectrum) and  $1.4\times 10^{-15}$ erg s$^{-1}$ cm$^{-2}$ \AA\/$^{-1}$(others)  erg s$^{-1}$ cm$^{-2}$ \AA$^{-1}$.} 
\label{fig:spec_ev96w}
\end{figure*}

%%%%%%%%%%%%%%%%%%%%%%%%%%%%%%%%%%%%%%%
\subsubsection{\ww}\label{sssp:96w}
The observations (cfr. Fig.~\ref{fig:spec_ev96w}) cover the first couple of months after the explosion, until the SN went behind the Sun. The object was recovered during the nebular phase ($252-312$ days). 

The evolution matches fairly well that of normal type IIP SNe, although the line velocities and the continuum temperatures are somewhat higher than in canonical SNe IIP at a comparable phase (see Table \ref{table:main} and Section \ref{sec:spevol}). \Ha\/, \Hb\/ and He~{\sc i} are visible in the early spectra and metal lines are detected since the 19d spectrum. 
 Also for this object the \Ha\/ profile appears flat-topped. However,  contrary to \ad, in this case we have detected HV features of \Ha\/ and \Hb\/ (cfr. Sect.~\ref{sec:interact}).
Fig.~\ref{fig:lineid} shows the line identifications on the spectrum of day 67, which turn out very similar to those of \ad\/ at a similar epoch.

Three late-time spectra sample the evolution in the nebular phase with the progressive growth of forbidden lines over permitted counterparts (e.g. [Ca~{\sc ii}] $\lambda\lambda$7291-7324 doublet vs. Ca~{\sc ii} IR triplet).

%In the spectra from 19d to 67d there are the classical Fe~{sc ii} lines between 5000 and 5200 \AA\/ as well as other metal lines such Sc~{sc ii} $\lambda$5527 and the Sc~{sc ii} $\lambda$6246 although weaker than in \ad\/. The same is true for Ba~{sc ii}  5450 \AA\/ and 6060 \AA\/ that become clearly visible only in the spectrum of 67d, then later with respect to the previous SNe.
%Also in this SN the spectrum of 67d shows O~{sc i}  $\lambda$7774, N~I at 8110 \AA\/ and O~{sc i} $\lambda$9260 {\bf (see Fig.~\ref{fig:lineid})}.
%Forbidden emissions of the [Ca~{sc ii}] $\lambda\lambda$7291-7324 doublet, [O~{sc i}] $\lambda\lambda$6300-6364 doublet and [Fe~{sc ii}] in the region of 7000--8200 \AA\/ are observed in the nebular spectra as the [S~II] doublet of the HII region, but the Na ID and near-IR Ca~{sc ii} features are clearly detected until $\sim$10 months after the explosion.

\subsection{Velocity and temperature evolution}\label{sec:spevol}

\begin{figure*}
\centering
\includegraphics[width=18cm]{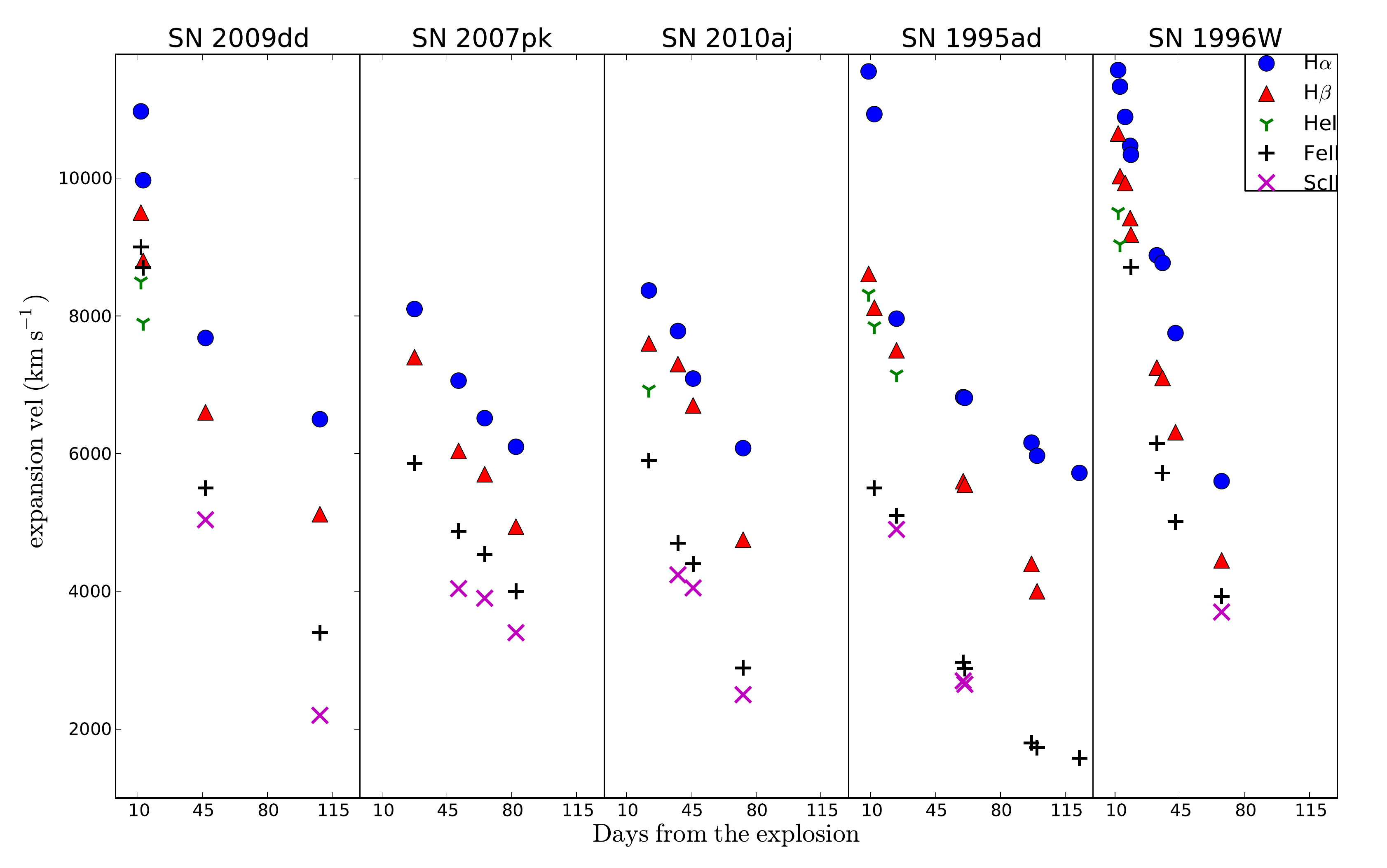}
\caption{
Expansion velocities measured for the SNe of our sample, derived from the position of the  P-Cygni minima for the following lines: \Ha, \Hb, He~{\sc i} $\lambda$5876, Fe~{\sc ii} $\lambda$5169 and Sc~{\sc ii} $\lambda$6246. The velocities of the Balmer lines of \pk\/ up to day +8, derived by the FWHM of the narrow emissions (cfr. Tab.~\ref{table:op}), are not shown here.}
\label{fig:vel}
\end{figure*}

The expansion velocities provide key information on the energetics of the explosion and the position of the photosphere inside the ejecta.
In Fig.~\ref{fig:vel} we show the evolution of the expansion velocities for our SN sample as inferred from the positions of the minima of the P-Cygni profiles for some representative lines (\Ha, \Hb, He~{\sc i} $\lambda$ 5876, Fe~{\sc ii} $\lambda$5169 and Sc~{\sc ii} $\lambda$6246).  The measured expansion velocities are listed in Tab. \ref{table:op}.
In all cases the velocities of \Ha\/ are systematically higher than those of \Hb.
At very early epochs the \Ha\/ velocities are over 10000 \kms\/ in SNe 1995ad, 1996W and 2009dd, and decline very fast.
The determination of the expansion velocity is problematic in \pk\/ because of the lack of absorptions (cfr. Sect. \ref{sec:indiv}). The early velocities of the Balmer lines reported in Tab. \ref{table:op} are derived from the FWHM of the emissions visible up to day +8.

The velocity of He~{\sc i} line, when visible, is significantly smaller than that of \Ha.
Fe~{\sc ii} and Sc~{\sc ii} are good indicators of the photospheric velocity and their values are the smallest among those measured. 
Sc~{\sc ii} remains always slower than Fe~{\sc ii} with v$_{\rm Sc~{\sc II}}\sim 0.9 \times$ v$_{\rm Fe~{\sc II}}$.

\ww\/ has metal line (photospheric) velocities that are a factor 1.5 higher than those of other SNe
whilst the velocities deduced from H and He~{\sc i} (outer ejecta) are not remarkably different.
The presence of a weak interaction could explain this behaviour \citep[for details see][]{phoenix07od} and is supported by the presence of HV \Ha\/  at 11500 \kms\/ during the entire photospheric evolution (cfr. Sect.~\ref{sec:interact}).
%In the evolution of \ad\/ we noted as the velocity of \Ha\/ decreases rapidly in the first 20 days, after the decrease is slower than before. Fe~{sc ii} reaches $\sim$2000 \kms\/ at about four moth from explosion, while Sc~{sc ii} reaches $\sim$1200 after five moth.
%Even \ww\/ shows as the velocities of \Ha\/ are higher than those of \Hb\/. The difference enhance at about 20 days, early respect to the other. For this object the only photospheric tracer available is the Fe~{sc ii}, that starts at velocity $\sim$8700 \kms\/ (19d) and reaches $\sim$4000  \kms\/ at about two months post explosion.

\begin{figure}[!ht]
\includegraphics[width=\columnwidth]{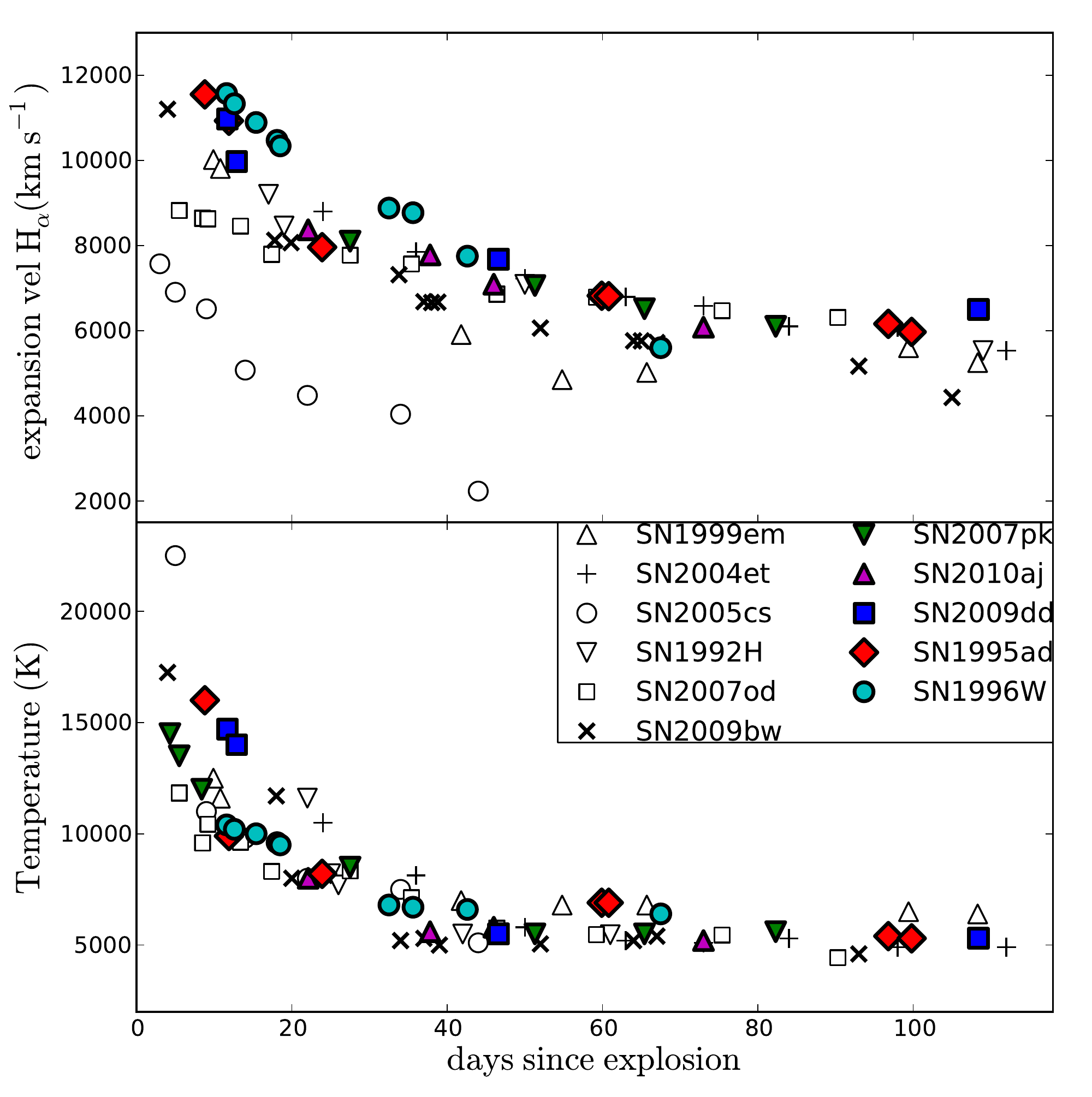}
\caption{Top: Comparison of the \Ha\/ velocities of our SNe with those of other SNe II. Bottom: Comparison of the continuum temperature evolutions}
% of our SNe sample with those of SNe 1999em, 2004et, 2005cs, 1992H, 2007od and 2009bw.}
\label{fig:cfr_velT}
\end{figure}

In Fig.~\ref{fig:cfr_velT} we show the comparison of the \Ha\/ velocity evolutions in our SN sample with those of other classical SNe II.
All SNe show a similar behavior, with the exception of the low-luminosity SN~2005cs, that is clearly an outlier.
In any case \ww\/ and \dd\/ show the highest velocities at early epochs while
during the mid-late photospheric phase all objects appear to be confined in a narrow strip of $\pm1000$ \kms. 
%The objects of our sample, in particular SNe 2009dd, 1995ad and 1996W, seem more similar to the weakly interacting \bw. 

%On the top of Fig.~\ref{fig:cfr_velT} is compared the \Ha\/ evolution of our sample with those of the type II SNe.
%The initial velocities of \dd\/ and \ad\/ are comparable to those of SNe 2009bw, 1992H and 2004et, while  the \ww\/ appear faster than that.
%The \ad\/ behaviour remembers that of \bw\/ and SN 1999em but shifted of some days.
%The only SN distinct is the \pk\/, for which the differences in the first 8 days from the type II are remarkable.
%After the first 20d, all the SNe follow the behaviour of the other SNe, even if the \dd\/ and \ww\/ values are greater than the other SNe, especially the \dd\/ value at $\sim$108d.
%In fact, the velocities of \ww\/ until $\sim$40 days are higher than the other SNe chosen for the comparison, presenting a monotonic decline different from the other objects, while the last photospheric point appears slower than the sample measurements.
%The \pk\/, \aj\/ and \ad\/ velocities slightly decline from $\sim$8000\kms\/ at 20d to $\sim$6000\kms\/ at 60d.

The bottom panel of the same figure reports the temperature evolution, as derived from the black-body fits to the spectral continua (see Tab.~\ref{table:op}).
The objects showing possible evidence of early interaction have,  during the first 20d, higher temperatures than the other SNe II.  
%The temperatures of \dd\/ and \ad\/ are the highest of the sample. 
After 40d the measured temperature remains constant (T$\sim$5500 K) for all objects.

\begin{table*}
  \caption{Observed continuum temperatures and line velocities for the objects of our sample.}
  \begin{center}
  \begin{tabular}{cclccccc}
  \hline
  \hline
  JD & Phase$\dagger$ & T & v(\Ha) & v(\Hb) & v(He~{\sc i}) & v(Fe~{\sc ii}) & v(Sc~{\sc ii})  \\
   +2400000 & (days) & (K) & (\kms)& (\kms)&(\kms) &(\kms) & (\kms) \\
 \hline
  \multicolumn{8}{c}{\dd}\\
 \hline
  54936.7 & 11.2 & 14700 $\pm$ 2000 & 10970 $\pm$ 150  & 9500 $\pm$ 370& 8500 $\pm$ 700& 9000 $\pm$ 500& \\
   54938.4 & 12.9 & 14000 $\pm$ 2000 & 9970 $\pm$ 150  & 8800 $\pm$ 400& 7900 $\pm$ 600& 8700 $\pm$ 500& \\
  54971.6 & 46.1 & 5500 $\pm$ 200& 7680 $\pm$ 250& 6600  $\pm$ 100& & 5500 $\pm$ 500& 5040 $\pm$ 100\\
  55033.4 & 107.9 & 5300 $\pm$ 700& 6500 $\pm$ 300& 5120 $\pm$ 230& &3400 $\pm$ 1000& 2200 $\pm$ 200\\
 \hline
 \multicolumn{8}{c}{\pk}\\
 \hline
  54416.3 & 4.3 & 14500 $\pm$ 2000 & 1800 $\pm$ 200$^*$  & 1800 $\pm$ 200$^*$& && \\
  54417.5 & 5.5& 13500 $\pm$ 1000& 2300 $\pm$ 100$^*$ & &&&\\
   54421.4 & 8.4& 12000 $\pm$ 1000& 2800 $\pm$ 500$^*$ & 2000 $\pm$ 200$^*$ &  &&\\
  54439.5 & 27.5 & 8500 $\pm$ 300& 8100 $\pm$ 800& 7400 $\pm$ 100& & 5860 $\pm$ 250 & \\
 54463.3 & 51.3 & 5500 $\pm$ 500 & 7060 $\pm$ 100 & 6040 $\pm$ 200 &  & 4875 $\pm$ 400 & 4040 $\pm$ 100\\
  54477.4 & 65.4 & 5500 $\pm$ 300 & 6515 $\pm$ 180 & 5700 $\pm$ 300 & & 4540 $\pm$ 100 & 3900 $\pm$ 180\\
  54494.3 & 82.3 & 5600 $\pm$ 600& 6100 $\pm$ 100& 4940 $\pm$ 500 & & 4000 $\pm$ 500 & 3400 $\pm$ 400  \\
 \hline
 \multicolumn{8}{c}{\aj}\\
 \hline
  55287.6 & 22.1 & 8000 $\pm$ 500 & 8370 $\pm$ 200  & 7600 $\pm$ 300 & 6930 $\pm$ 1000& 5900 $\pm$ 700& \\
  55303.3 & 37.8 & 5600 $\pm$ 500& 7780 $\pm$ 700 & 7300  $\pm$ 350 & & 4700 $\pm$ 200& 4240 $\pm$ 500\\
  55311.5 & 46.0& 5800 $\pm$ 500& 7090 $\pm$ 300 & 6700 $\pm$ 300&  &4400 $\pm$ 200& 4050 $\pm$ 200\\
   55338.5 & 73.0 & 5200 $\pm$ 400& 6080 $\pm$ 300& 4750 $\pm$ 380& & 2890 $\pm$ 250 &2500 $\pm$ 200 \\
 \hline
 \multicolumn{8}{c}{\ad}\\
 \hline
  49989.8 & 8.8 & 16000 $\pm$ 1500 & 11500 $\pm$ 1100  & 8610 $\pm$ 600& 8321$\pm$ 600& & \\
  49992.9 & 11.9 & 9900 $\pm$ 400& 10930 $\pm$ 580& 8120  $\pm$ 520& 7850 $\pm$ 400& 5500 $\pm$ 1500 &\\
   50004.9 & 23.9 & 8200 $\pm$ 600& 7960 $\pm$ 270& 7500 $\pm$ 200& 7150 $\pm$ 300 &5100 $\pm$ 350 & 4900 $\pm$ 300\\
   50040.8 & 59.9 & 6900 $\pm$ 500& 6820 $\pm$ 200& 5600 $\pm$ 100 & & 2970 $\pm$ 100 & 2700 $\pm$ 100  \\
  50041.8 & 60.8 & 6900 $\pm$ 600 & 6810 $\pm$ 210 & 5550 $\pm$ 100 & & 2880 $\pm$ 120 & 2650 $\pm$ 100 \\
  50077.8 & 96.8 & 5400 $\pm$ 600 & 6160 $\pm$ 190 & 4400 $\pm$ 300 &  &1800 $\pm$ 200 &\\
 50080.8 & 99.8 & 5300$\pm$ 900& 5970 $\pm$ 360 & 4000 $\pm$ 500 &  & 1730 $\pm$ 400&\\
  50103.7 & 122.7 & 5700 $\pm$ 400& 5720 $\pm$ 380&   & & 1580 $\pm$ 210 & \\
  50131.7 & 150.7 & 5500 $\pm$ 900& 5570 $\pm$ 220&  & & 1280 $\pm$ 190 &\\
  50132.7 & 151.7 & 6700 $\pm$ 1500 & 5470 $\pm$ 290&   &&1210 $\pm$ 200 &\\
  50133.7 & 152.7& 7200 $\pm$ 1300& 5320 $\pm$ 205 &  & & & \\
    \hline
 \multicolumn{8}{c}{\ww}\\
 \hline
  50191.6 & 11.6 & 10400 $\pm$ 300 & 11570 $\pm$ 500  & 10650 $\pm$ 500& 9510 $\pm$ 500& &\\
  50192.6 & 12.6 & 10200 $\pm$ 300& 11330 $\pm$ 500& 10030  $\pm$ 500& 9040  $\pm$ 500&&\\
  50195.4 & 15.4& 10000 $\pm$ 200& 10890 $\pm$ 500& 9930 $\pm$ 400&  &&\\
  50198.4 & 18.1 & 9600 $\pm$ 300& 10470 $\pm$ 500& 9420 $\pm$ 400& && \\
  50198.6 & 18.5& 9500 $\pm$ 300 & 10340 $\pm$ 500 & 9180 $\pm$ 400 &  & 8710 $\pm$ 400 &\\
  50212.5 & 32.5& 6800 $\pm$ 300 & 8880 $\pm$ 300 & 7250 $\pm$ 300 & &  6150 $\pm$ 300&\\
  50215.6 & 35.6& 6700 $\pm$ 300& 8770 $\pm$ 300 & 7100 $\pm$ 300 & &5720 $\pm$ 300 & \\
   50222.6 & 42.6 & 6600 $\pm$ 300& 7750 $\pm$ 300& 6310 $\pm$ 300 & & 5010 $\pm$ 300 & \\
  50247.5 & 67.5 & 6400 $\pm$ 500& 5600 $\pm$ 300& 4450 $\pm$ 300& &3930 $\pm$ 250 & 3700 $\pm$ 200 \\
  50432.8 & 252.8 &  & 3700 $\pm$ 200 & 2830 $\pm$ 200& &&\\
  50478.5 & 298.5&  & 3600 $\pm$ 200 && &  \\
     \hline
\end{tabular}
\begin{flushleft} 
$ \dagger$ with respect to the explosion epochs (cfr. Tab.~\ref{table:main})\\
$^*$ the velocities are the FWHM of the narrow emissions
\end{flushleft} 
\end{center}
\label{table:op}
\end{table*}

\subsection{Comparative evolution}\label{sec:comp}

%In Fig~\ref{fig:cfr} we compare the spectra of our sample with those of SNe showing similar spectroscopic features in order to better understand their nature and strengthening the explosion epochs derived previously.
%In this Section we highlight possible spectroscopic interaction signatures through a comparison of the early and photospheric spectra with those of other type II SNe. Possible indirect evidences of early, weak\footnote{We define as weak interaction a phenomenon that does not alter significantly the light curve and the spectral evolution.} interaction are flat topped line profiles as shown by  \od\/ \citep[CSM mass $\sim10^{-4}$ \M,][]{07od}, and the presence of HV features, originating in the ejecta just below the shock front. {\bf The flat-topped emission is the consequence of a multicomponent profile impossible to fit with a single Gaussian/Lorentzian (see Fig.~\ref{fig:mc}).} %The ratio between emission and absorption component is  $p=v_{emission}/v_{absorption}\sim1$ in a normal ejecta. Higher values than that imply a boxy shape of the emission profile.}

\begin{figure*}[!ht]
%\vspace{174pt}
\centering
\includegraphics[width=18cm]{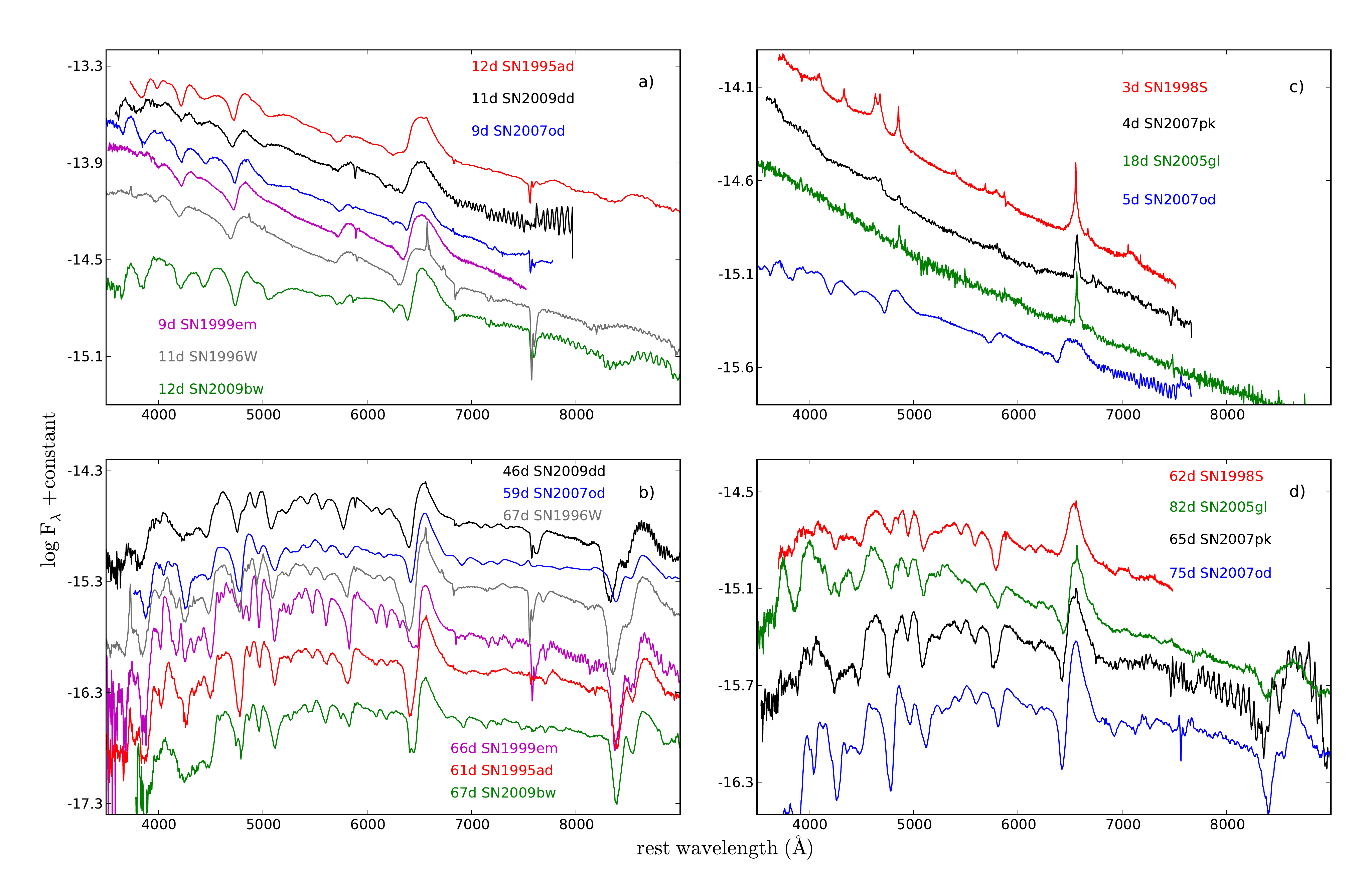}
\caption{
Panel a): Comparison between early spectra of \ad, \ww\/ and \dd, and those of the interacting SNe 1999em, 2007od and 2009bw, all showing signs of weak early interaction. Panel b): Comparison of the same objects during the plateau phase. Panel c): Comparison among early spectra of \pk\/, with those of other strongly- (SNe 1998S and 2005gl) and weakly-interacting (\od) SNe. Panel d): Comparison of the same objects during the plateau phase. The phase of SN 2005gl has been adopted from \citet{gy07}. } 
\label{fig:cfr}
\end{figure*}

In panel a) of Fig.~\ref{fig:cfr} we compare the early spectra of the SNe in our sample with those of other type II events. 
SNe 1995ad and 1996W show flat-topped profiles which may be due to contamination of the narrow emission from underlying H~II regions (favoured for \ad, Sect.~\ref{sec:indiv}) or to interaction \citep[as in \od, ][]{07od,phoenix07od}. In the case of \ww\/ the latter is supported by the simultaneous presence of HV features (cfr. Sect.~\ref{sec:interact}). The emission profile of \dd\/ is roundish and similar to that seen in \em.
%The multicomponent \Ha\/ profiles previously claimed for \ad\/  and \ww\/ in Sect.~\ref{sec:indiv} appear less consistent compared with that of \od\/. In this case the multicomponent profile is probably due to the contamination of the \Ha\/ emission of the H~II region on the blueshifted maximum of the SN emission for both the objects. This conclusion is strengthened by the higher velocity and stronger intensity of their \Ha\/ absorption lines compared to that of \od\/ \citep[see][for further details about interaction and lower Balmer velocities]{phoenix07od} as shown in Sect.~\ref{sec:spevol}. The \dd\/ \Ha\/ emission profile is similar to those of \em\/ and \bw\/ without flat component.
%All the SNe show metal lines, the only exception is the \ww\/ while they are more pronounced in \bw\/ as expected from the more advanced phase. 
However, on the blue side of the \Ha\/ emission in SN~2009dd spectrum,
three absorptions are clearly detected. To our knowledge, similar features have not been seen in other type II SNe (at least with this strength).
The bluest component was identified with Si~{\sc ii} $\lambda$6355 as in \od\/ \citep{07od,phoenix07od} and \bw\/ \citep{09bw}.
The central component might be a HV feature of \Ha\/ with $v=13800$ \kms\/ or, alternatively, Fe~{\sc ii} $\lambda 6456$ with an unusually high optical depth.
% compared to other Fe~{sc ii} lines of the objects and of the other SNe of the comparison. 
Although there are no other clear spectral evidences, such as HV \Hb\/, the X-ray detection at early times (cfr. Sect.~\ref{sec:sne}) may support the interaction scenario. 
Also, the weak features on the blue side of the \Ha\/ and \Hb\/  of  \ww\/ could be related to  HV layers at $\sim$11500 \kms.
%For \ad\/ there is a  possible HV \Ha\/ at 6280\AA\/ with $v\sim13000$ \kms\/. It is barely visible in the 24d spectrum and disappears thereafter.

In panel (b) we compare the plateau phase spectra of the same objects. At this stage also \bw\/ and \em\/  show HV \Ha\/ at $\sim$7300 \kms\/ \citep{09bw} and $\sim$8200 \kms\/ \citep{chu07}, respectively.
The weak HV \Ha\/  components of \dd\/ and \ww\/   appear at  velocity  $\sim$13800 and $\sim$11500 \kms, respectively, remaining steady during the early phases.

%!!! NO WAY, in \ad\/ NON SI VEDE NULLA !!! MT !!!
%, $\sim$10900 \kms\/ for \ad.   

%%  ILLEGGIBILE, MANCANO ALCUNE COMPONENTI, INUTILE ANZI MISLEADING
%\begin{figure}
%%\vspace{174pt}
%\includegraphics[width=\columnwidth]{mc95ad07pk.pdf}
%\caption{Left panel: \Ha\/ multicomponent profile of +11.9d spectrum of \ad\/, white dotted line is the combination of the two gaussian profiles (red and green dotted lines). Right panel: \Ha\/ multicomponent profile of +27.5d spectrum of \pk\/. \Ha\/ emission is the combination of three gaussian profiles (green, red and cyan dotted lines).} 
%\label{fig:mc}
%\end{figure}

Panels c) and d) show the early and late photospheric spectra of \pk, the only SN of the sample with a clear evidence of interaction at early times.
We compare the spectra with those of the type IIn SN 1998S, the weakly interacting type IIP \od\/ and the type IIn SN 2005gl, which have similar spectral evolution.
The narrow H Balmer lines of \pk\/ and the double-horned line at $\sim$4600\AA, possibly related to highly ionised elements such as C~{\sc iii}/N~{\sc iii} or  C~{\sc iv},  stand out over a very blue continuum. 
The similarity with SN 1998S  is remarkable, but also SN 2005gl shared the same features.
The weakly interacting \od\/  had instead a definitely different spectral features. 
%There is also a good correspondence with the earliest spectrum of \bw\/  \citep{09bw}. 
%On the contrary, the early spectrum is rather different from normal SNII like the IIP \em. 

As mentioned in Sect. \ref{sec:indiv}, the spectrum of \pk\/ evolves rapidly and already on day 50, during the plateau, it resembles closely those of standard type IIPs. 
In panel (d) we compare late photospheric spectra of the same objects about two months past explosion. 
\pk\/ and SN 2005gl have evolved in a similar fashion. Both show well-developed absorption components for the lines of all ions and  deeper absorptions of the Balmer lines with respect to SN 1998S, while the lines of the ions of the inner ejecta like Fe~{\sc ii} and Sc~{\sc ii} seem to have the same strengths.
The absorption component of \Ha\/ is significantly stronger in \od\/ than in other objects.

%\begin{figure}
%\begin{center}
%\includegraphics[width=8.5cm]{HV_all.pdf}
%\caption{{\bf illegible. sbatti tutte le fasi una sopra dell'altra senza criterio. Non so ... mettile in un ordine cronologico di fase in modo che il decremento di velocita' non confonda. - Inoltre la terza componente di assorbimento di \dd\/ esce dal grafico.}  Zoom of the 6200\AA\/ (right-hand panel) spectral region during the photospheric phase. The {\it x}-axes are in expansion velocity coordinates with respect to the rest-frame position of \Ha\/. The HV feature of the Balmer are in the range from $\sim$7000 \kms\/(\bw) to $\sim$14000 \kms\/(\dd). In the brackets are reported the phase of the spectrum and the FWHM.}
%\label{fig:hvall}
%\end{center}
%\end{figure}

\begin{table*}
\caption{Main information on the SN sample.}
\begin{center}
\begin{tabular}{lccccc}
\hline
 & 2009dd & 2007pk & 2010aj & 1995ad & 1996W\\
\hline
$\alpha$ (J2000.0)	&12$^{h}$05$^{m}$34$^{s}$.10	&01$^{h}$31$^{m}$47$^{s}$.07 & 12$^{h}$40$^{m}$15$^{s}$.16& 06$^{h}$01$^{m}$06$^{s}$.21 & 11$^{h}$59$^{m}$28$^{s}$.98\\
$\delta$  (J2000.0)	&+50$^{o}$32'19".40 	&+33$^{o}$36'54".70 & -09$^{o}$18'14".30 &-23$^{o}$40'28".9 & -19$^{o}$15'21".9 \\
host galaxy		& NGC~4088& NGC~579& MGC~-01-32-035&  NGC 2139& NGC 4027\\
galaxy type 		&Sbc&  Scd & Sc&  SBc& SBc\\
offset from nucleus 		& 1".5 W, 4".0 S  & 7".4 E, 1".6 S& 12".4 W, 11".7 S &  25" W, 5" S& 17" W, 34" N\\
\\
local metallicity [12+log(O/H)] & 8.59 & 8.50 & 8.63   &  8.60 & 8.60\\
adopted recession velocity$^{\dagger}$ [\kms]	&  $1025\pm15$ & $5116\pm16$ & $6386\pm20$ &   $1674\pm14$ & $1779\pm29$ 	\\
adopted distance modulus ($\mu$)	&  $30.74\pm0.15$	&$34.23\pm0.15$ & $34.71\pm0.15$&   $31.80\pm0.15$ &$31.93\pm0.15$	\\
adopted reddening, E$_{\rm tot}$(B-V)		& 0.45 & 0.11&0.04& 0.04&0.23\\
\\
observed B peak magnitude & $15.07 \pm 0.06$ & $16.00 \pm 0.15$ & $17.92 \pm $ 0.06& $14.77 \pm 0.20$ & $15.45 \pm 0.03$  \\
absolute B peak magnitude & $\leq -17.61 \pm 0.27$& $-18.70 \pm 0.23$ & $\leq -16.95 \pm 0.18$ & $-17.18 \pm 0.26$ & $-17.59 \pm 0.26$  \\
explosion epoch (JD)   & $2454925.5\pm5$ & $2454412\pm5$& $2455265.5\pm4$&  $2449981.0\pm3$&$2450180.0\pm3$\\
%peak abs mag & & & & & \\
L$_{\rm bol}$ peak [$\times10^{42}$ erg s$^{-1}$]& 2.16 & 6.26 &2.68 &1.55& 1.85 \\
light curve peak (JD) & $2454937\pm4$& $2454420\pm2$& $2455269\pm4$ & $2449989\pm4$ & $2450186\pm4$ \\
late decline $\gamma_V$ [\mcento]& 1.15 &  -- & 3.0$^*$&0.93 &0.86 \\
%interval days 		& & & & & \\
M(\ni)[\M]			&  0.029  & -- & $<$0.007 &   0.028 & 0.14 \\
\\
M(ejecta)[\M]			                                & 8.0                   & -- & 9.5          &  5.0           & --  \\
explosion energy ($\times 10^{51}$  erg)	& 0.20                & -- & 0.55           & 0.20           & --   \\
progenitor radius [cm]                              & $5\times10^{13}$ & --&  $2\times10^{13}$& $4\times10^{13}$ & --   \\
\\
interaction evidences & X, HV\Ha\/ & X, blue continuum, & -- & --  & HV \Ha\/ \& \Hb\\
                                   &                  & narrow CS \Ha       &     &     &                          \\
 mass loss [M$_{\odot}$ yr$^{-1}$] & $\gtrsim10^{-6}$  & $\gtrsim10^{-5}$  & --& --  & $\sim10^{-6}$  \\
\hline
\end{tabular}
\end{center}
%$^*$ provided by NED\\  !!!  discende direttamente da Vvirgo !!! MT !!!
$^{\dagger}$ corrected for Virgo infall\\
$^*$ in R band, short baseline
\label{table:main}
\end{table*}

\section{Discussion}\label{sec:dis}

In previous Sections we presented new photometric and spectroscopic data of five SNe II and discussed their evolution from the photospheric to the nebular phase.

The SNe of our sample are relatively bright, with an average absolute peak magnitude M$_{V}\leq-16.95$ (i.e. L$_{\rm peak}\sim3\times10^{42}$ erg s$^{-1}$), above the average for SN II  \citep{pa94,li11}.
The light curves are characterized by extended plateaus, although \pk\/  may be considered a transitional object between IIP and IIL.
The expansion velocities of the ejecta, ranging from $10000-12000$ \kms\/ at early phases to $\sim$5000 \kms at the end of the photospheric phase, are similar to those of other SNe II, 
with the objects showing signs of interaction being moderately faster. The same is true for the temperature.
There is a high degree of individuality in the shape of the light curve with the transition to the nebular stage occurring at different epochs and with different magnitude drops from the plateau.
Also the inferred  \ni\/ masses span over one order of magnitude (1.4$\times10^{-1}$ to 7$\times10^{-3}$\M, cfr. Sect.~\ref{sec:bol}).

%Generally these properties are consistent with progenitor stars having an initial (ZAMS) mass of 8--15 \M\/ \citep{sm09}. This mass range consists of two kinds of stars having a different evolutionary path prior to the core collapse event. The first one is formed by the massive super-asymptotic giant branch (SAGB) stars which, after H-, He-, and C-burning, form a degenerate Neon-Oxygen core \citep[e.g.][and references therein]{pumo06,pumo07}, where the physical conditions can be suitable for triggering electron-capture reactions on $^{24}$Mg and other nuclei which are present in {\bf small} amounts in the core, leading to a so-called electron-capture supernova (EC-SN) event \citep[e.g.][and references therein]{pumo09}. The other kind of progenitors are ``standard'' massive stars at the lower mass range that can produce iron core collapse SNe (Fe CC-SNe), after having ignited all nuclear burnings up to the formation of an iron core \citep[e.g.][]{ww86,wo02,he03}.

\subsection{Signatures of interaction}\label{sec:interact}

\dd\/ shows three absorption components in the blue side of  \Ha\/ emission. The middle one 6235 \AA\/ is likely an HV\Ha\/ component with v$\sim$13800 \kms. 
Alternative identifications with metal lines (e.g. Si~{\sc ii}, Sc~{\sc ii}, Ba~{\sc ii}) are unlikely because the position of the absorption minimum would imply too low velocities 
(e.g. half of the Fe~{\sc ii} velocity). The putative HV\Ha\/ feature, between the Si $\lambda$6355 (v$\sim$9000 \kms) and the main \Ha\/ absorption, mimic an unusual triple absorption profile.
%A further support to the hypothesis of interaction comes from the X-ray detection of a source at the SN coordinates during the first month after the explosion, with an increasing X-ray luminosity.

%\pk\/ has a somehow different behaviour. It  can be properly classified as a type IIn SN at early stages for the detection of narrow H emission lines without evident absorption components.
%The fourth spectrum (27d) shows a more complex structure in the Balmer profiles, with a broad blue shoulder,  possibly due to interaction of the ejecta with an asymmetric medium, and a narrow component due to the un-shocked CSM. 

Two of the SNe of our sample, 2007pk and 2009dd  were detected in X-ray,  that is a bona fide indicator of ejecta-CSM interaction. The spectral evolution of SN 2007pk, 
with the detection of a narrow \Ha\/ in pure emission, suggests that  interaction began very early after explosion but ended one month later. To our knowledge the only other object 
having a similar  evolution was  SN 2005gl \citep{gy07,gy09}.  Considering the strong X-ray emission of SN 2007pk, 10 times larger than those of \dd\/ and \em\/ \citep{im09,elm03}, we argue that the ejecta-CSM interaction was also quite strong.

\begin{figure}[hb!]
%\vspace{174pt}
\includegraphics[width=\columnwidth]{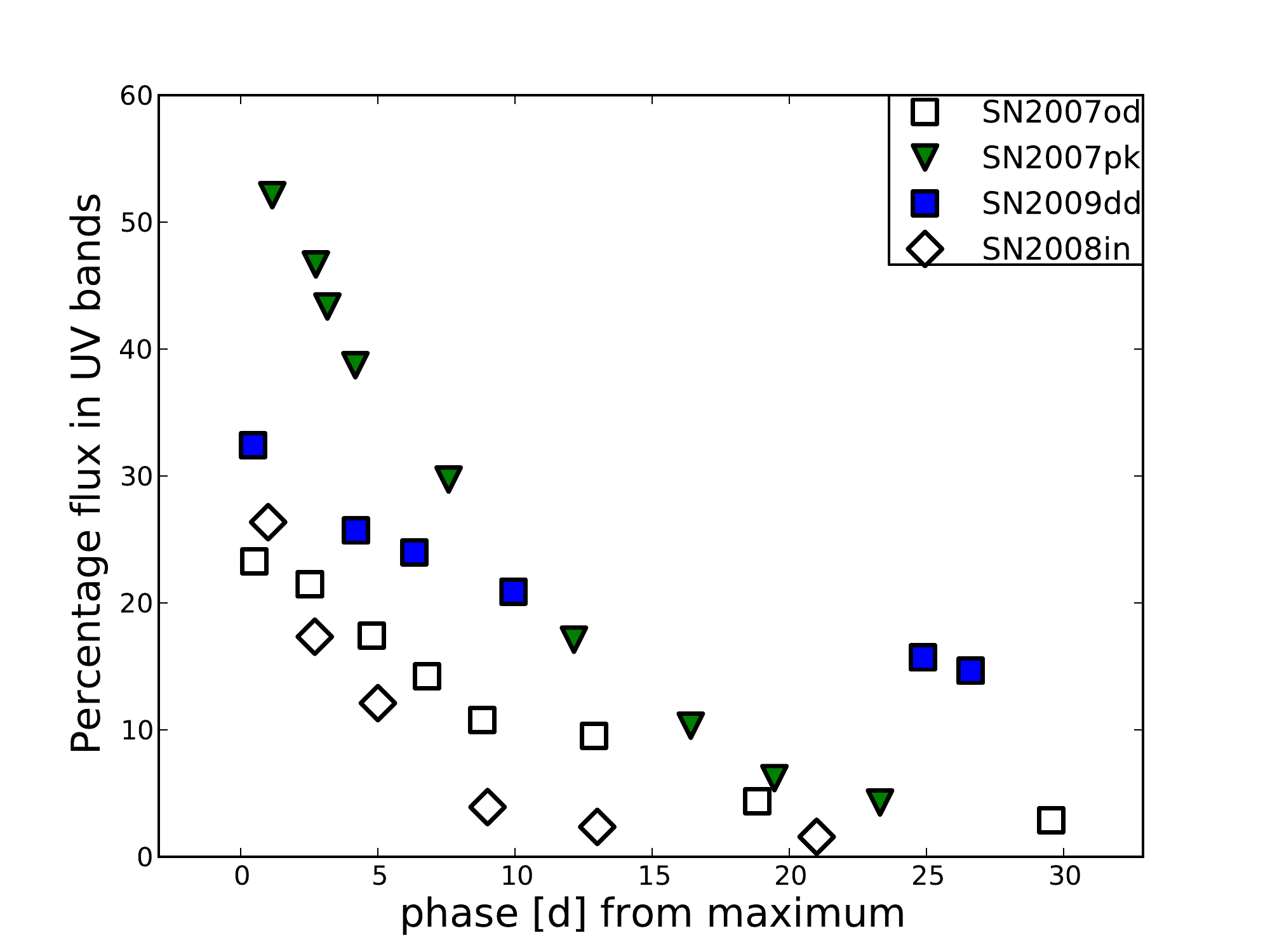}
\caption{Flux contribution of the UV (uvw2, uvm2, uvw1) to the total (UV-to-I) bolometric light curves of \pk\/ and \dd\/ compared to the weakly interacting \od\/ and the non-interacting SN 2008in.} 
\label{fig:uv}
\end{figure}

In general, the shock produces soft X-ray or far-UV emission depending on the ejecta density profile, with flatter profiles leading to higher shock velocity, higher post-shock temperature and hence greater emission at higher frequency \citep{ch94}. Exploiting the fact that both \dd\/ and \pk\/ have been monitored by SWIFT, we could compute the UV contribution to the bolometric emission. This is shown in Fig.~\ref{fig:uv} in which the corresponding values for the mildly interacting \od\/ and the non-interacting SN~2008in \citep{roy11} are reported for comparison.
At early times, when the X-ray flux was higher, the UV contribution of \pk\/  was twice as strong as that of \dd, that can be attributed to a flatter density profile.
However, the UV flux decreased rapidly, and on day 20 was comparable to that of other objects, a sign that the dense CSM material probably was already swept away at that time.
Much slower is the evolution of the UV flux in \dd, that always remained significantly higher than in the other objects.

\begin{figure}[hb!]
\begin{center}
\includegraphics[width=8.5cm]{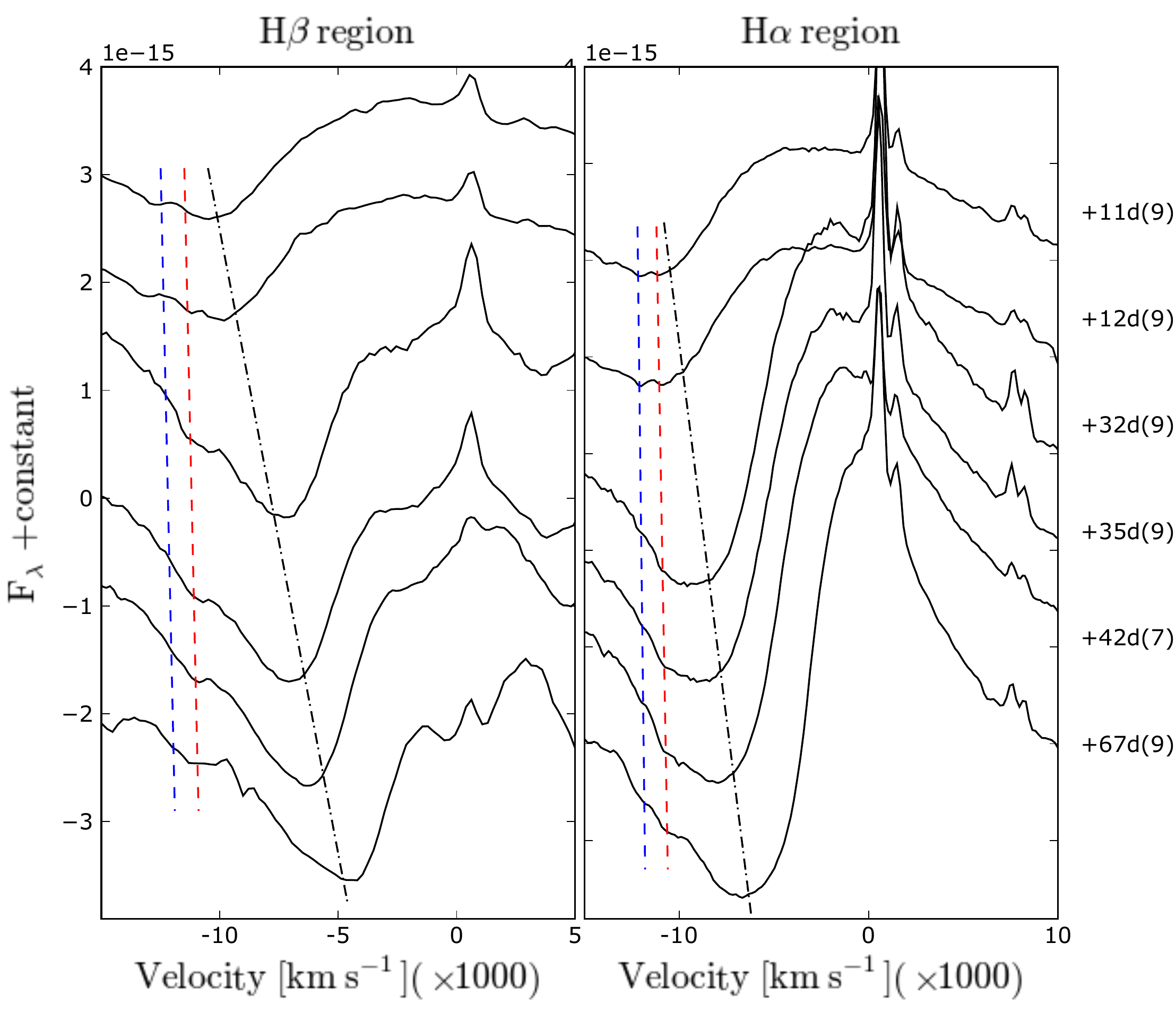}
\caption{Blow-up of the 4600\AA\/ (left-hand panel) and 6200\AA\/ (right-hand panel) spectral regions during the plateau phase of \ww. The {\it x}-axes are in expansion velocity coordinates with respect to the rest-frame position of \Hb\/ and \Ha\/, respectively. To guide the eye, two dash-dotted lines are drawn in the spectra marking the position of minima of the strongest absorption features, while red dashed lines, at comparable velocities, follow the HV \Ha\/ and \Hb\/ features ($\sim$11500 \kms\/). The blue dashed lines are tied to the second HV feature at $\sim$12500 \kms, which is visible only in the \Ha\/ region.}
\label{fig:hv96w}
\end{center}
\end{figure}

HV absorption components were identified on the spectra of a few SNe, e.g. SNe 1999em, 2004dj, 2007od and 2009bw. 
\citet{chu07} have shown that the interaction of the ejecta of a SN IIP with an average RSG wind can produce absorptions (shoulders) on the blue wings of  H lines due to enhanced excitation of the gas in the outer un-shocked ejecta.
In the previous Sections we noted that similar features are present in  SN~1996W. Their velocity evolution as shown in Fig.~\ref{fig:hv96w} remains constant with time. While the line positions at the earliest epochs are compatible with an Fe~{\sc ii} identification, the lack of velocity evolution,  contrasting with the behaviour of the main \Ha\/   absorption, favours the HV identification.

Finally, two other objects, SNe 1995ad and 2010aj, did not show unequivocal signs of interaction.

\subsection{Explosion and progenitor parameters}\label{sec:m}

Using the well-tested approach applied to other CC-SNe (e.g.~SNe 2007od, 2009bw, and 2009E; 
see Inserra et al. 2011, 2012, and Pastorello et al. 2012), we estimate the physical 
parameters describing the progenitors at the explosion (e.g.~the ejected mass, the 
progenitor radius, and the explosion energy) by performing a model/observation comparison 
for our sample of CC-SNe. This is based on a simultaneous $\chi^{2}$ fit of the main CC-SN 
observables (namely, the bolometric light curve, the evolution of the photospheric velocity and 
the continuum temperature at the photosphere). We do not considered SNe 2007pk and 1996W  because the observational data are not sufficient for obtaining reliable 
estimates of the ejecta parameters. In particular, their light curves do not cover the critical transition 
from the plateau to the radioactive tail.

Two codes are employed to calculate the parameters. The first is the semi-analytic code 
described in \citet{za03}, which is used to perform preparatory studies to explore the 
parameter space describing the CC-SN progenitor at the explosion. The second, used to compute 
a denser grid of more accurate models, is a new general-relativistic, radiation-hydrodynamics 
Lagrangian code \citep[for details see][]{pumo10,pumo11}. Its main features are: 
1) an accurate treatment of radiative transfer coupled to relativistic hydrodynamics, 2) a self-consistent
treatment of the evolution of ejected material taking into account both the gravitational effects 
of the compact remnant and the heating effects linked to the decays of the radioactive isotopes 
synthesized during the CC-SN explosion, and 3) a fully implicit Lagrangian approach to the solution 
of the coupled non-linear finite difference system of relativistic radiation-hydro equations.

%{\bf *** MLP - seguente paragrafo: non so se ``if present'' basta. Ho considerato che nel paper, pur non credendo piu' alla interazione per la 10aj e 95ad, si esclude solo l'evidenza di chiari segni di interazione, mentre si ``specula'' sulla possibilita' che l'interazione possa esserci o, cmq, che non possa essere totalemente esclusa, considerato che:  1) per la 10aj ci potrebbe essere polvere a early phase che potrebbe essere un segno di CSM e, quindi, di possibile interazione (sect. 5.1, pg 20); 2) per la 95ad c'e' una analogia coi profili Ha della 07od che potrebbe di nuovo far pensare a interazione (Sect. 4.1, pg 15). ***}

The modelling with these two codes is appropriate if the emission from the CC-SN is dominated 
by the expanding ejecta. As we mentioned before for some of the SNe of our sample, there may be contamination from interaction. In the following we assume that this
effect does not dominate the quasi-bolometric luminosity and therefore the modelling produces reasonable estimates of the main physical parameters of the explosive events, although it may not be possible to precisely 
reproduce all the observed features (see below).

\begin{figure}[hb!]
\begin{center}
\includegraphics[width=\columnwidth]{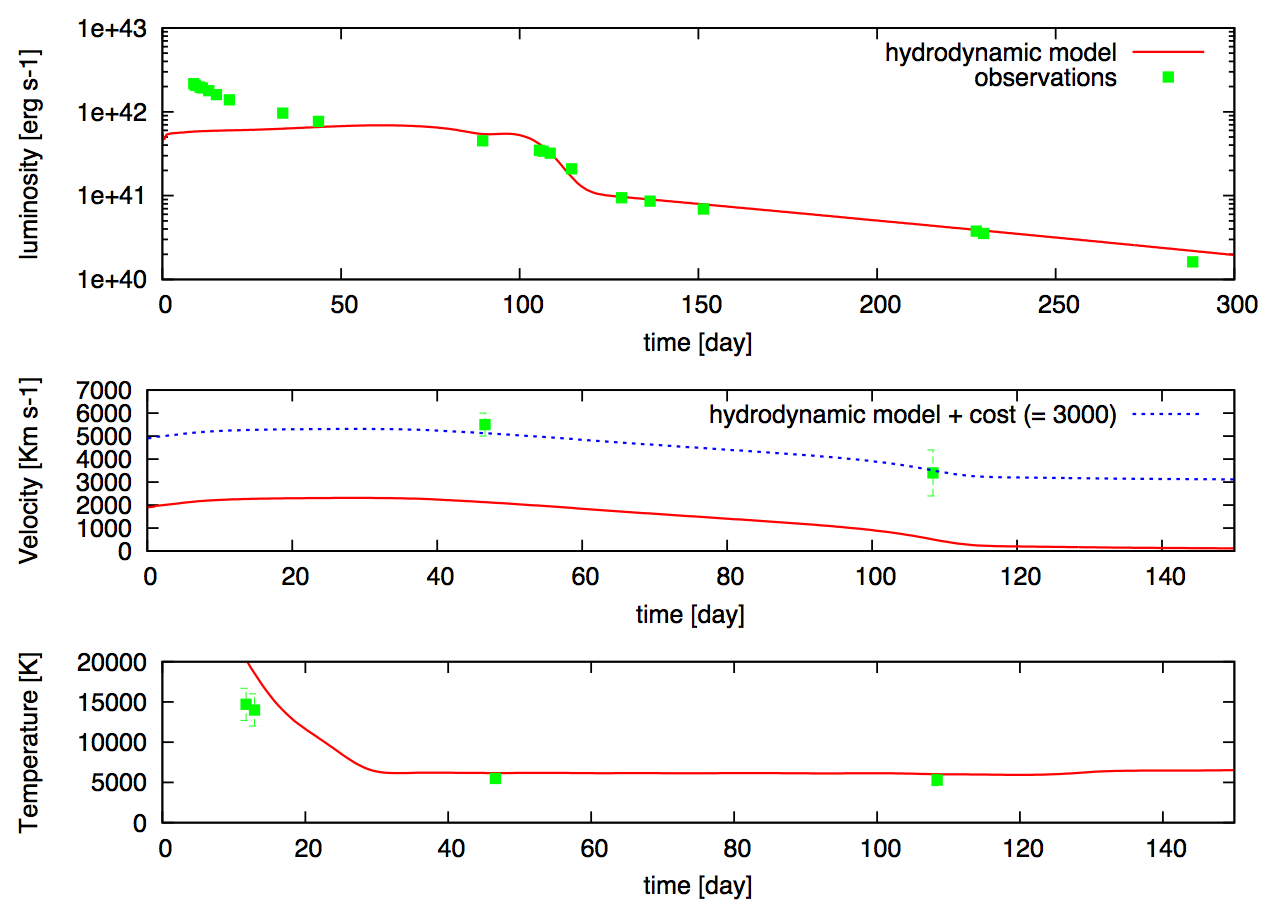}
\caption{Comparison of the evolution of the main observables of SN 2009dd with the best-fit models                                                            
computed with the general-relativistic, radiation-hydrodynamics code (total energy $\sim 0.2$ foe, 
initial radius $5 \times 10^{13}$ cm, envelope mass  $8$ \M).                                                                                                     
Top, middle, and bottom panels show the quasi-bolometric light curve, the photospheric velocity (the blue dashed line refers to the model plus an additive constant to match the data), 
and the photospheric temperature as a function of time. To estimate the photospheric velocity                                                      
from observations, we used the value inferred from the Fe~{\sc ii} lines (see text for further details). The $x-$axes refer to days since explosion.}
\label{fig:mod09dd}
\end{center}
\end{figure}

\begin{figure}[hb!]
\begin{center}
\includegraphics[width=\columnwidth]{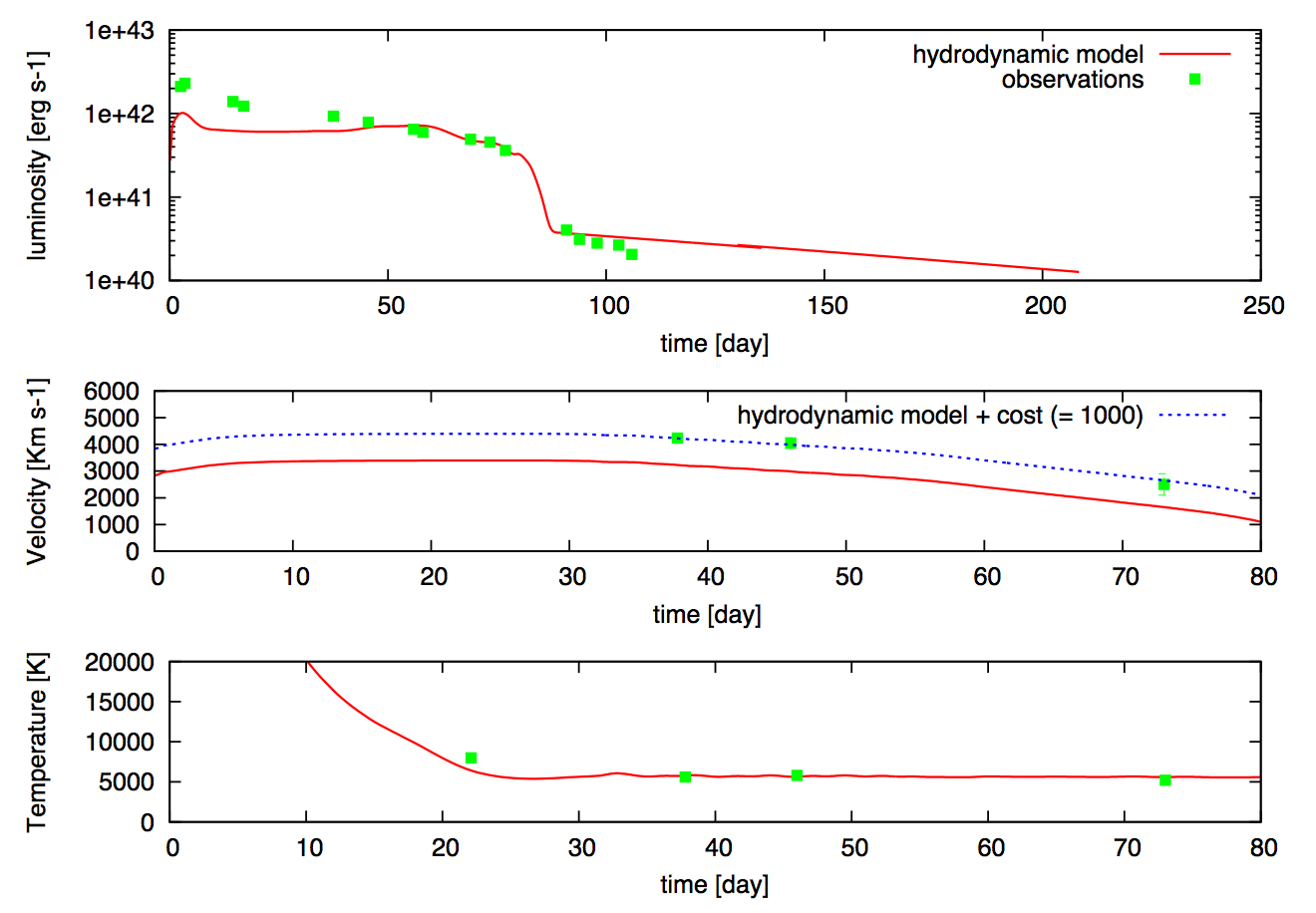}
\caption{As for Fig.~\ref{fig:mod09dd}, but for SN 2010aj. The best-fit model evaluated using the 
general-relativistic, radiation-hydrodynamics code has an initial radius of $2 \times 10^{13}$ cm, 
a total energy $\sim 0.55$ foe, and an envelope mass of $9.5$ \M.}
\label{fig:mod10aj}
\end{center}
\end{figure}

\begin{figure}[hb!]
\begin{center}
\includegraphics[width=\columnwidth]{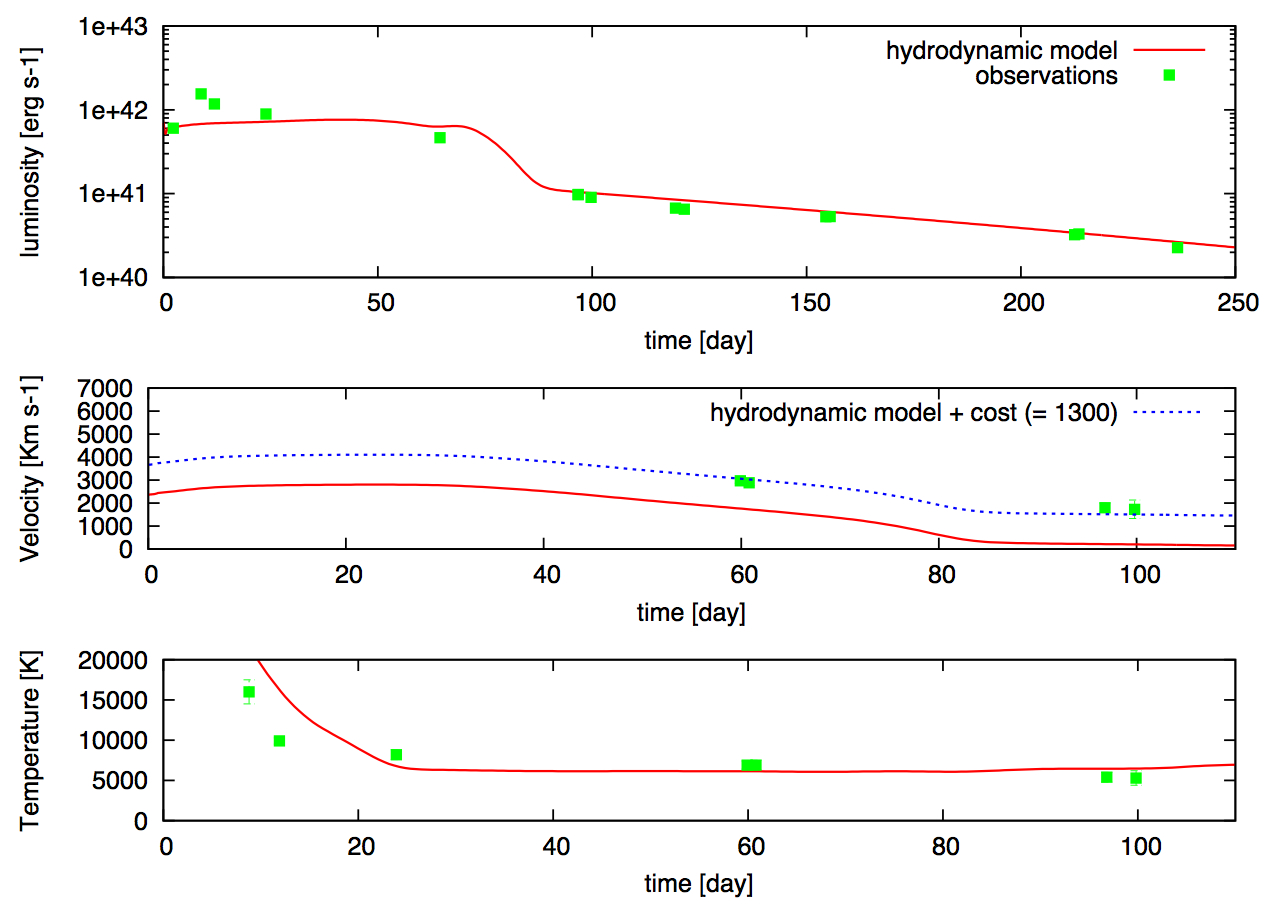}
\caption{As for Fig.~\ref{fig:mod09dd}, but for SN 1995ad. The best-fit model evaluated using the 
general-relativistic, radiation-hydrodynamics code has an initial radius of $4 \times 10^{13}$ cm, 
a total energy $\sim 0.2$ foe, and an envelope mass of $5$ \M.}
\label{fig:mod95ad}
\end{center}
\end{figure}

The best-fitting models for each SN are reported in Figs.~\ref{fig:mod09dd}, \ref{fig:mod10aj}, \ref{fig:mod95ad}, and their parameters are listed in Tab.~\ref{table:main}.

The agreement between the models and the observed light curves and photospheric temperatures 
are quite satisfactory, except at early epochs ($\lesssim$ 20-40 d). Such discrepancies at 
early times are caused both by approximate initial density profiles used in the simulations, 
which do not reproduce accurately the radial profiles of the outermost high-velocity shells of 
the ejecta formed after shock breakout \citep[cfr.][]{pumo11} and, possibly, by ejecta-CSM 
interaction leading to luminosity excesses \citep[see also][and references therein]{09bw}. 
For these reasons, we do not include in the $\chi^{2}$ fit the line velocity measurements taken 
during the first 20-40 d.

The agreement between the observed velocity evolution and our best-fit models is less 
satisfactory. This may be related to a systematic shift between the true photospheric velocity 
and the value estimated from the observed P-Cygni line profiles \citep[see][]{de05},
according to which the optical depth in the lines seems to be higher than that in the continuum, 
moving the line photosphere to a larger radius. Such effect may be enhanced in case of
ejecta/CSM interaction. We noticed that  in the case of  \dd\/ in which evidence for interaction is stronger, the discrepancy is  larger, $\sim3000$ \kms
(blue dashed line in Fig.~\ref{fig:mod09dd}). 

Taking in mind these caveats, we may elaborate on the physical parameters for the modelled SNe. 
The parameters of SN 2010aj (moderate ejecta mass, low amount of \ni, and low explosion energy), may be consistent  with a scenario of explosion and mass loss from a massive super-asymptotic giant  branch\footnote{We remind that these stars, after having ignited H-, He-, and C-burning, are able  to form a degenerate Neon-Oxygen core \citep[e.g.][and references therein]{pumo06,pumo07}, where the 
physical conditions can be suitable for triggering electron-capture reactions on $^{24}$Mg and 
other nuclei which are present in trace amounts in the core, leading to a so-called electron-capture 
supernova (EC-SN) event \citep[e.g.][and references therein]{pumo09}.} (SAGB) stars with an 
initial (ZAMS) mass close to the upper limit for this class of stars \citep[$\sim 11$\M; see][for details]{pumo09}. 
%MLP scrive: It is to note that such value is also consistent with that inferred from the ratio $R$ (cfr. Sect.~\ref{sec:interact}).
% MT: non abbiamo nessuna determinai di R x 2010aj !!??
For SN 1995ad, the values of the inferred parameters may be consistent with both SAGB progenitors and Fe CC 
progenitors with initial masses close to the lower limit. Both scenarios may be plausible, even if the 
former poorly explains the relatively high amounts of \ni\/ \citep[compared to what expected in SNe from SAGB 
progenitors; e.g.][]{wa09} and the latter may not account for the relatively low ejecta 
mass of \ad\/ and the CO molecules observed during the nebular phase  \citep{sp96}. 

In this context it can be useful to compare the results of the SN data modelling with those derived from an
independent diagnostic. 
In nebular spectra of CC-SNe, the flux ratio $R =$ [Ca~{\sc ii}] $\lambda\lambda$7291,7324 / [O~{\sc i}] $\lambda\lambda$6300,6364 is a 
useful tool for guessing the mass of the core and, consequently, that of the progenitor. 
As a genearal trend, small $R$ ratios correspond to higher core masses and, hence, to higher main sequence masses \citep{fr89,fr87}.
%and {\bf hence to higher envelope masses} as recently confirmed in the model of \citet{an12}. 
For the two peculiar type IIP SNe 1987A and 2005cs the $R$ ratios are $R\sim3$ \citep{elm03} and $R\sim4.2$
\citep{pa09}, respectively, while for the canonical type IIP SNe 1992H and 1999em we find $R\sim1.6$ and $R\sim4.7$, respectively. 
These measurements would suggest higher core masses for SNe 1987A and 1992H, and lower masses for SNe 2005cs and 1999em.
In addition, other indicators (e.g. SN data modelling, study of the progenitors in pre-explosion images) suggest that the two former 
objects have M$_{\rm ZAMS}$ of about 20 \M\/, whilst the latter (SNe 2005cs and 1999em) have lower mass precursors \citep[$\sim$10 \M,][]{sm09a}.  
We estimated the ratio $R$ also for \ad\/ and \ww,  obtaining $R\sim4.0$ and $R\sim2.3$, respectively. This would point toward a somewhat more massive
progenitor for  \ww\/, and a lower mass for the precursor of \ad\/, that would be quite similar to those of SNe 2005cs and 1999em, and
in good agreement with that indicated by the above modelling.
We have to remark, however, that these measurements have to be taken with caution, since 
 \citet{mag12} have recently questioned the robustness of the relation between the progenitor mass 
and the observed flux ratio $R$.%, because these lines have significant contributions from primordial %abundances. 

\section{Conclusions}\label{sec:final}

In this paper we have presented photometric and spectroscopic observations of three recent type II SNe (2007pk, 2009dd and 2010aj) and previously unpublished data of SNe 1995ad and 1996W. Together with SNe 2007od and 2009bw  \citep[recently studied by our group,][]{07od, 09bw} they belong to a group of moderately bright type II SNe.  
%{\bf A reasonable understanding of the early photospheric evolution} of these transients requires good timing because of the short period of visibility of the spectral features linked to the interaction. For this reason, we may consider the temporal coverage of our supernova sample not ideal. {\bf Although the full sample shows a few different properties, we noticed that the objects share a bright peak and plateau. This similarity allowed us to threat them as a small group of H-rich CC-SNe and provide a reasonable complete picture of their photometric and spectroscopic evolution.} 

The light curves show peak magnitudes between M$_{B}$~$\leq-16.95$ and durations of the plateau, between 30 and 70 days.
At late times three objects follow the decline rates of \co\/ to \fe, allowing us to determine the ejected masses of \ni\/ (between 0.028 and 0.140 \M). For another object, \aj, the observations ended before the final onset of the radioactive tail, thus allowing us to derive only an upper limit M(\ni)$<0.007$ \M. No late-time observations are available for \pk.

With the noticeable exception of \pk\/ during the earliest epochs,
the spectra of all the objects evolve like normal type II SNe in terms of spectral features, expansion velocities, and  blackbody temperatures.
 \pk\/ was somewhat different, showing after the discovery a featureless blue continuum, resolved narrow Balmer line in emission, significant X-ray flux and UV excess, 
all pointing toward relatively strong ejecta-CSM interaction. After one month the evidences of interaction ceased,  and the SN followed afterward a normal evolution.
\pk\/ reached the brightest absolute magnitude (M$_B=-18.70$)
among the objects studied here and showed the fastest luminosity decline, intermediate between SNe IIL and SNe IIP.

Hints of interaction with the CSM have been found for other two objects, \dd\/ and \ww. 
\dd\/ was detected by Swift XRT in the 0.2--10 keV range, at a luminosity lower than that of \pk. It also showed an high UV contribution to the bolometric luminosity. 
%A notch on the blue side of \Ha\/ could be associated to an HV feature formed in the CDS behind %the reverse shock due to the interaction with a RSG wind, in analogy to those found in SN 2004dj %and SN 1999em \citep{chu07}.
%Features due to interaction of the SN ejecta with a standard RSG wind, have been identified on the %blue wings of \Ha\/ and \Hb\/ of \ww. 
SN~1996W was photometrically and spectroscopically a normal SN IIP. 
The extended coverage at late time has allowed us to derive an ejected M(\ni)$=0.14$ \M\/ for this object, which is the largest in our sample.

The remaining two objects, \aj\/ and \ad, did not show evident signs of interaction. \aj\/ showed the largest drop from the plateau and a decline rate in the earliest part of the radioactive tail that may be attributed to early dust formation. Unfortunately, the available observations can not prove the dust formation hypothesis. 
If the observed luminosity at about 100d is governed only by the \ni\/ decay, the ejected mass of \ni\/ is M(\ni)$\leq0.007$ \M.
\ad\/ has been extensively observed both at early and late time and behaved like a normal non-interacting SN~II. Dust formation has been suggested  at  late times (t$>200d$) because of an increase in the slope of the light curve and of the detection of CO.

The modeling with our code (Sect.~\ref{sec:m})  has allowed the determination of the main parameters of the explosions and of the progenitors for SNe 2009dd, 2010aj and 1995ad, (reported in Tab.~\ref{table:main}).
The masses of the ejecta are in the range of 5.0--9.5 \M, corresponding to total masses at the moment of the explosion of the order of 7.0--11.5 \M. 
A low mass progenitor for \ad\/ is also indicated by the high value of flux ratio $R$ of [Ca~{\sc ii}] over [O~{\sc i}] (Sect.~\ref{sec:interact}).
Despite some discrepancy with the fits (see Sect.~\ref{sec:m}), the results
are consistent with either SNe from SAGB stars or RSGs exploded as Fe CC-SNe.
The SAGB scenario seems more appropriate for \aj\/ because of the moderate ejected mass, the low amount of  \ni\/ and the low explosion energy. 
%The possible dust formation at the end of the plateau is also consistent.
Both progenitor scenarios are equally plausible for the other two objects.
While there is evidence of interaction for \dd, some weak interaction for \ad\/ and \aj\/ could explain the disagreement found in the velocity evolution with respect to the theoretical models.

Our models could not be applied to \pk\/ and \ww\/ because the available data do not cover the crucial feature of the light curves, i.e. the transition from the plateau to the radioactive tail that marks the end of the recombination phase. Observations of \pk\/ cover the first few months when the object transformed from a strongly interacting type IIn with a $linear$ light curve to a normal type II. Nothing can be said on the nature of the progenitor because the interaction with the CSM is governed by recent episodes of mass-loss.
Indication on the mass of the progenitor of \ww\/ comes from the large ejected \ni\/ mass ($0.14$\M) and from the flux ratio of [Ca~{\sc ii}] over [O~{\sc i}], both pointing to a relatively large initial mass ($15-20$\M). 

%Other objects with similar moderately bright luminosity have been used here as reference.
Altogether these objects  share the common property of having relatively bright peak luminosity, which is not -anyway-  as extreme as the peculiar SN~2009kf \citep{bot10} that, indeed, pointed toward different explosion mechanisms.
The light curves of these mildly luminous objects can be different as to shape (linear vs. plateau), drop to the radioactive tail and late time luminosity. In addition, their spectral characteristics show some 
individuality, e.g. transition from type IIn to normal type II or peculiar line profiles. In a few cases, thanks to a deeper investigation, the presence of HV features or dust formation has been revealed. 
%All exploded in locations with nearly solar metallicity.
All of this indicates a significant heterogeneity in luminous SNe~II, and the only possible common property seems to be some minor signs of weak ejecta-CSM interaction.

\section*{Acknowledgments}
We thank an anonymous referee for the useful suggestion that help us for improving the paper.
C.I., S.B., F.B., E.C. and M.T. are partially supported by the PRIN-INAF 2011 with the project "Transient Universe".
M.L.P. acknowledges the financial support by the Bonino-Pulejo Foundation 
and from the contract ASI-INAF n. I/009/10/0 (CUP:F71J10000020005). The TriGrid VL project, the {\it ``consorzio COMETA''}  and the INAF - Padua Astronomical Observatory are also acknowledged for computer facilities.
This work is partially based on observations of the European supernova
collaboration involved in the ESO-NTT large programme 184.D-1140 led by
Stefano Benetti.
F.B. acknowledges support from FONDECYT through Postdoctoral grant 3120227 and
from the Millennium Center for Supernova Science through grant P10-064-F (funded by ``Programa Bicentenario de Ciencia y Tecnolog\`ia de CONICYT" and ``Programa Iniciativa Cient\`ifica Milenio de MIDEPLAN")
S.T. acknowledges support by the  Transregional Collaborative Research Cebter TRR33 \textquotedblleft The Dark Universe" of the German Research Fondation (DFG).
N.E.R. acknowledges support by the MICINN grant AYA11-24704/ESP, by the ESF EUROCORES Program EuroGENESIS (MICINN grant EUI2009-04170), by SGR grants of the Generalitat de Catalunya, and by
EU-FEDER funds.
%We thank the support astronomers at the Telescopio Nazionale Galileo, the Copernico Telescope, the 2.2m Telescope at Calar Alto, the Liverpool Telescope, the Nordic Optical Telescope, the ESO telescopes, the Hale Telescope on Palomar Observatory and the Danish Telescope for performing the follow-up observations of the SNe presented. 
This research has made use of the NASA/IPAC Extragalactic Database (NED) which is operated by the Jet Propulsion Laboratory, California Institute of Technology, under contract with the National Aeronautics and Space Administration. 
We acknowledge the usage of the HyperLeda database (http://leda.univ-lyon1.fr). We also thank the High Energy Astrophysics Science Archive Research Center (HEASARC), provided by NASA's Goddard Space Flight Center, for the SWIFT data.

\bibliographystyle{aa}

\Online

\begin{appendix}
\section{Online material}
\begin{table*}
 %\begin{minipage}{140mm}
  \caption{Magnitudes of the local sequence stars in the field of SN 2009dd (cfr. Fig.~\ref{fig:09dd}). The errors in brackets are the r.m.s.}
 \begin{center}
  \begin{tabular}{cccccc}
  \hline
  \hline
   ID & U & B & V & R & I \\
 \hline
 1 & 16.24 (.07) & 17.01 (.02) & 16.34 (.01) & 16.11 (.01) & 15.76 (.02)  \\
 2 & 15.64 (.05)  & 15.23 (.01) & 14.26 (.02) & 13.63 (.01) & 13.19 (.01) \\
 3 &  -& 19.07 (.02) & 17.72 (.02) & 16.89 (.01) & 16.28 (.02) \\
 4 & 16.97 (.01) &  16.41 (.02) & 15.48 (.02) & 15.04 (.01) & 14.59 (.01) \\
 5 & -& 19.75 (.02) & 19.09 (.02) & 18.68 (.02) & 18.48 (.03) \\
 6 & -& 19.91 (.03) & 19.06 (.02) & 18.32 (.02) & 17.77 (.02) \\
 7  &  -& 21.07 (.03) & 19.60 (.02) & 18.39 (.01) & 16.84 (.02)  \\
 8 &  -& 20.47 (.02) & 19.57 (.02) & 18.99 (.02) & 18.40 (.02)\\
 9 &  18.58( - ) & 17.66  (.02) & 16.25 (.03) & 15.61 (.01) & 14.96 (.02) \\
 10 & 19.80 ( - ) & 19.00 (.02) & 17.95 (.03) & 17.52 (.01) & 17.12 (.02) \\
 \hline
\end{tabular}
\end{center}
\label{table:ls09dd}%\end{minipage}
\end{table*}

\begin{table*}
 %\begin{minipage}{140mm}
  \caption{Magnitudes of the local sequence stars in the field of SN 2007pk (cfr. Fig.~\ref{fig:07pk}).}
 \begin{center}
  \begin{tabular}{cccccc}
  \hline
  \hline
   ID & U & B & V & R & I \\
 \hline
 1 & 17.00 (.02) & 17.11 (.02) & 16.53 (.02) & 16.21 (.02) & 15.82 (.02)  \\
 2 & 19.12 (.02) & 18.61 (.03) & 17.71 (.01) & 17.08 (.03) & 16.70 (.02) \\
 3 & 17.26 (.03) & 17.08 (.02) & 16.34 (.02) & 15.80 (.03) & 15.52 (.02) \\
 4 & 14.70 (.02) &  14.79 (.02) & 14.35 (.03) & 13.85 (.01) & 13.54 (.02) \\
 5 & 16.16 (.02) & 16.20 (.02) & 15.56 (.02) & 15.22 (.03) & 14.77 (.03) \\
 6 & 18.22 (.02) & 17.89 (.03) & 16.95 (.02) & 16.56 (.03) & 16.13 (.02) \\
 7  & 15.81 (.03) & 15.86 (.03) & 15.26 (.01) & 14.93 (.02) & 14.69 (.01)  \\
 8 &  15.05 (.02) & 14.40 (.03) & 13.42 (.03) & 12.84 (.02) & 12.38 (.01)\\
 9 &  17.14 (.02) & 16.92  (.02) & 16.41 (.02) & 16.13 (.02) & 15.86 (.01) \\
 10 & 17.54 (.02) & 17.48 (.04) & 16.79 (.01) & 16.45 (.02) & 16.14 (.02) \\
 11 & 17.40 (.02) & 17.40 (.03) & 16.68 (.01) & 16.37 (.02) & 16.18 (.05)\\
 12 & 20.50 (.03) & 19.42 (.03) & 18.20 (.02) & 17.47 (.03) & 16.82 (.02)\\
 13 &  19.20 (.03) & 18.76 (.02) & 17.90 (.01) & 17.42 (.02) & 16.90 (.03) \\
 \hline
\end{tabular}
\end{center}
\label{table:ls07pk}%\end{minipage}
\end{table*}

\begin{table*}
 %\begin{minipage}{140mm}
  \caption{As Tab.~\ref{table:ls09dd} but for SN 2010aj (cfr. Fig.~\ref{fig:10aj}).}
 \begin{center}
  \begin{tabular}{cccccc}
  \hline
  \hline
   ID & U & B & V & R & I \\
 \hline
 1 & 19.14 (.06) & 18.52 (.03) & 17.62 (.01) & 16.99 (.01) & 16.48 (.01)  \\
 2 & 17.72 (.06)  & 17.68 (.02) & 17.09 (.04) & 16.80 (.01) & 16.42 (.02) \\
 3 &  19.00 ( - ) & 17.91 (.05) & 16.39 (.02) & 16.31 (.01) & 16.07 (.02) \\
 4 & 18.72 ( - ) &  17.74 (.04) & 16.58 (.04) & 15.94 (.01) & 15.37 (.01) \\
 5 & 16.90 (.01) & 16.77 (.01) & 16.05 (.02) & 15.62 (.02) & 15.24 (.01) \\
 6 & 14.97 ( - ) & 15.03 (.02) & 14.48 (.01) & 14.17 (.01) & 13.81 (.01) \\
 7  & 18.89 ( - ) & 18.72 (.01) & 17.69 (.07) & 17.26 (.01) & 16.66 (.07)  \\
 8 & 18.95 ( - ) & 19.27 (.02) & 18.66 (.02) & 18.37 (.01) & 17.83 (.08)\\
 9 &  18.87 ( - ) & 18.06  (.02) & 17.35 (.05) & 16.76 (.01) & 16.64 (.01) \\
 10 & 17.91 ( - ) & 18.44 (.04) & 17.92 (.01) & 17.63 (.01) & 17.01 (.08) \\
 \hline
\end{tabular}
\end{center}
\label{table:ls10aj}%\end{minipage}
\end{table*}

\begin{table*}
 %\begin{minipage}{140mm}
  \caption{Magnitudes of the local sequence stars in the field of \ad\/ (cfr. Fig.~\ref{fig:95ad}).}
 \begin{center}
  \begin{tabular}{cccccc}
  \hline
  \hline
   ID & B & V & R & I \\
 \hline
 1 &  17.92 (.02) & 17.31 (.01) & 16.96 (.01) & 16.65 (.01)  \\
 2 &  17.98 (.01) & 17.61 (.01) & 17.28 (.01) & 16.89 (.01) \\
 3 &  16.25 (.01) & 15.29 (.01) & 14.78 (.01) & 14.29 (.01) \\
 4 &   15.88 (.01) & 15.39 (.01) & 15.11 (.01) & 14.78 (.01) \\
 5 &  17.93 (.02) & 17.33 (.01) & 16.97 (.01) & 16.62 (.04) \\
 6 &  18.63 (.02) & 18.08 (.01) & 17.74 (.01) & 17.43 (.01) \\
 7  &  19.21 (.06) & 17.72 (.08) & 16.56 (.07) & 15.17 (.06)  \\
 8 &   18.80 (.02) & 18.30 (.07) & 18.01 (.05) & 17.73 (.02)\\
 9 &   19.59  (.04) & 18.14 (.08) & 17.23 (.05) & 16.32 (.04) \\
 11 &  17.78 (.03) & 17.16 (.05) & 16.80 (.03) & 16.40 (.07)\\
 12 &  16.74 (.01) & 16.18 (.05) & 15.86 (.06) & 15.47 (.10)\\
 \hline
\end{tabular}
\end{center}
\label{table:ls95ad}%\end{minipage}
\end{table*}

\begin{table*}
 %\begin{minipage}{140mm}
  \caption{Magnitudes of the local sequence stars in the field of \ww\/ (cfr. Fig.~\ref{fig:96w}).}
 \begin{center}
  \begin{tabular}{cccccc}
  \hline
  \hline
   ID & U & B & V & R & I \\
 \hline
 2 & 19.18 (.02) & 18.28 (.01) & 17.29 (.01) & 16.71 (.01) & 16.22 (.01)  \\
 3 & 16.90 (.04)  & 15.94 (.01) & 14.90 (.01) & 14.28 (.01) & 13.76 (.01) \\
 4 &  19.08 (.02) & 18.94 (.02) & 18.14 (.01) & 17.76 (.01) & 17.39 (.02) \\
 5 & 19.22 (.02) &  19.36 (.03) & 18.77 (.02) & 18.40 (.01) & 18.04 (.02) \\
 6 & 20.03 (.03) & 19.88 (.04) & 19.29 (.03) & 18.83 (.01) & 18.47 (.02) \\
 \hline
\end{tabular}
\end{center}
\label{table:ls96w}%\end{minipage}
\end{table*}

\begin{table*}
  \caption{Journal of spectroscopic observations of the SNe in our sample.}
  \begin{center}
  \begin{tabular}{cccccc}
  \hline
  \hline
  Date & JD & Phase$^{*}$ & Instrumental $^\diamond$& Range & Resolution$^{\dagger}$ \\
  & +2400000 & (days) & configuration &(\AA) & (\AA) \\
 \hline
 \multicolumn{6}{c}{\dd}\\
 \hline
 09/04/14 & 54936.7 & 11.2 & NOT+ALFOSC+gm4 & 3480-7500 & 13\\
 09/04/16 & 54938.4 & 12.9 & CAHA+CAFOS+b200 & 3400-8700 & 11.3\\
 09/05/20	& 54971.6 & 46.1 & TNG+DOLORES+LRB,LRR & 3700-9220 & 15\\
 09/07/20	& 55033.4 & 107.9 & TNG+DOLORES+LRB,LRR & 3800-10100 & 15\\
 09/11/19 & 55155.7 & 230.2 & CAHA+CAFOS+g200 & 4000-9700 & 9.5 \\
 09/11/21 & 55157.7 & 232.2 & TNG+DOLORES+LRR & 5100-9300 & 10.3\\
 10/05/18 & 55334.5 & 409.0 & TNG+DOLORES+LRR & 5030-9270 & 9.8\\
 \hline
 \multicolumn{6}{c}{\pk}\\
 \hline
  07/11/11 & 54416.3 & 4.3 & Copernico + AFOSC+gm4 & 3650-7800 & 24\\
 07/11/12 & 54417.5 & 5.5 & Copernico + AFOSC+gm2 & 5320-9080 & 36\\
 07/11/16 & 54421.4 & 8.4 & TNG+DOLORES+LRB,LRR & 3400-9000 & 14 \\
 07/12/05 & 54439.5 & 27.5 & Copernico +AFOSC+gm4,gm2 & 3700-9050 & 25  \\
 07/12/28 & 54463.3 & 51.3 & Copernico +AFOSC+gm4,gm2 & 3780-9180 & 23 \\
 08/01/11 & 54477.4 & 65.4 & NOT+ALFOSC+gm4 & 3580-9120 & 14\\
 08/01/12 & 54478.4 & 66.4 & TNG+NICS+IJ & 8660-13480 & 18\\
 08/01/28 & 54494.3 & 82.3 & Copernico +AFOSC+gm4 & 3670-7770& 23\\
 08/02/01 & 54497.4 &  85.4 & WHT+ISIS+R300B,R158R & 3500-9800 & 10 \\
 08/09/05 & 54714.5 & 302.5 & TNG+DOLORES+LRR & 5150-10230& 16\\
 08/10/02 & 54741.7 & 329.7 & Palomar+DBSP+red & 5800-9990& 17\\
 \hline
 \multicolumn{6}{c}{\aj}\\
 \hline
  10/03/30 & 55287.6 & 22.1 & WHT+ISIS+R300B,R158R & 3130-11130 & 5.4-6.3\\
 10/04/17 & 55302.8 & 37.8 & NTT+EFOSC2+gm11 & 3800-7500 & 13 \\
 10/04/24 & 55311.5 & 46.0 & TNG+DOLORES+LRB,LRR & 3380-9900 & 10\\
 10/05/21 & 55338.5 & 73.0 & TNG+DOLORES+LRB,LRR & 3500-9700 & 10\\
 11/02/12 & 55604.8 & 338.8 & NTT+EFOSC2+gm13 & 3650-9300 & 17\\
 \hline
 \multicolumn{6}{c}{\ad}\\
 \hline
  95/09/29 & 49989.8 & 8.8 & ESO 1.5+B$\&$C+gr2 & 3930-7790 & 5 \\
 95/10/02 & 49992.9 & 11.9 & ESO 3.6+EFOSC1+B300,R300& 3750-9920 & 14+17 \\
 95/10/14 & 50004.9 & 23.9 & ESO 3.6+EFOSC1+B300& 3750-6940 & 19\\
 95/11/19 & 50040.8 & 59.9 & ESO 1.5+B$\&$C+gr2& 3100-10710 & 10 \\
 95/11/20 & 50041.8 & 60.8 & ESO 1.5+B$\&$C+gr2& 3100-10710 & 14 \\
  95/12/26 & 50077.8 & 96.8 & MPG-ESO 2.2+EFOSC2+gr3,gr5,gr1& 3720-9040 & 11+11+60\\
   95/12/29 &50080.8 & 99.8 & ESO 3.6+EFOSC1+B300,R300& 3740-9910 & 14+17\\
    96/01/21 & 50103.7 & 122.7 & ESO 1.5+B$\&$C+gr2& 3460-11100 & 9 \\
      96/02/18 & 50131.7 & 150.7 & MPG-ESO 2.2+EFOSC2+gr5,gr6& 3350-8980 & 11+11\\
      96/02/19 & 50132.7 & 151.7 & ESO 1.5+B$\&$C+gr2& 3040-10000 & 9 \\
      96/02/20 & 50133.7 & 152.7 & ESO 1.5+B$\&$C+gr2& 3040-10000 & 9 \\
       97/02/11 & 50490.6 & 509.6 & ESO 3.6+EFOSC1+R300& 6000-9850 & 23\\
    \hline
 \multicolumn{6}{c}{\ww}\\
 \hline
 96/04/18 & 50191.6 & 11.6 & ESO 1.5+B$\&$C+gr2 & 3160-10650 & 9 \\
 96/04/19 & 50192.6 & 12.6 & ESO 1.5+B$\&$C+gr2 & 3160-10650 & 9 \\
 96/06/21 & 50195.4 & 15.4 & Copernico+B$\&$C+150tr. & 4250-8600 & 29\\
 96/04/24 & 50198.4 & 18.1 & ESO 1.5+B$\&$C+gr2 & 3130-10430 & 9 \\
 96/04/25 & 50198.6 & 18.5 & ESO 1.5+B$\&$C+gr2 & 3130-10430& 9 \\
  96/05/09 & 50212.5 & 32.5 & ESO 1.5+B$\&$C+gr2 & 3120-10590& 9\\
   96/05/12 &50215.6 & 35.6 & ESO 1.5+B$\&$C+gr2  & 3120-10590 & 9\\
    96/05/19 & 50222.6 & 42.6 & MPG-ESO 2.2+EFOSC2+gr5,gr6 & 2850-9050 & 7+7 \\
      96/06/12 & 50247.5 & 67.5 & ESO 1.5+B$\&$C+gr2 & 2900-10460 & 9\\
      96/12/15 & 50432.8 & 252.8 & MPG-ESO 2.2+EFOSC2+gr6 & 3840-7980 & 10 \\
      97/01/30 & 50478.5 & 298.5 & Danish 1.54+DFOSC+gr5 & 5000-10180 & 8 \\
       97/02/13 & 50492.7 & 312.7 & MPG-ESO 2.2+EFOSC2+gr5 & 5220-9280 & 9\\
     \hline
\end{tabular}
\begin{flushleft}
 * with respect to the explosion epochs (cfr. Tab.~\ref{table:main})\\
$^\diamond$ coded as in Tab.~\ref{table:cht}\\
 $\dagger$ measured from the full-width at half maximum (FWHM) of the night sky lines\\
 \end{flushleft}
\end{center}
\label{table:sp}
\end{table*}

\end{appendix}
\end{document}